\documentclass[a4paper,11pt]{article}
\usepackage{amsfonts}

\usepackage{graphicx}
 \graphicspath{{Figures/}}

\usepackage{amsmath}
\usepackage{amssymb}
\usepackage{latexsym}
\usepackage[colorlinks]{hyperref}
\usepackage{color}
\usepackage{float}
\usepackage{cite}
\usepackage{makeidx}
\usepackage{colortbl}
\usepackage{csquotes}
\usepackage[squaren]{SIunits}

\usepackage[textfont={footnotesize,sf},labelfont={color=blue,bf,sf},labelsep=endash]{caption}
\usepackage[position=top,labelfont={color=blue,bf,sf}]{subfig}
\usepackage{bm}

\usepackage[title]{appendix}

\newcommand{\bra}{\begin{array}}
    \newcommand{\era}{\end{array}}
\newcommand{\beq}{\begin{equation}}
\newcommand{\eeq}{\end{equation}}
\newcommand{\bqr}{\begin{eqnarray}}
\newcommand{\eqr}{\end{eqnarray}}

\def\BC{\bb C}
\def\_\BC{\bbi C}

\def\no2 {{\textstyle{n\over 2}}}

\begin{document}
    \begin{titlepage}
        \setcounter{page}{1}
        \renewcommand{\thefootnote}{\fnsymbol{footnote}}

        \begin{flushright}
        \end{flushright}

        \vspace{5mm}
        \begin{center}
{\Large \bf {Effect of Strain on Band Engineering in Gapped
Graphene}}

\vspace{6mm} {\bf Hasna Chnafa}$^{a}$, {\bf Miloud Mekkaoui}$^{a}$, {\bf Ahmed Jellal\footnote{\sf a.jellal@ucd.ac.ma}}$^{a,b}$ and {\bf Abdelhadi Bahaoui}$^{a}$

\vspace{5mm}

{$^a$\em Laboratory of Theoretical Physics,  
    Faculty of Sciences, Choua\"ib Doukkali University},\\
{\em PO Box 20, 24000 El Jadida, Morocco}

{$^b$\em Canadian Quantum Research Center, 204-3002 32 Ave Vernon},\\{\em BC V1T 2L7, Canada}

\vspace{3cm}

\begin{abstract}
We study the effect of  strain on the band engineering in gapped graphene subject to external sources. By applying the Floquet theory, we determine the effective Hamiltonian of electron dressed by a linearly, circularly and an elliptically polarized dressing field in the presence of strain  along armchair and zigzag directions. Our results show that the energy spectrum exhibits different symmetries and {for the strainless case it takes an isotropic and anisotropic forms whatever the values of irradiation intensity}, whereas it is linear as in the case of pristine graphene. It {increases} slowly when strain is applied along the armchair direction but {rapidly} for the zigzag case. Moreover, it is found that the renormalized band gap changes along  different strain magnitudes and does not change for the polarization phase $\theta$ compared to linear and circular polarizations where its values change oppositely.

    \vspace{5cm}

    \textbf{\noindent PACS numbers}: 72.80.Vp, 73.21.-b, 
    71.10.Pm, 03.65.Pm

    \textbf{\noindent Keywords}: Graphene, strain, Floquet theory,  energy spectrum, band gap.
\end{abstract}

\end{center}

\end{titlepage}

\section{Introduction}

The physics of low energy carriers in graphene is governed by a Dirac like-Hamiltonian and carriers are massless fermions having a linear dispersion relation in momentum space \cite{ref5}. Graphene has many electronic and mechanical properties \cite{ref8}, such as Hall effect \cite{ref9,ref11}, Klein tunneling \cite{ref13}, elastic strain engineering
\cite{ref14,ref15,ref16,ref17}, which would be too many to list. Graphene is considered as 
a gapless semiconductor  and some methods have been used to create band gap. Experimentally, it has been shown one can generate a gap 
by depositing graphene on substrate hexagonal boron nitride to have a gap of order $\sim 100$~\text{meV} \cite{ref18,ref19}. Moreover, the electronic properties of graphene based nanostructures can be adjusted by distorting a
deformation on the graphene sample \cite{ref20,ref21,ref22,ref23}. Indeed, since its discovery researchers have conducted extensive research on the influence of elastic strain on mechanical and physical properties of graphene \cite{ref24a,ref24b}. It showed that 
graphene has an effective young's modulus  and simultaneously can reversibly support elastic strain up to $25\%$ \cite{ref24}. It is found that  the mechanical strains in graphene can  change Dirac points, which causes  Dirac fermions to have asymmetrical effective Fermi velocities $v_{x}\neq v_{y}$ \cite{ref16,a5,a8}.

On the other hand, controllable quantum systems can be realized using  external fields \cite{111, 122} or mechanical deformations
\cite{444,144} allowing to generate novel states of matter. These can be described by  effective Hamiltonian based on the Floquet theory of periodically driven quantum systems. Additionally, the interaction between electron and electromagnetic field gives new physics that changes the electronic properties of a driven system. Such coupling actually is known as
electron dressed by field  or simply dressed electron \cite{666} and has been studied in different occasions. Indeed, the physical
properties of dressed electrons were studied in various systems such that quantum wells \cite{888,1333}, quantum rings \cite{1666,1777} and graphene \cite{1888,2000, 2333,2444} as well as others. 

Motivated by the results obtained in
\cite{Kibis}, we theoretically investigate the electron-field interaction in gapped graphene subject to the tensional strain within the minimal coupling approach. By applying the Floquet theory \cite{a11}, we end up with an
effective Hamiltonian as function of strain for the linearly, circularly and elliptically polarized dressing fields. The solutions of energy spectrum are separately obtained by solving  Dirac equation for the three considered dressing fields. Subsequently, we numerically study the effect of strain along armchair and zigzag directions on the energy spectrum as well as the 
renormalized electronic band gap for different values of the irradiation intensity $I \sim E_0^2$, with $E_0$ being the electric field. 
Consequently, we show that the  effect of strain  causes some changes 
on the energy spectrum and band gap along the armchair direction, but it produces remarkable influence along the zigzag direction. We conclude that, the energy spectrum can be controlled by adjusting the strain amplitude and $I$.

The present paper is organized as follows. In section 2, we present a theoretical model describing a gapped graphene subject to external sources. In section 3, we explicitly determine the solutions of  energy spectrum using Schr\"{o}dinger equation. To do,  we  apply  the Floquet theory of quantum system driven by an oscillating fields to obtain the effective Hamiltonian as function of strain, band gap. Then we calculate the energy
spectrum of {electron dressed} by a linearly, circularly and elliptically polarized electromagnetic wave in terms of the physical parameters characterizing our system. To give a better understanding, we numerically analyze and discuss our results under suitable conditions
in section 4. Our conclusions are given in the final section.

\section{Theoretical model}
\begin{figure}[H]
		\qquad\qquad\quad \centering
		\includegraphics[scale=0.45]{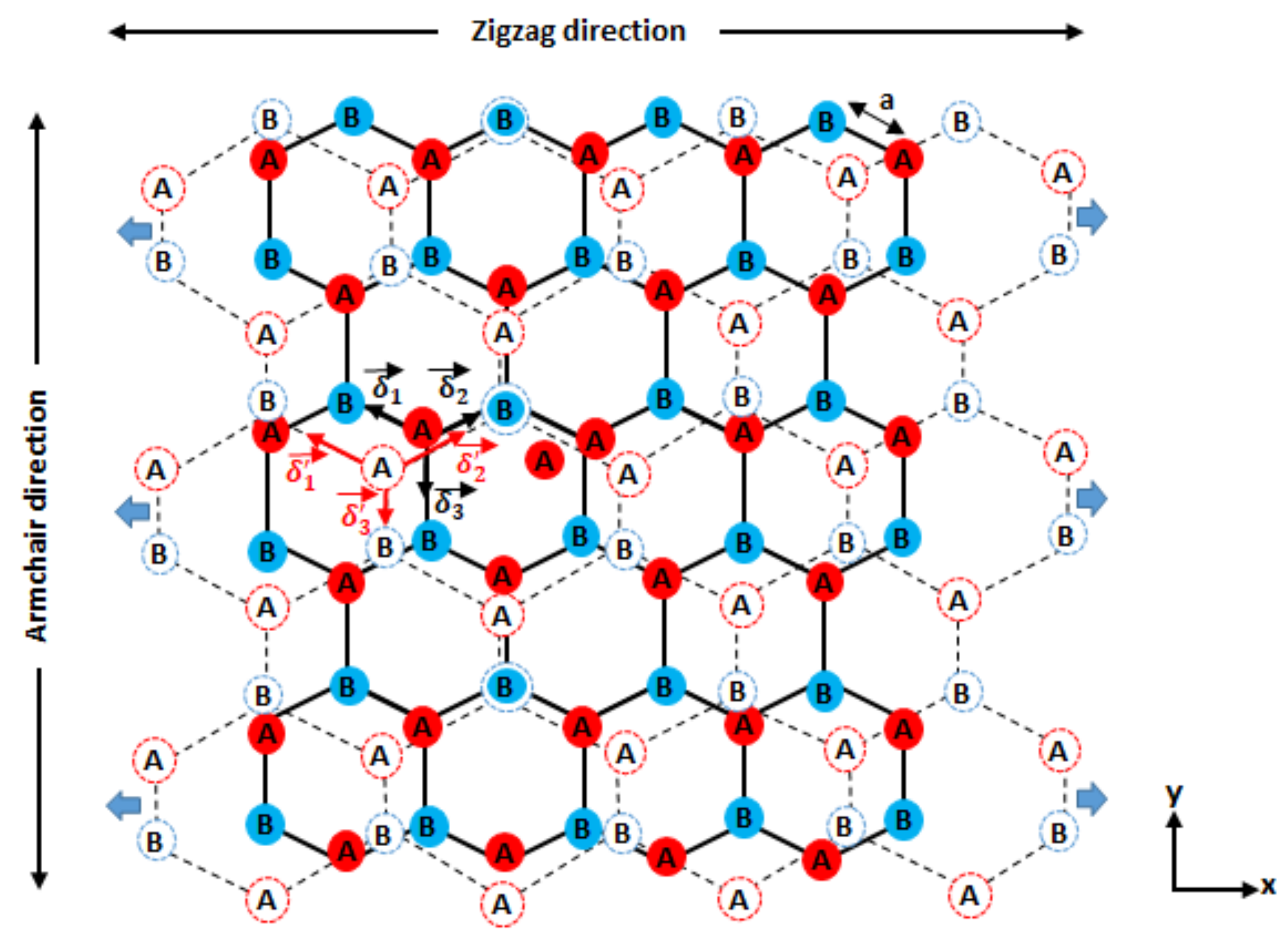}
	\caption{\sf{(color online) {Structural deformations of graphene for tensile strain along zigzag ($x$-axis) and armchair ($y$-axis) directions}.}}\label{db.5}
\end{figure}

{To do our task}, we study the effect of a tensional strain in gapped graphene {along armchair and zigzag directions} illuminated by a continuous wave propagate along the $z$-axis with frequency $\omega$ as shown in Figure \ref{db.5}. The electromagnetic wave can be neither absorbed nor emitted by the electrons and considered as a dressing field \cite{Kibis}. {Figure \ref{db.5} presents the graphene atomic with the solid and dashed circles denote sublattices $A$ (red) and $B$ (blue) in undeformed and deformed configurations, respectively, representing three nearest neighbor vectors ${\delta}_{i}$ and ${\delta}^{'}_{i}$ with $(i=1,2,3)$}. When armchair and zigzag directions is under tension and for small strain, ${\delta}^{'}_{i}$ can be written as
\begin{eqnarray}\label{xtzr}
&&|\delta_{1}^{'}|_{(A)}=|\delta_{2}^{'}|_{(A)}=a\left( 1-\frac{3}{4}\sigma S+\frac{1}{4}S\right), \qquad |\delta_{3}^{'}|_{(A)}=a\left( 1+S\right)\\\label{xtt}
&&|\delta_{1}^{'}|_{(Z)}=|\delta_{2}^{'}|_{(Z)}=a\left(
1+\frac{3}{4} S-\frac{1}{4}\sigma S\right), \qquad
|\delta_{3}^{'}|_{(Z)}=a\left( 1-\sigma S\right).
\end{eqnarray}  where the Poisson ratio is $\sigma\approx 0.165$ for graphene, $S$ is the strain and $a$ is the distance between neighboring atoms. {In the tight binding approximation, the only effect of strain is to modify the altered hopping integral parameter $t_{i}^{'}$} which described by a empirical relation
\begin{eqnarray}\label{xt}
t_{i}^{'}=t_{0}e^{-3.37\left(|\delta_{i}^{'}|/a-1\right)}, \qquad i=1,2,3
\end{eqnarray} due to stretching or shrinking of the distance vectors between the nearest neighbor carbon atoms \cite{ref14} and $t_{0}\approx 2.7$\text{eV} \cite{ref8} is being the hopping energy without deformation. {From (\ref{xtzr}-\ref{xt}) it is clearly seen that $t_{1}^{'}$ as well as $t_{2}^{'}$ change with the same value because  $|\delta_{1}^{'}|=|\delta_{2}^{'}|$ and whatever the value of strain along zigzag direction, $\{t_{1}^{'}=t_{2}^{'}\}$ decreases and $t_{3}^{'}$ increases, i.e. $t_{1}^{'}=t_{2}^{'}<t_{3}^{'}$ but  for strain along armchair direction,  $t_{3}^{'}$ becomes small compared to $t_{1}^{'}$ and $t_{2}^{'}$ \cite{ref14}}.
 By introducing a vector potential $\textbf{A}=(A_x,A_y)$ of the dressing field, the electronic properties of our system can be described by the two-band Hamiltonian 
\begin{equation}\label{xtr}
\mathcal{H}=
\begin{pmatrix}
\frac{\Delta_{g}}{2} & \tau v_{x}(S)\left( p_{x}+{|e|}A_{x}\right) -iv_{y}(S)\left(p_{y}+{|e|}A_{y}\right) \\
\tau v_{x}(S)\left(p_{x}+{|e|}A_{x}\right) +iv_{y}(S)\left(p_{y}+{|e|}A_{y}\right) & -\frac{\Delta_{g}}{2} \\
\end{pmatrix}%
\end{equation}
where $\Delta_{g}$ is the band gap between the conduction and the valence bands, $\textbf{p}=(p_x,p_y)$ is the momentum operator and $\tau=\pm 1$ is the valleys index corresponds to the inequivalent valleys centered at the high-symmetry points $K$ and $K^{'}$.  
The effective Fermi velocities $v_x(S)$ and $v_y(S)$ are tuned by the tensional strain 
\cite{a6,a5,a7,a8} and in our study we distinguish two cases such that the strain is along either armchair direction 
\begin{eqnarray}\label{xtyr}
v_{x}(S)=\frac{\sqrt{3}}{2\hbar}a(1-\sigma
S)\sqrt{4t_{1}^{'2}-t_{3}^{'2}}, \qquad  
v_{y}(S)=\frac{3}{2\hbar}a(1+ S)t_{3}^{'}
\end{eqnarray} or zigzag one 
\begin{eqnarray}\label{xrtr}
v_{x}(S)=\frac{\sqrt{3}}{2\hbar}a\left(1+S\right)\sqrt{4t_{1}^{'2}-t_{3}^{'2}},\qquad
v_{y}(S)=\frac{3}{2\hbar}a(1-\sigma S)t_{3}^{'} 
\end{eqnarray}
In the forthcoming analysis, we fix the introduced potential vector by considering three cases of dressing fields. 
For each case, we will use the Floquet approach
to determine the eigenenergies and eigenspinors.
\section{Electron dressing field}
Our main goal here is to derive  the  solution of energy spectrum  of an electron dressing field by incident light with linear, circular and elliptical polarizations.

\subsection{Linearly polarized dressing field}
We consider in the case of a linearly polarized electromagnetic wave along the $x$-axis, the vector potential $\textbf{A}=\frac{E_0}{\omega} \left(\cos\omega t, 0\right)$
with $E_0$ is the electric field of the
electromagnetic wave and $\omega$ is  its frequency. For this, we write
(\ref{xtr}) as
\begin{equation}\label{H1g}
\mathcal{H}=\left(%
\begin{array}{cc}
\frac{\Delta_{g}}{2} & \tau v_{x}(S) \left( p_{x}+\frac{E_{0}{|e|}}{{\omega}}\cos\omega t\right)-i v_{y}(S) p_{y} \\
\tau v_{x}(S) \left( p_{x}+\frac{E_{0}{|e|}}{{\omega}}\cos\omega t\right)+i v_{y}(S) p_{y} & -\frac{\Delta_{g}}{2} \\
\end{array}%
\right)
\end{equation}
To seek eigenspinors of the full Hamiltonian (\ref{H1g}), we introduce the ansatz
\begin{equation}\label{at3}
\psi=\frac{{\Xi}_{1}(t)}{\sqrt{2}}\left(\begin{array}{cc}
1\\
1 \\
\end{array}
\right)e^{-i\tau \frac{E_{0}v_{x}(S){|e|}}{{\hbar}\omega^{2}}\sin{\omega t}}  + \frac{{\Xi}_{2}(t)}{\sqrt{2}} \left(\begin{array}{cc}
1\\
-1 \\
\end{array}
\right)e^{i\tau \frac{E_{0}v_{x}(S){|e|}}{{\hbar}\omega^{2}}\sin{\omega t}}
\end{equation}
{that can be injected into the Schr\"{o}dinger equation,
	$i\hbar\frac{\partial\psi}{\partial t}=\mathcal{H}\psi$, to obtain two differential equation describing the quantum dynamics of our system in similar way to the case of  when $S = 0$ done in \cite{Kibis}}. Based on the  Floquet theory of periodically driven quantum systems field \cite{a1,a3,a4} and taking into account of such periodicity, we can develop $\Xi_{1,2}(t)$ (\ref{at3}) in Fourier series
\begin{equation}\label{r1}
{\Xi}_{1,2}(t)= e^{-i\frac{\varepsilon}{ {\hbar}}t}  \sum
_{n=-\infty}^{+\infty}{\Xi}_{1,2}^{(n)}e^{i n{\omega t}}.
\end{equation}
{By introducing the Bessel functions $J_{n}(z)$ of the first kind and requiring   
	 the high frequency $\hbar\omega$ assumption \cite{Kibis}, we  show that the eigenvalues for the linear polarization case take the form}
\begin{equation}\label{rt18}
\varepsilon=\gamma\left[\left(\frac{\Lambda_{g}}{2}\right)^{2}+v_{x}^{2}(S){\hbar}^{2}{k_{x}^{2}}+v_{y}^{2}(S)J^{2}_{0}{\left[\frac{2E_{0}v_{x}(S){|e|}}{{\hbar}\color{black}{\omega^{2}}}\right]}{\hbar}^{2}{k_{y}^{2}}\right]^{\frac{1}{2}}
\end{equation}
 {such that  the band gap is the strain amplitude-dependent
 	\begin{equation}
 		\Lambda_{g}(S)=\Delta_{g}J_{0}\left[\frac{2E_{0}v_{x}(S){|e|}}{{\hbar}\omega^{2}}\right]
 	\end{equation} 
 and $\gamma=\pm 1$ is the sign function. Note  that for $S=0$, the velocities $v_x$
 and $v_y$ reduce to the Fermi one and then 
 we recover the eigenvalues obtained in \cite{Kibis}}.

\subsection{Circularly polarized dressing field}
{Let us consider a
dressing field to be circularly polarized along the $x,y$-axes, 
which is
characterized by a vector potential of the form} $\textbf{A}=\frac{E_{0}}{\omega }\left(\cos \xi \omega t,\sin \xi\omega t\right)$  and the chirality index $\xi=\pm 1$ describes the
clockwise/counter-clockwise circular polarizations. {This can be implemented in the Hamiltonian (\ref{xtr}) to obtain}
\begin{align}\label{ru4}
\nonumber\mathcal{{{H}}}=&\left(%
\begin{array}{cc}
\frac{\Delta_{g}}{2} & \tau  v_{x}(S) p_{x}-i v_{y}(S) p_{y} \\
\tau v_{x}(S) p_{x}+i v_{y}(S)p_{y} & -\frac{\Delta_{g}}{2} \\
\end{array}%
\right)\\
& +\frac{E_{0}v_{x}(S){|e|}}{2\omega}\left[\begin{pmatrix}
0 & \tau -\frac{v_{y}(S)}{v_{x}(S)}{\xi}  \\
\tau +\frac{v_{y}(S)}{v_{x}(S)}{\xi}  & 0
\end{pmatrix}e^{i\color{black}{{\omega t}}}+\begin{pmatrix}
0 & \tau+\frac{v_{y}(S)}{v_{x}(S)}{\xi} \\
\tau-\frac{v_{y}(S)}{v_{x}(S)}{\xi}  & 0
\end{pmatrix}e^{-i\color{black}{{\omega t}}}\right].
\end{align} 
{To determine the corresponding eigenvalues one can use  the Floquet-Magnus approach  \cite{a18} to renormalize the time-dependent Hamiltonian (\ref{ru4}) and 
	consider an expansion to the second order of $1/\hbar\omega^{2}$ \cite{Kibis}. This process yields
\begin{equation}\label{rt8}
\varepsilon=
\gamma\left[\left(\frac{\Lambda_{g}}{2}\right)^{2}+v^{2}_{x}(S)\left(1-\frac{E_{0}^{2}{v_{y}^{2}(S)}{|e|}^{2}}{{\hbar}^{2}{\omega^{4}}}\right)^{2} {\hbar}^{2}{k_{x}^{2}}+ v^{2}_{y}(S) \left(
1-\frac{E_{0}^{2}{v_{x}^{2}(S)}{|e|}^{2}}{{\hbar}^{2}{\omega^{4}}}\right)^{2}{\hbar}^{2} {k_{y}^{2}}\right]^{\frac{1}{2}}
\end{equation}
 and the band gap takes the form
	\begin{equation}
		\Lambda_{g}=\Delta _{g}\left[ 1-\frac{E_{0}^{2}v_{x}^{2}(S){|e|}^{2}}{{\hbar}^{2}{\omega^{4}}}\left(1+\frac{v_{y}^{2}(S)}{v_{x}^{2}(S)}\right)\right]-2\tau{\xi} \frac{E_{0}^{2}{v_{x}(S)}{v_{y}(S)}{|e|}^{2}}{\hbar}{\omega^{3}}.
	\end{equation}
	 At this levels we have some comments in order. Firstly, we notice that
	 \eqref{rt8} is encoding more information compared to   \eqref{rt18} because it involves in addition to the electric field amplitude $E_{0}$, the frequency $\hbar\omega$ and valley indices, the chirality of polarization. Of course for strainless case ($S=0$), we reproduce the results obtained in \cite{Kibis}. 
	 Secondly, based on the Hamiltonian form in \eqref{ru4} we can define an effective vector potential
	 \begin{equation}
	 	\textbf{A}_{\sf eff}=\frac{E_{0}}{\omega}v_{x}(S)\left(\cos \omega t,\frac{v_{y}(S)}{v_{x}(S)}\xi\sin \omega t\right)
	 \end{equation}
 which is actually sharing some common features with elliptically one (see next subsection) under the following mapping
 \begin{equation}
 	\frac{E_{0}}{\omega }v_{x}(S)\longrightarrow \frac{E_{0}}{\omega }, \qquad
 	\frac{v_{y}(S)}{v_{x}(S)}\xi\longrightarrow\sin \theta.
 \end{equation}
 Such similarity tells us that one can reproduce the effect of elliptically dressing field simply by considering the strain effect. This remains among the 
interesting results derived so far, which will  be numerically analyzed in the next}.


\subsection{Elliptically polarized dressing field}

For the case of an electromagnetic wave elliptically polarized dressing field  oriented along the $x$-axis, we consider the vector potential $\textbf{A}=\frac{E_{0}}{\omega } \left(\cos \omega t,\sin \theta \sin\omega t\right)$ such that angle $\theta$ defines its polarization phase. Note that such case  involves the two previous potentials because simply by requiring $\theta=0$ and $\theta=\pm\pi/2$
we recover the linear and circular polarization cases, respectively.
As before, let us write the time-dependent total Hamiltonian (\ref{xtr})
\begin{align}\label{run4}
\nonumber\mathcal{{{H}}}=&\left(%
\begin{array}{cc}
\frac{\Delta_{g}}{2} & \tau  v_{x}(S) p_{x}-i v_{y}(S) p_{y} \\
\tau v_{x}(S) p_{x}+i v_{y}(S)p_{y} & -\frac{\Delta_{g}}{2} \\
\end{array}%
\right)\\
& +\frac{E_{0}v_{x}(S){|e|}}{2\omega}\left[\begin{pmatrix}
0 & \tau -\frac{v_{y}(S)}{v_{x}(S)}\sin \theta  \\
\tau +\frac{v_{y}(S)}{v_{x}(S)}\sin \theta  & 0
\end{pmatrix}e^{i\omega t}+\begin{pmatrix}
0 & \tau+\frac{v_{y}(S)}{v_{x}(S)}\sin \theta \\
\tau-\frac{v_{y}(S)}{v_{x}(S)}\sin \theta  & 0
\end{pmatrix}e^{-i\omega t}\right].
\end{align} 
{
	Using again 
the Floquet-Magnus approach for a periodically driven quantum system in similar way to the previous case of circular polarization, to end up with  the eigenvalues
\begin{equation}\label{rtk8}
\varepsilon=
\gamma\left[\left(\frac{\Lambda_{g}}{2}\right)^{2}+v^{2}_{x}(S)\left(1-\frac{E_{0}^{2}{v_{y}^{2}(S)}{|e|}^{2}}{{{\hbar}^{2}{\omega^{4}}}}\sin ^{2}\theta\right)^{2}{\hbar}^{2}{k_{x}^{2}}+ v^{2}_{y}(S)\left(
1-\frac{E_{0}^{2}{v_{x}^{2}(S)}{|e|}^{2}}{{{\hbar}^{2}{\omega^{4}}}}\right)^{2}{\hbar}^{2} {k_{y}^{2}}\right]^{\frac{1}{2}}
\end{equation} 
where the band gap reads as 
\begin{equation}
	\Lambda_{g}=\Delta _{g}\left[ 1-\frac{E_{0}^{2}v_{x}^{2}(S){|e|}^{2}}{{\hbar}^{2}\color{black}{\omega^{4}}}\left(1+\frac{v_{y}^{2}(S)}{v_{x}^{2}(S)}\sin ^{2}\theta\right)\right]-2\tau \frac{ E_{0}^{2}v_{x}(S){v_{y}(S)}{|e|}}{{\hbar}{\omega^{3}}}\sin \theta.
\end{equation}
 Actually
  (\ref{rtk8}) is involving different physical parameters and also generalizing the two former dispersion relations derived for the cases of the linearly and circularly polarized dressing fields. More precisely, for  $\frac{2E_{0}v_{x}(S)|e|}{\hbar\omega^{2}}\ll1$,  $\theta=0,\pm \pi/2$, and high frequency $\hbar\omega\gg\Delta_{g}$ \cite{Kibis}, (\ref{rtk8}) turns into the eigenvalues
	(\ref{rt18}) and (\ref{rt8}). 
	Also note in passing that   (\ref{rtk8}) reduces to those obtained in \cite{Kibis} for  the case $S=0$.}

Consequently, we will investigate the  behavior of our system based on different configurations of the involved physical parameters. Indeed, the numerical implementation of our theoretical model will be used to study the energy spectrum $\varepsilon$, the renormalized band gap $|\Lambda_g/\Delta_g|$ under suitable conditions of the wave vector components $(k_x,k_y)$, irradiation intensity $I \sim E_0^2$, $E_{0}$ being the electric field, strain amplitude $S$ and some particular values of polarisation phase $\theta$.
{{

		\section{Numerical results}
Figure \ref{fig4} presents the energy spectrum $\varepsilon$ of electron dressed by the linearly polarized field versus the wave vector component $k_{x}$. We choose {${\Delta_{g}}=2$  \text{meV}, $\hbar\omega=10$  \text{meV} and three values of the irradiation intensities 
$I=(0.0,13.3 \ \text{kW/\text{cm$^{2}$}},26.7\ \text{kW/\text{cm$^{2}$}})$ with $S=(0.0, 0.5, 0.7)$ for armchair (A), $S=(0.1,0.15)$ for zigzag (Z)}. For the case without strain ($S=0.0$), {we reproduce} the results obtained in \cite{Kibis}, where the up and down bands are symmetrical and the energy spectrum is {isotropic for zero dressing field and anisotropic for different values of $I$}. In Figures
\text{\ref{fig4}}\textbf{\color{blue}{(a)}}, \text{\ref{fig4}}\textbf{\color{blue}{(b)}}  \color{black}{when} strain is applied along the armchair direction {with $S=(0.5,0.7)$}, we observe that it causes some changes on $\varepsilon$ and we still have the same behavior {of zero strain except that the band gap is increased slowly by increasing the values of the strain $S$}. Figures \text{\ref{fig4}}\textbf{\color{blue}{(c)}}, \text{\ref{fig4}}\textbf{\color{blue}{(d)}} show that the strain along zigzag direction produces remarkable influence {on $\varepsilon$ because it increases dramatically for small values of $S$. Particularly, for $S=0.15$ and in the presence of the dressing field (orange and black lines), we observe a different behavior of the energy spectrum compared to previous cases where  it becomes almost constant and its band gap is very large. It is interesting to note that $\varepsilon$ increases for different strain amplitude $S$ and decreases by increasing the irradiation intensity $I$}.}
		\begin{figure}[H]
			\centering
			\subfloat[]{
				\centering
				\includegraphics[scale=0.23]{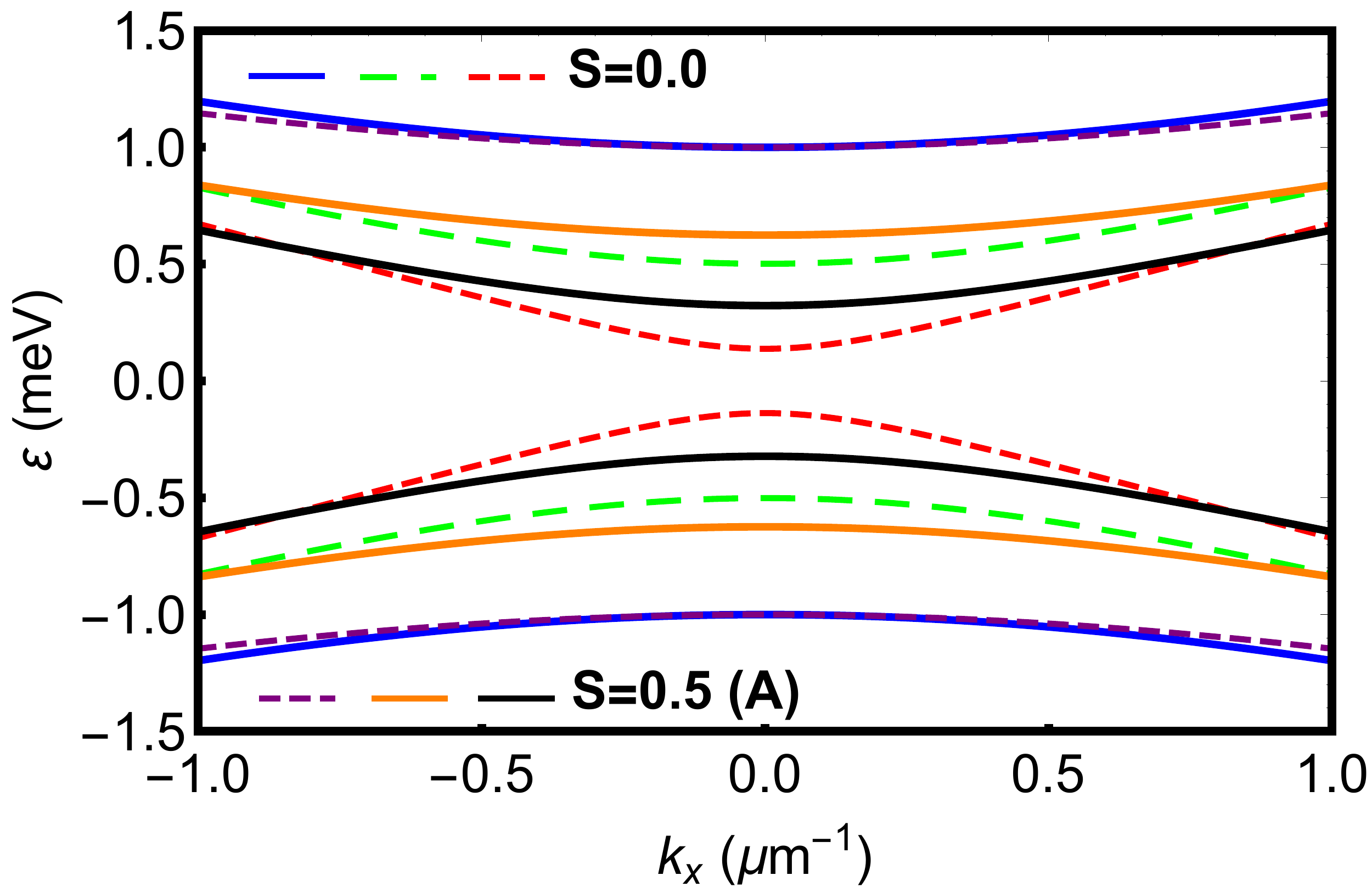}
				\label{g1}}\hspace{2cm}
			\subfloat[]{
				\centering
				\includegraphics[scale=0.23]{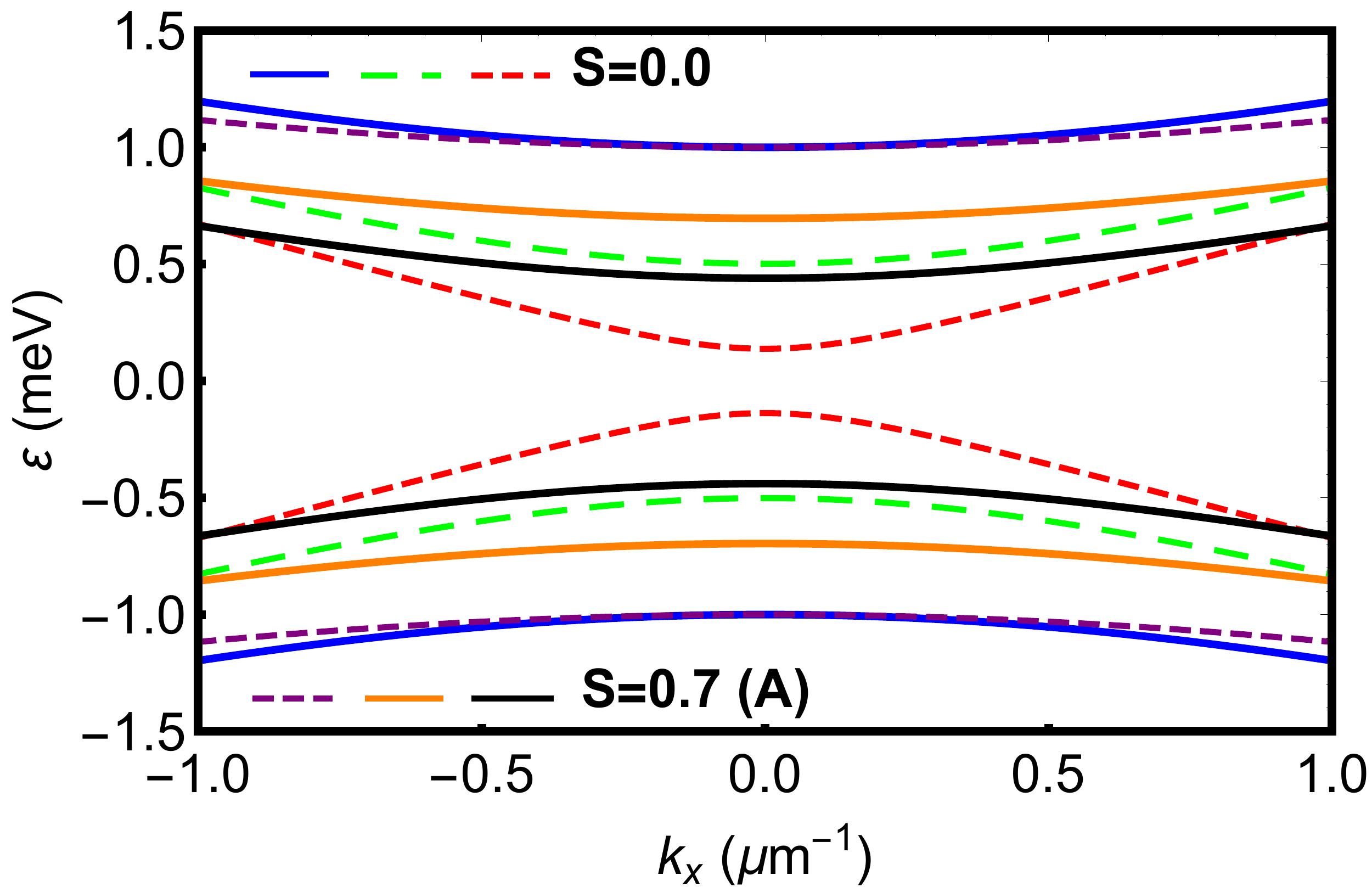}
				\label{g3}}\\
			\subfloat[]{
				\centering
				\includegraphics[scale=0.23]{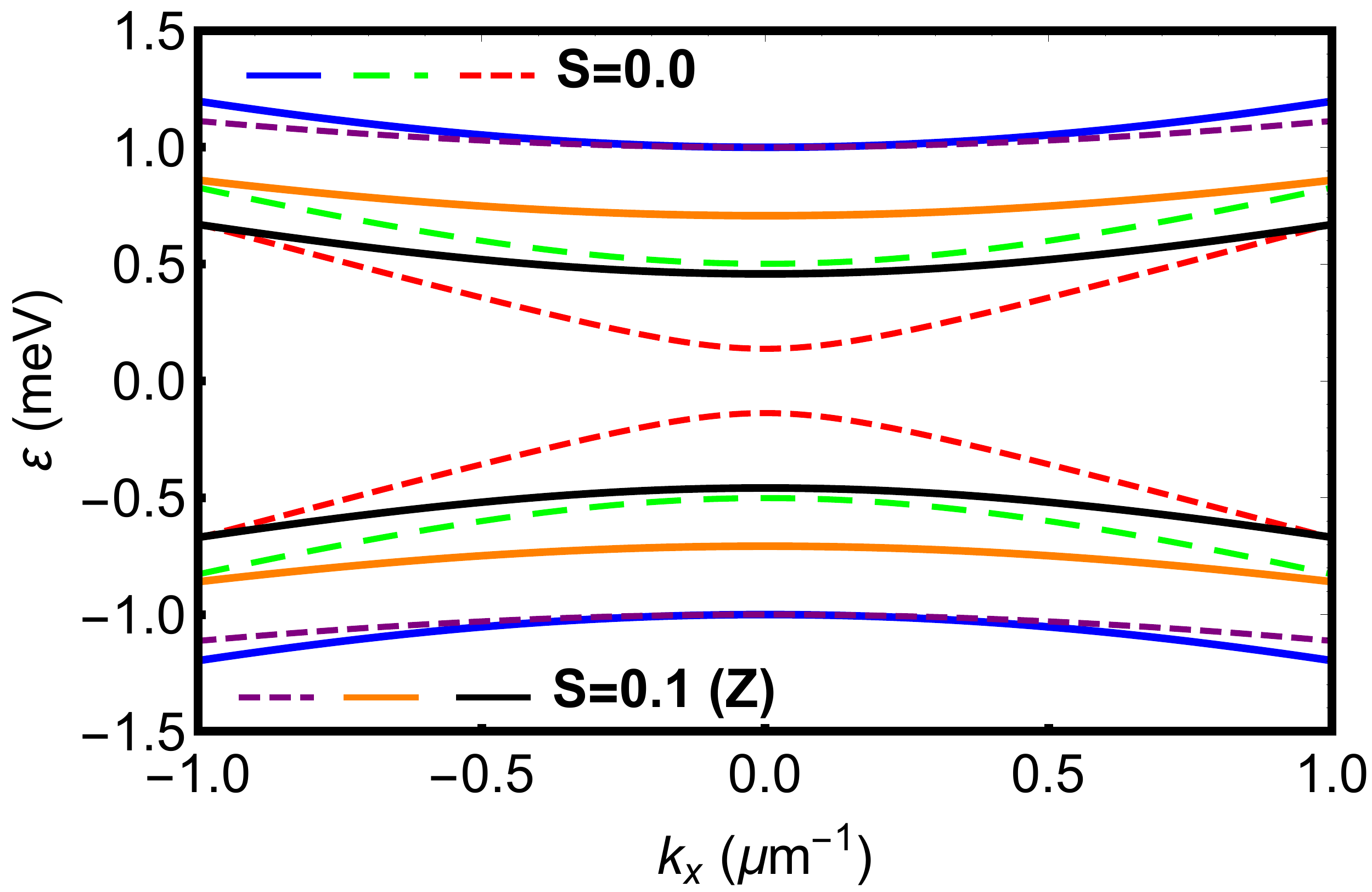}
				\label{g4}}\hspace{2cm}\subfloat[]{
				\centering
				\includegraphics[scale=0.23]{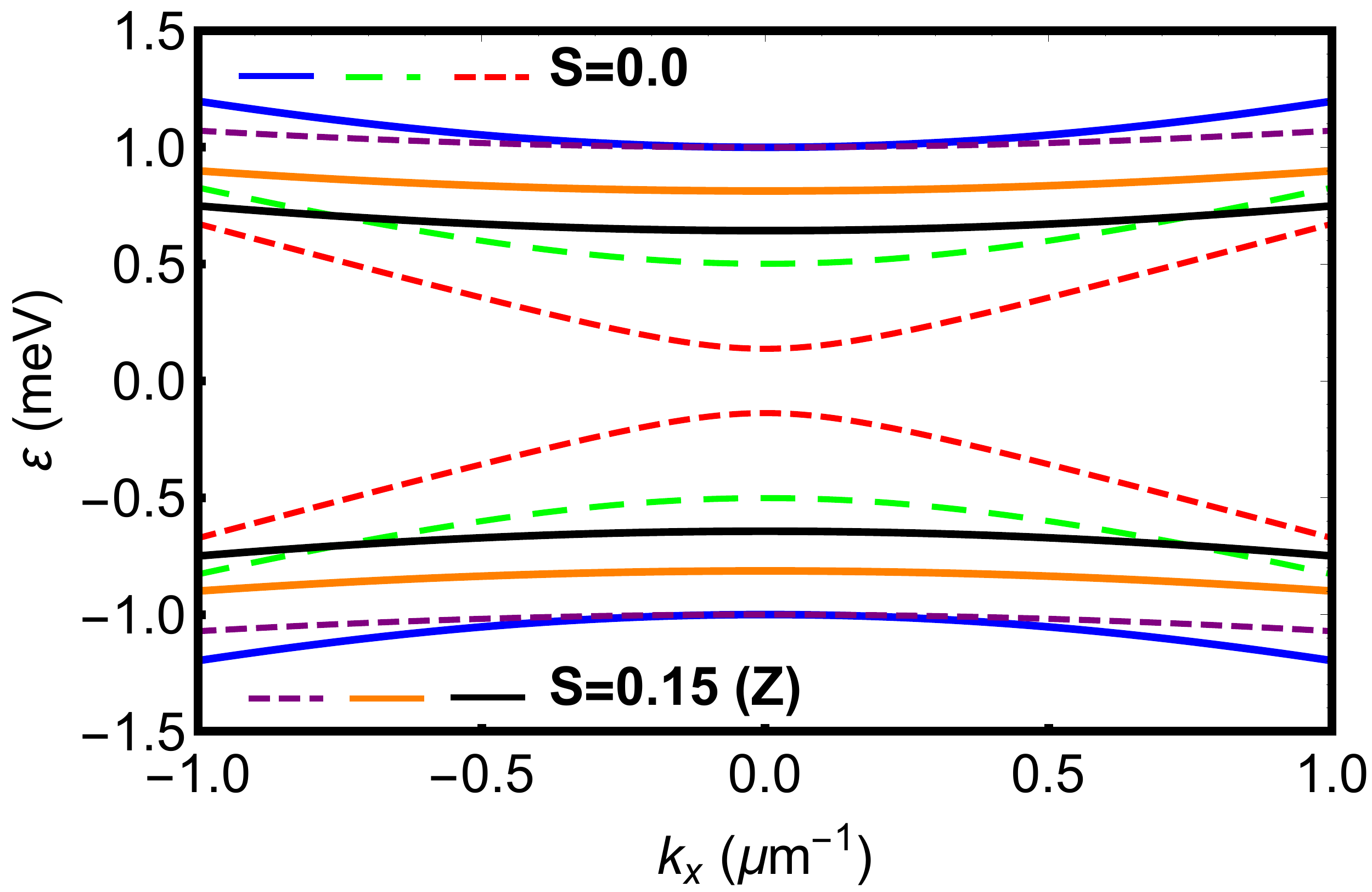}
				\label{g5}}
\caption{\sf{(color online) The energy spectrum $\varepsilon$ of electron dressed by the linearly polarized field versus the wave vector component $k_x$ for {${\Delta_{g}}=2$  \text{meV}, $\hbar\omega=10$  \text{meV} with three values of the irradiation intensities $I=0.0$  (blue and magenta lines), $I=13.3$ \text{kW/\text{cm$^{2}$}} (green and orange  lines), $I=26.7$ \text{kW/\text{cm$^{2}$}} (red and black lines).} \textbf{\color{blue}{(a),(b)}}\color{black}{:} {Effect of armchair strain direction with $S=(0.5,0.7)$.} \textbf{\color{blue}{(c),(d)}}\color{black}{:} {Effect of zigzag strain direction with $S=(0.1,0.15)$}.}}\label{fig4}
		\end{figure}
	
In Figure \ref{fig5}, we plot the energy spectrum $\varepsilon$ of dressed electron versus the wave vector component ${k_{y}}$ for {${\Delta_{g}}=2$  \text{meV}, $\hbar\omega=10$  \text{meV} with three values of the irradiation intensity $I=0.0$ (blue line),  $I=13.3$ \text{kW/\text{cm$^{2}$}} (green line), $I=26.7$ \text{kW/\text{cm$^{2}$}} (red line)}\color{black}{.} As far as $S= 0.0$ is concerned, {$\varepsilon$ is isotropic in the absence of the dressing field but anisotropic for the both values of $I$} and decreases up to near zero values {for $I=26.7$ \text{kW/\text{cm$^{2}$}} (red line) showing a similar  behavior  to that found in \cite{Kibis}}. {In Figures
\text{\ref{fig5}}\textbf{\color{blue}{(a)}}, \text{\ref{fig5}}\textbf{\color{blue}{(b)}} we observe that the armchair strain direction with $S=(0.5,0.7)$} affects 
$\varepsilon$, which 
becomes {anisotropic and increases rapidly by increasing the values of
$S$}. On the other hand, we observe that the zigzag strain direction produces remarkable influence on $\varepsilon$ 
as presented in Figures
\text{\ref{fig5}}\textbf{\color{blue}{(c)}}, \text{\ref{fig5}}\textbf{\color{blue}{(d)}}. More precisely, $\varepsilon$ is showing different behavior compared to the previous results because {it is the same with the case $S=0.0$ for $I=0.0$ but its band gap becomes very large for high values of $I$ and $S$. Also, we notice that the energy spectrum $\varepsilon$ decreases quickly as long as $I$ increases but increases dramatically for $S\neq0$.} 
Therefore, we conclude that the effects depend on the direction of applied strain.
\begin{figure}[H]
\centering
\subfloat[]{
\centering
\includegraphics[scale=0.23]{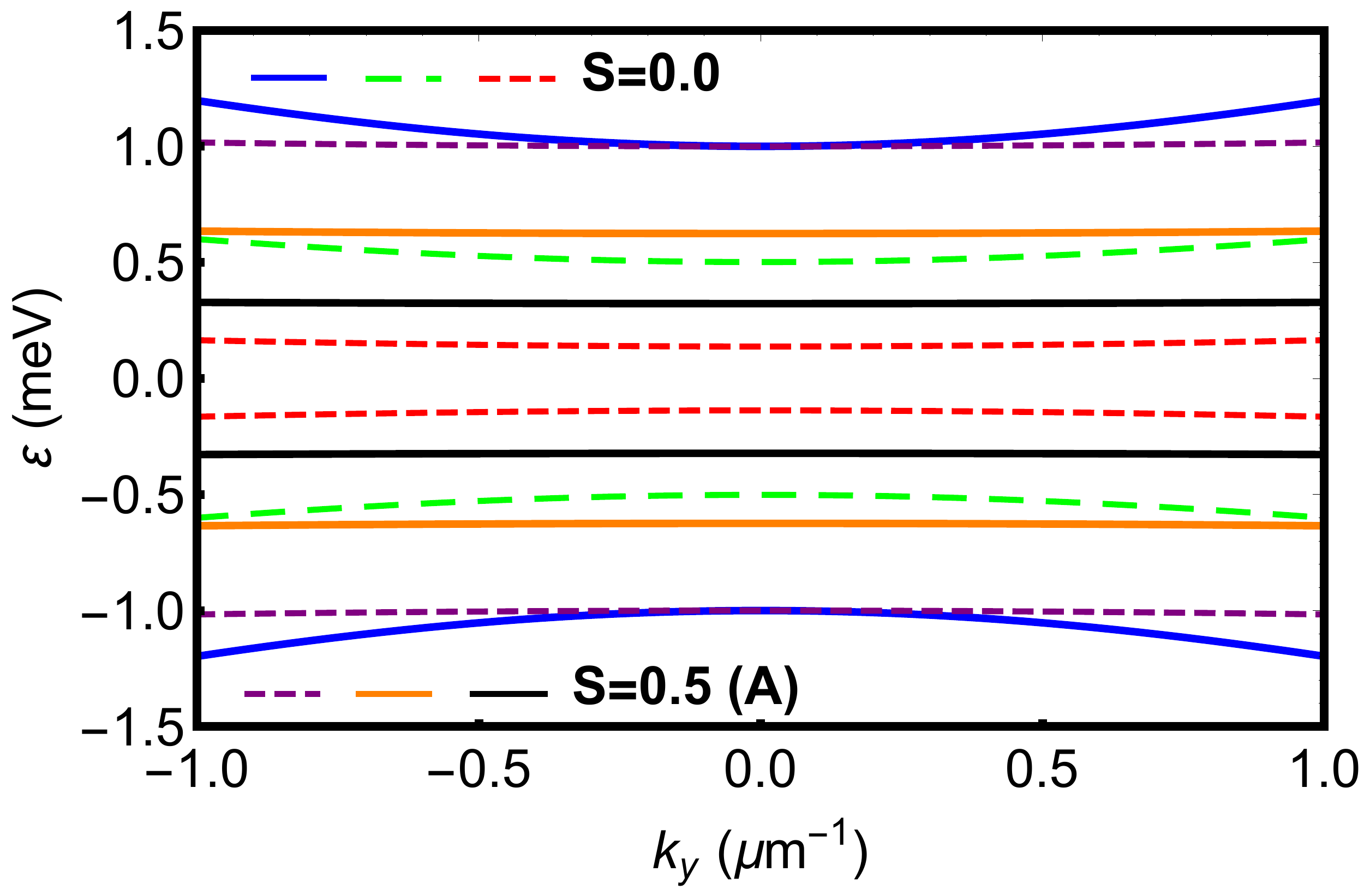}
\label{figr}}\hspace{2cm}
\subfloat[]{
\centering
\includegraphics[scale=0.23]{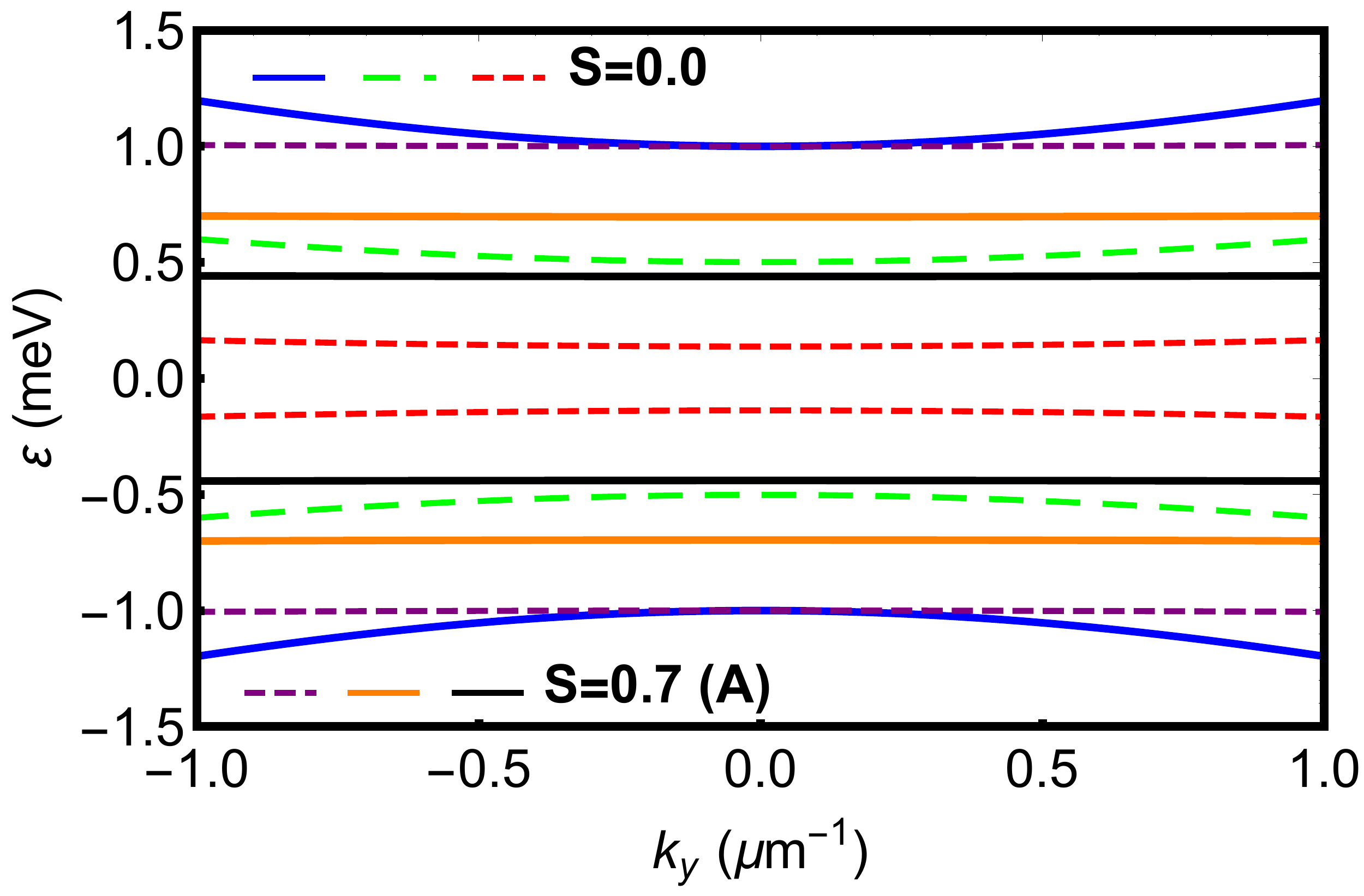}
\label{fgigjr}}\\
\subfloat[]{
\centering
\includegraphics[scale=0.23]{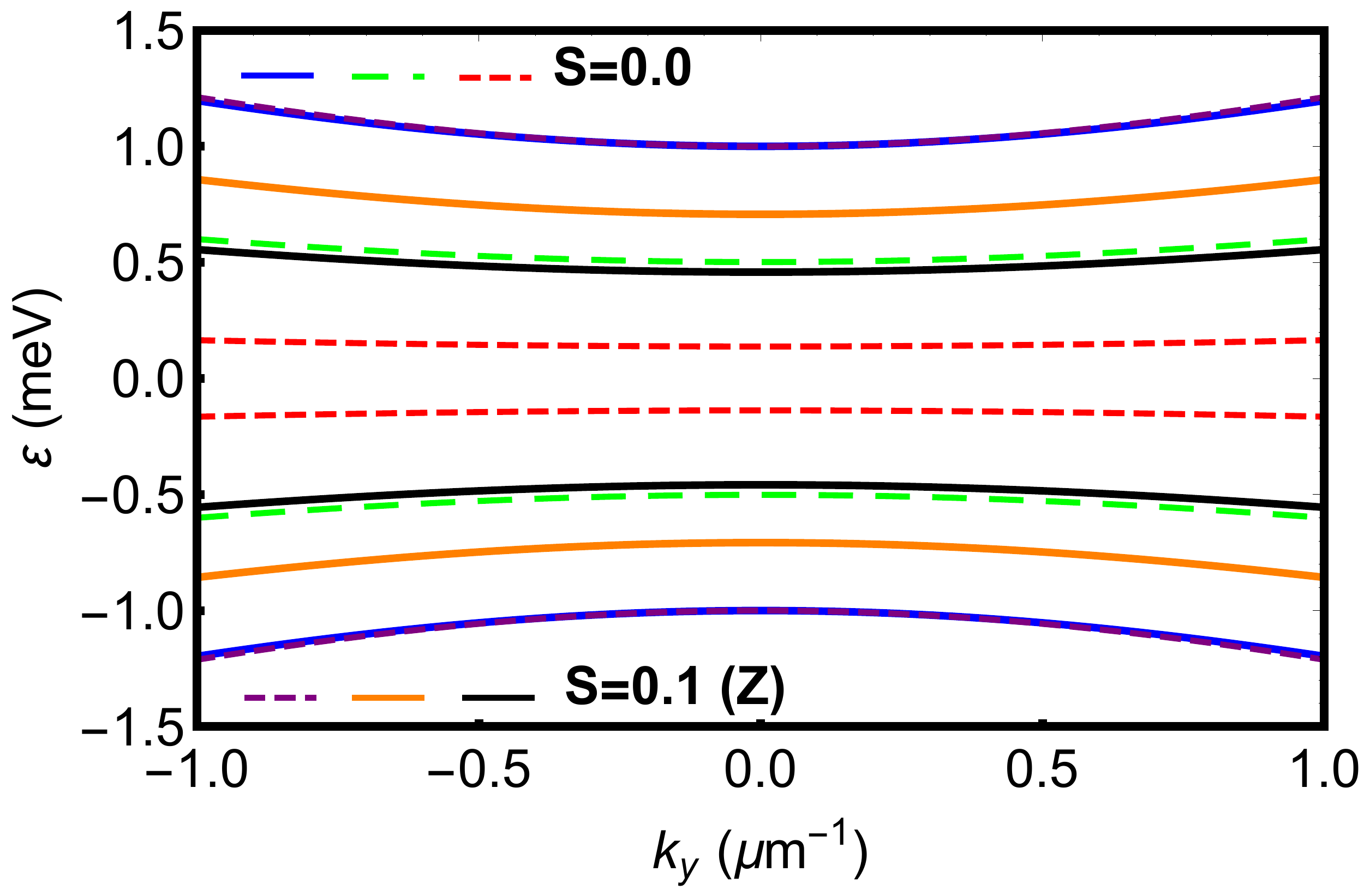}
\label{igr}}\hspace{2cm}
\subfloat[]{
\centering
\includegraphics[scale=0.23]{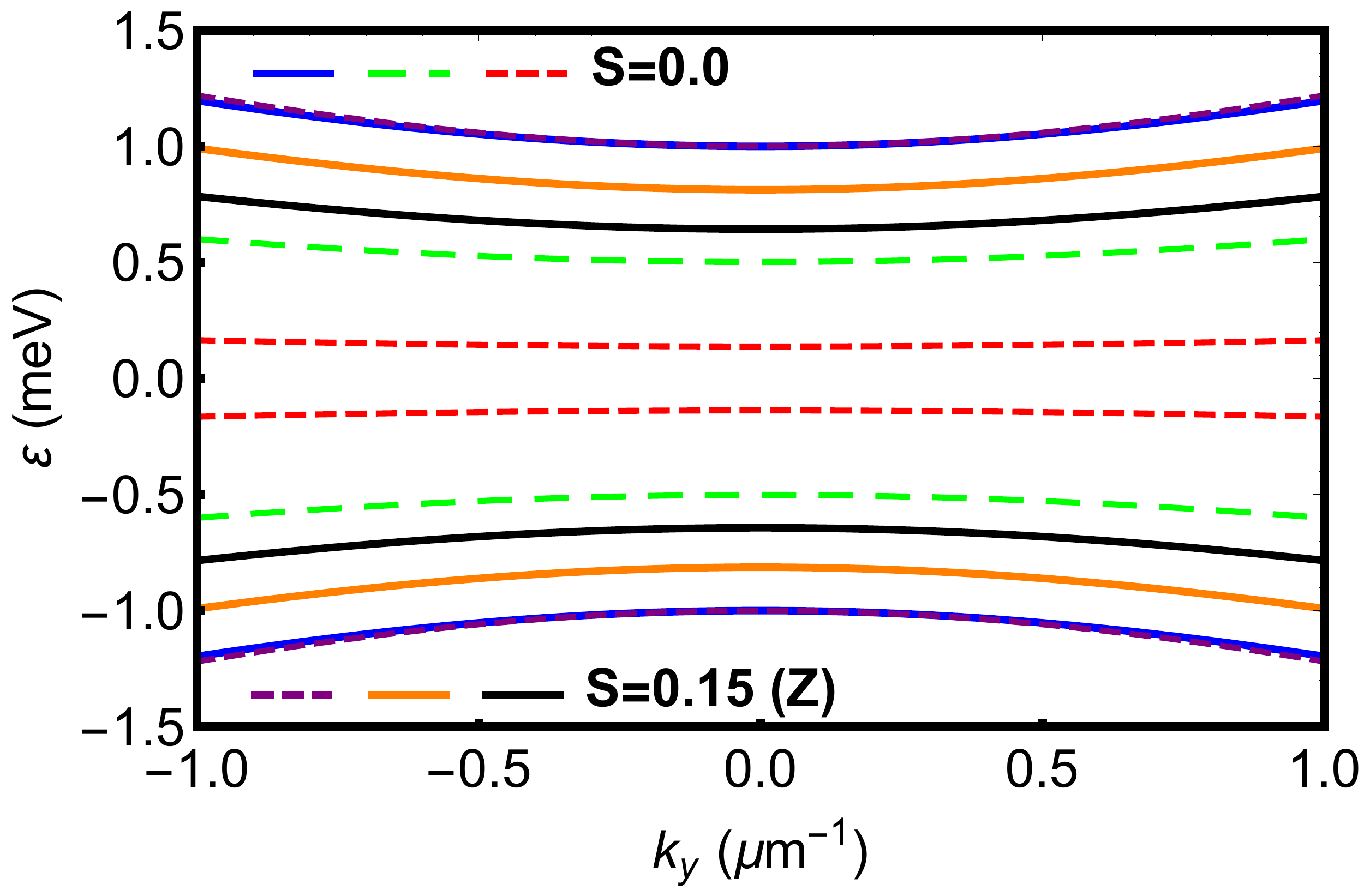}
\label{fgigr}}
\caption{\sf{(color online) The energy spectrum $\varepsilon$ of electron dressed by the linearly polarized field versus the wave vector component $k_y$ for {${\Delta_{g}}=2$  \text{meV}, $\hbar\omega=10$  \text{meV} with three values of the irradiation intensity  {$I=0.0$}  \color{black}{(blue and magenta lines)}, {$I=13.3$ \text{kW/\text{cm$^{2}$}}} \color{black}{(green and orange  lines)}, {$I=26.7$ \text{kW/\text{cm$^{2}$}}} \color{black}{(red and black lines)}}.  \textbf{\color{blue}{(a),(b)}}\color{black}{:} {Effect of armchair strain direction with $S=(0.5,0.7)$}\color{black}{.}
		\textbf{\color{blue}{(c),(d)}}\color{black}{:} {Effect of zigzag strain direction with $S=(0.1,0.15)$}\color{black}{.}}}\label{fig5}
\end{figure}
Figure \ref{fig6} illustrates the energy spectrum {$\varepsilon$ of electron dressed by the linearly polarized field versus the strain amplitude $S$ for ${\Delta_{g}}=2$  \text{meV}, $\hbar\omega=10$  \text{meV with $I=0.0$ (blue line)}}{, $I=13.3$ \text{kW/\text{cm$^{2}$}} (green line), $I=26.7$ \text{kW/\text{cm$^{2}$}} (red line) and different values of the wave vector components ($k_{x},k_{y}$)}. It is clearly seen that $\varepsilon$  is showing different
 behaviors for armchair and zigzag directions and it decreases by increasing the values of  $I$ and  $S$. For a strain applied along armchair direction for the parameters $k_{x}=0.0/0.6$ \text{$\mu$m$^{-1}$}, $k_{y}=1$ \text{$\mu$m$^{-1}$} and  $I=0.0$, we observe  that $\varepsilon$ decreases slowly as long as $S$ increases and converges to two values such as $\varepsilon=1.2$ \text{meV} for $k_{x}=0.0$ and $\varepsilon=1.3$ \text{meV} for $k_{x}=0.6$ \text{$\mu$m$^{-1}$}. In the presence of the dressing field ($I\neq0$), $\varepsilon$  decreases in the interval $0<S<0.25$ and  after that it increases rapidly (green and red lines) and its band gap becomes large for $k_{x}=0.6$ \text{$\mu$m$^{-1}$} as shown in Figures  \text{\ref{fig6}}\textbf{\color{blue}{(a)}}, \text{\ref{fig6}}\textbf{\color{blue}{(c)}}. On the other hand, when we change the values of $k_{x}$ and $k_{y}$ as  in Figures \text{\ref{fig6}}\textbf{\color{blue}{(b)}}, \text{\ref{fig6}}\textbf{\color{blue}{(d)}},    $\varepsilon$ takes an anisotropic form, which is mostly the same as in  the previous Figures but  becomes large.} However, we observe that there is a remarkable difference by switching the strain to the zigzag direction. Indeed according to {Figures \text{\ref{fig6}}\textbf{\color{blue}{(e)}}, \text{\ref{fig6}}\textbf{\color{blue}{(g)}} we notice that for $I=0.0$, $\varepsilon$ increases/decreases dramatically for $k_{x}=0/0.6$ \text{$\mu$m$^{-1}$}, $k_{y}=1$ \text{$\mu$m$^{-1}$} but it is reduced in the interval $0<S<0.23$ for $I\neq0.0$ and increased rapidly at the value $S=0.23$. The results of Figures \text{\ref{fig6}}\textbf{\color{blue}{(f)}}, \text{\ref{fig6}}\textbf{\color{blue}{(h)}} are similar to those of Figures  \text{\ref{fig6}}\textbf{\color{blue}{(e)}}, \text{\ref{fig6}}\textbf{\color{blue}{(g)}} except that for zero dressing field with $k_{x}=1$ \text{$\mu$m$^{-1}$}, $k_{y}=0/0.6$ \text{$\mu$m$^{-1}$}, $\varepsilon$ decreases quickly until becomes null at $S=0.77/0.86$  and starts from large values when $k_{x}$ and $k_{y}$ are modified}\color{black}{.} In addition, there is a symmetry separating positive and negative  behavior of $\varepsilon$. Then, we emphasis that the energy spectrum can be controlled by tuning the strain amplitude $S$, irradiation intensity $I$ {and wave vector components ($k_{x}$,$k_{y}$)}\color{black}{.}

\begin{figure}[H]
	\centering
	\subfloat[]{
		\centering
		\includegraphics[scale=0.23]{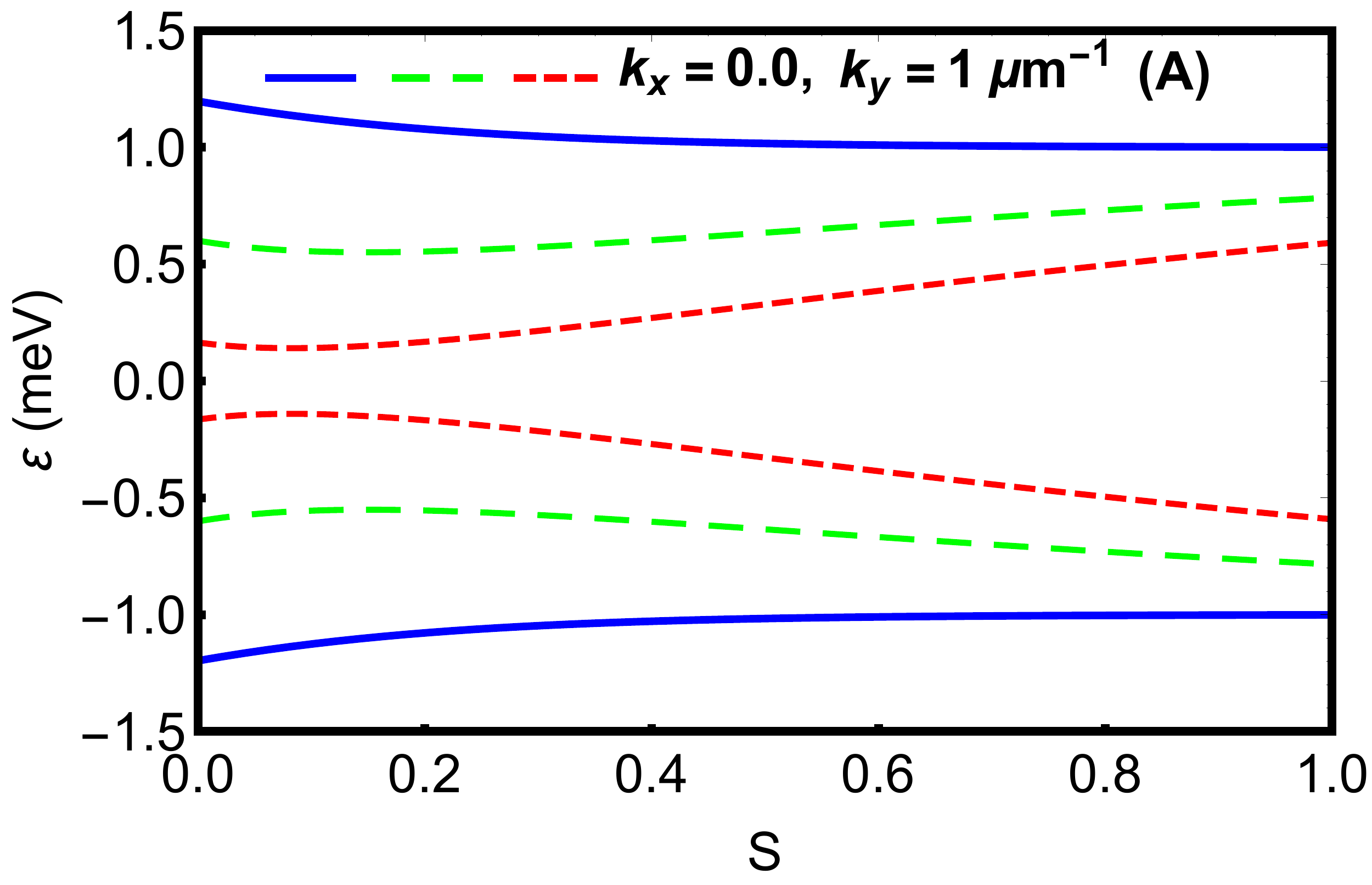}
		\label{figunr5}}\hspace{2cm}
	\subfloat[]{
		\centering
		\includegraphics[scale=0.23]{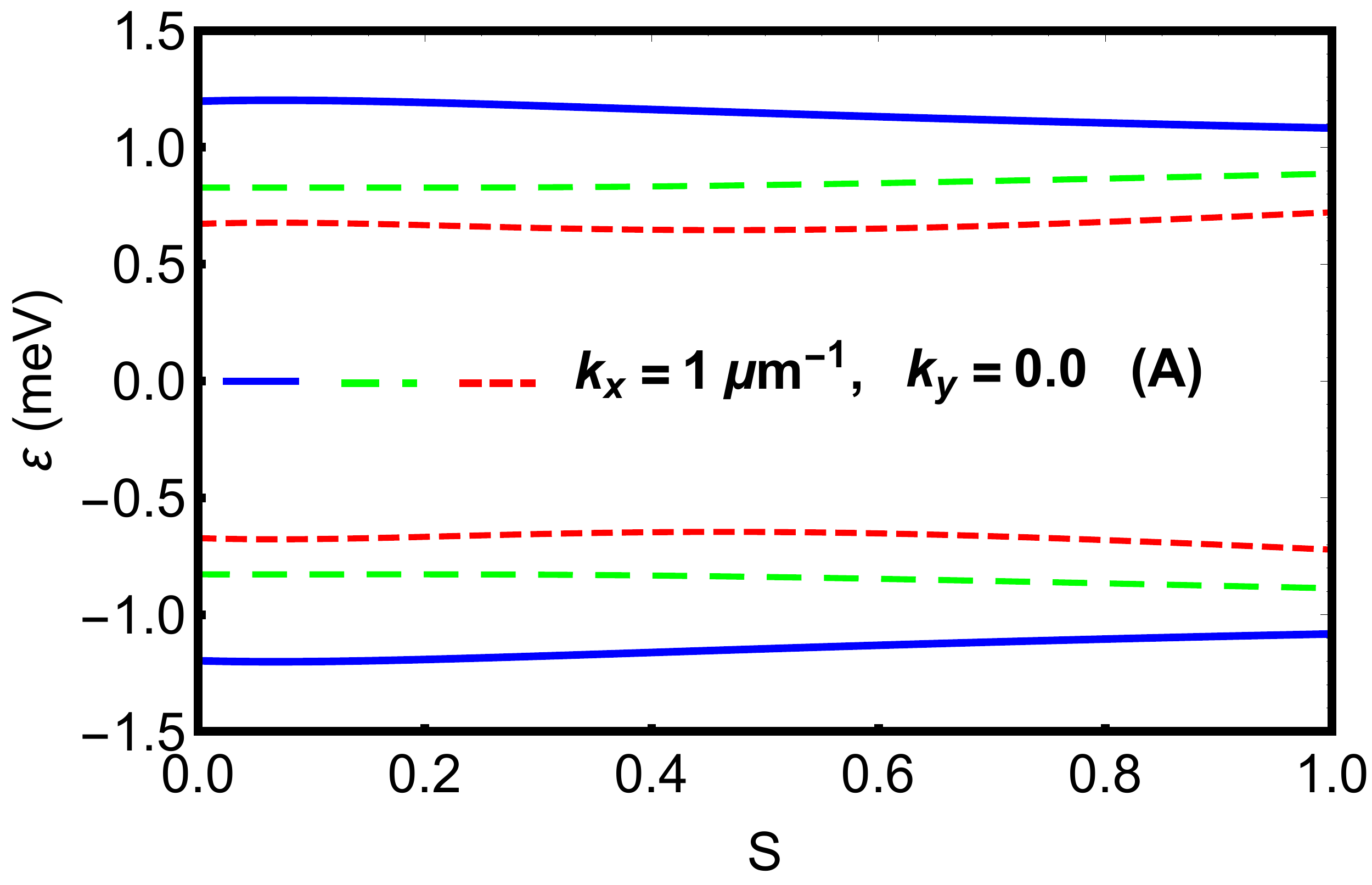}
		\label{figunre}}	\\
	\subfloat[]{
		\centering
		\includegraphics[scale=0.23]{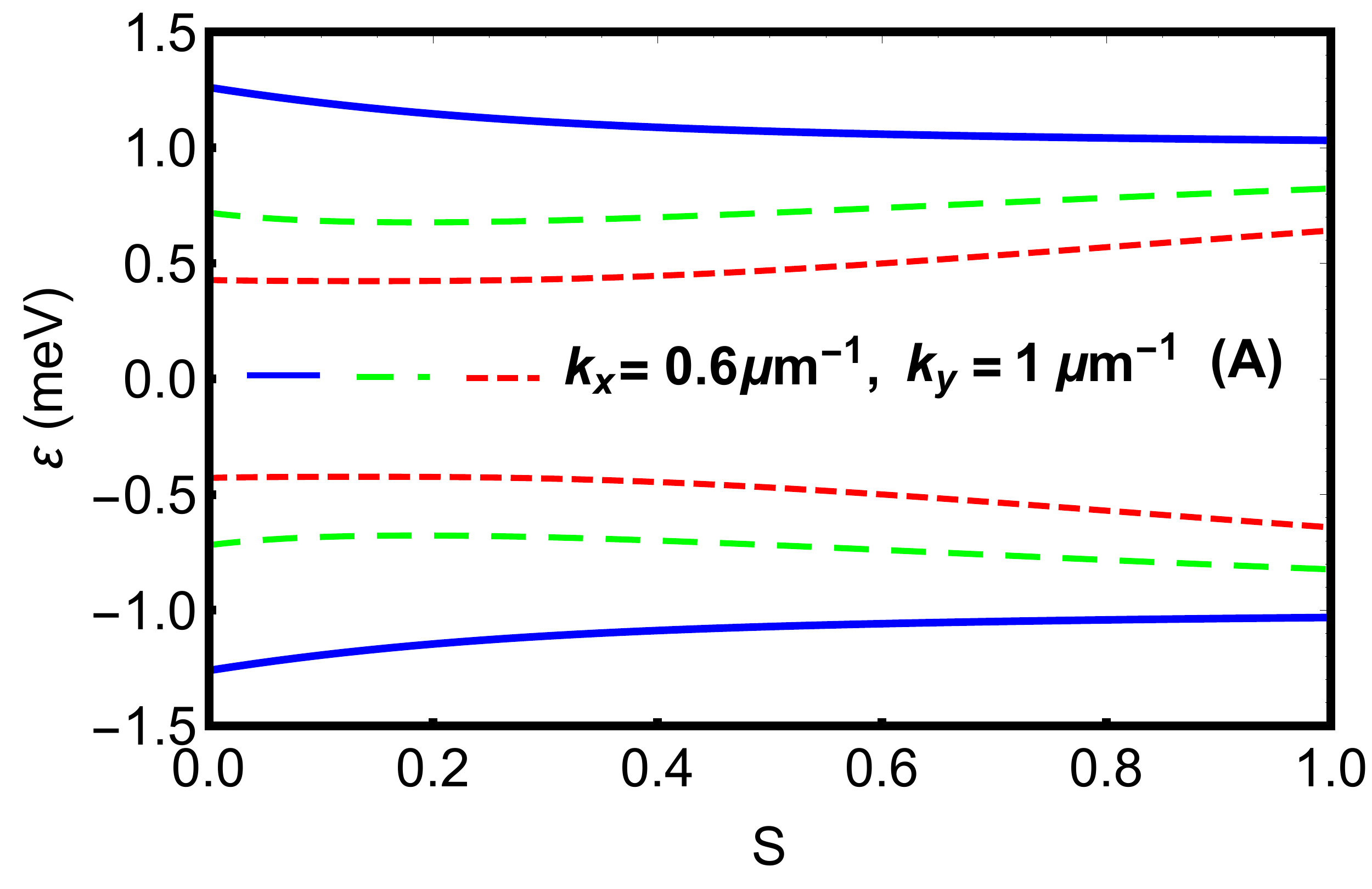}		\label{figurn5}}\hspace{2cm}
	\subfloat[]{
		\centering
		\includegraphics[scale=0.23]{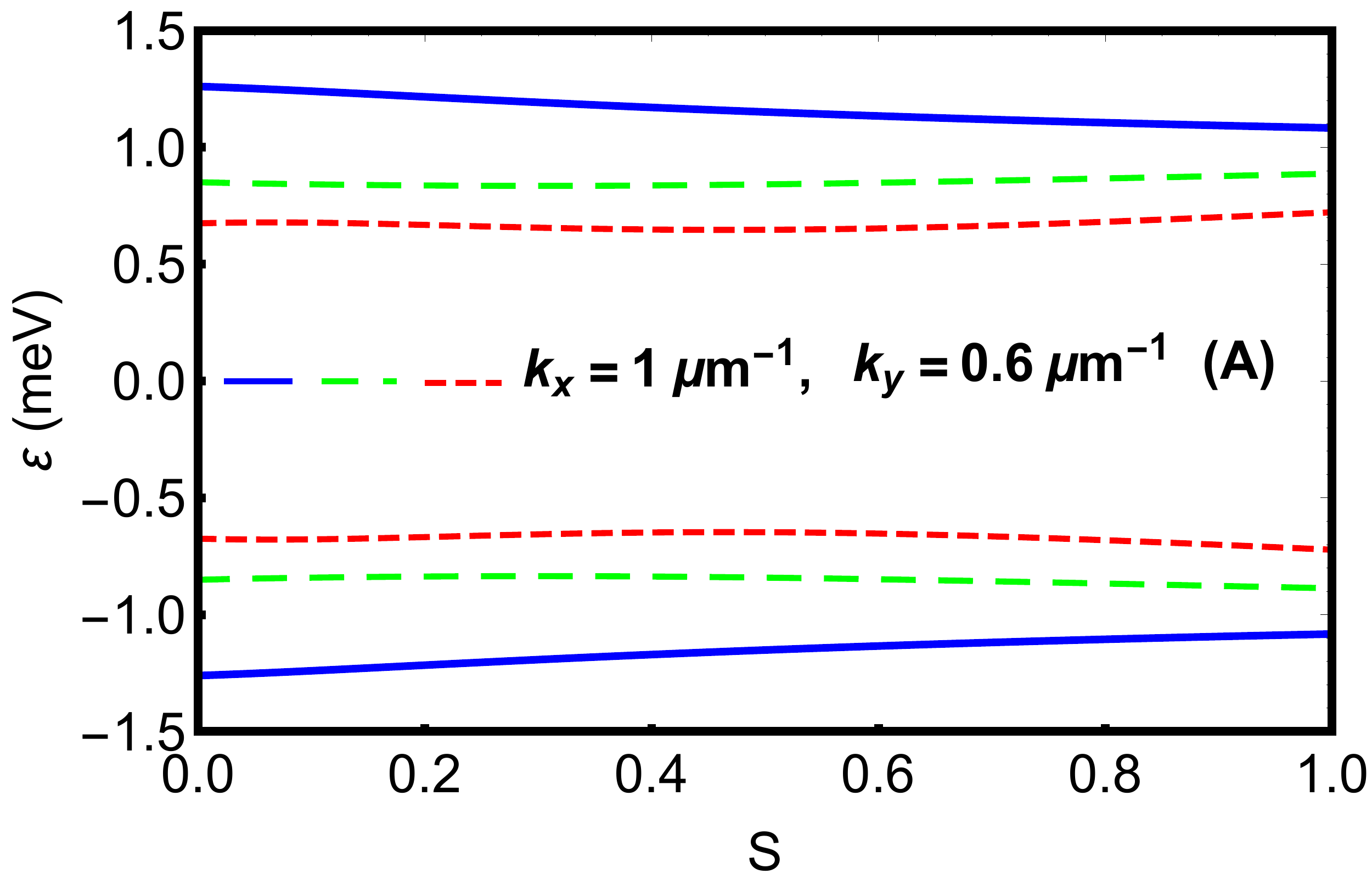}
		\label{figurne}}\\
	\subfloat[]{
		\centering
		\includegraphics[scale=0.23]{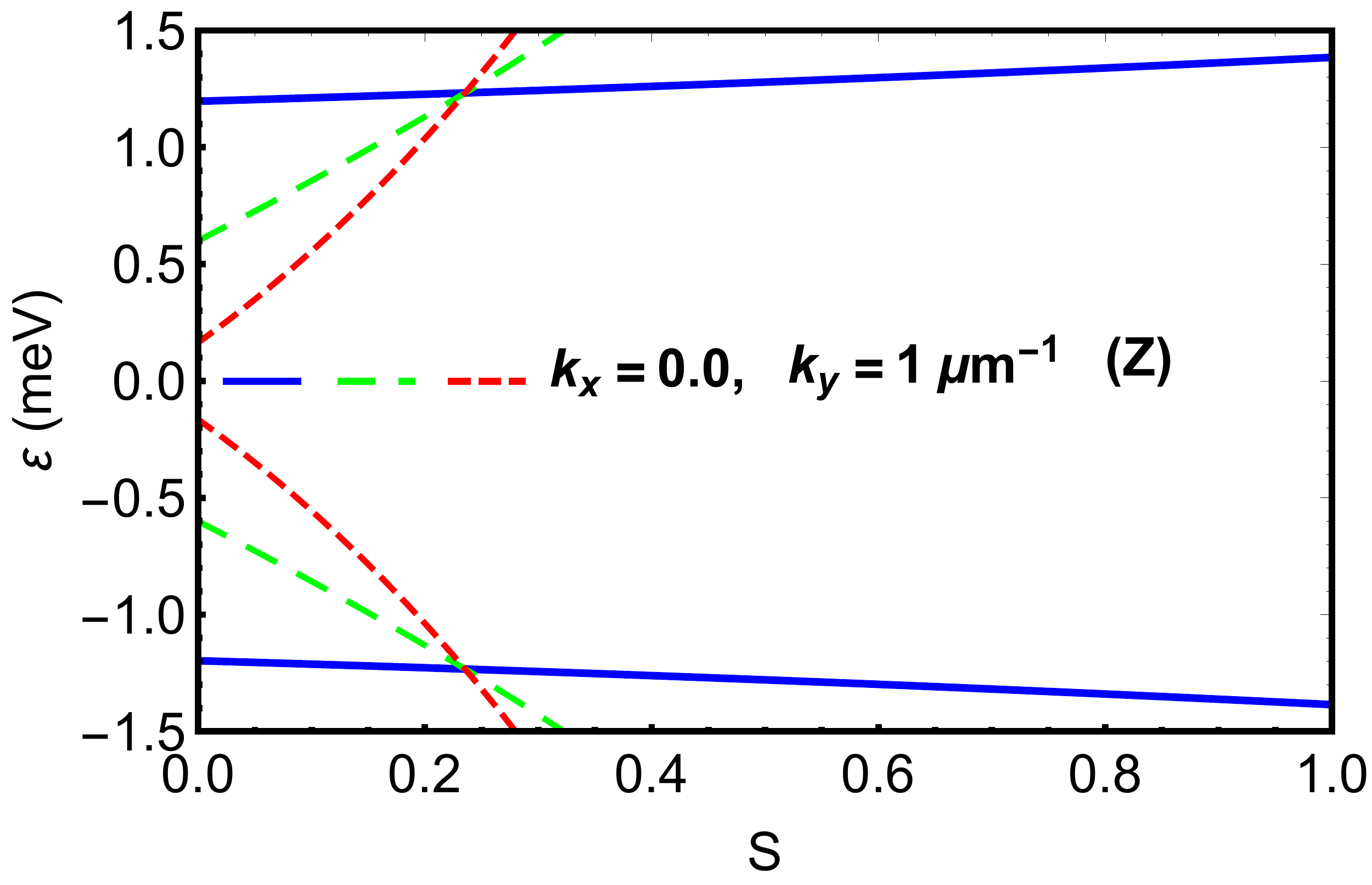}
		\label{figurc5}}\hspace{2cm}
	\subfloat[]{
		\centering
		\includegraphics[scale=0.23]{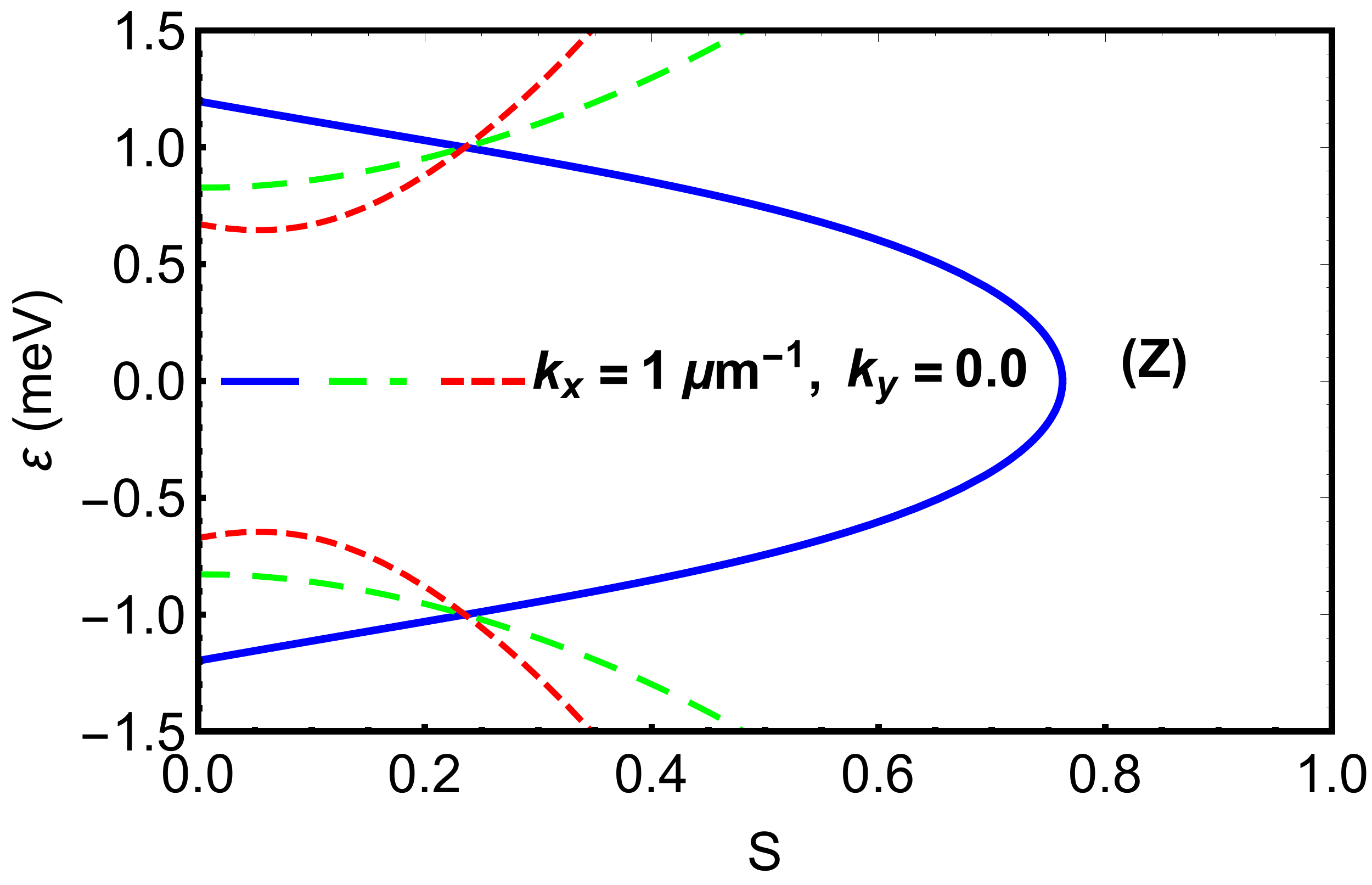}
		\label{figurce}}	\\
	\subfloat[]{
		\centering
		\includegraphics[scale=0.23]{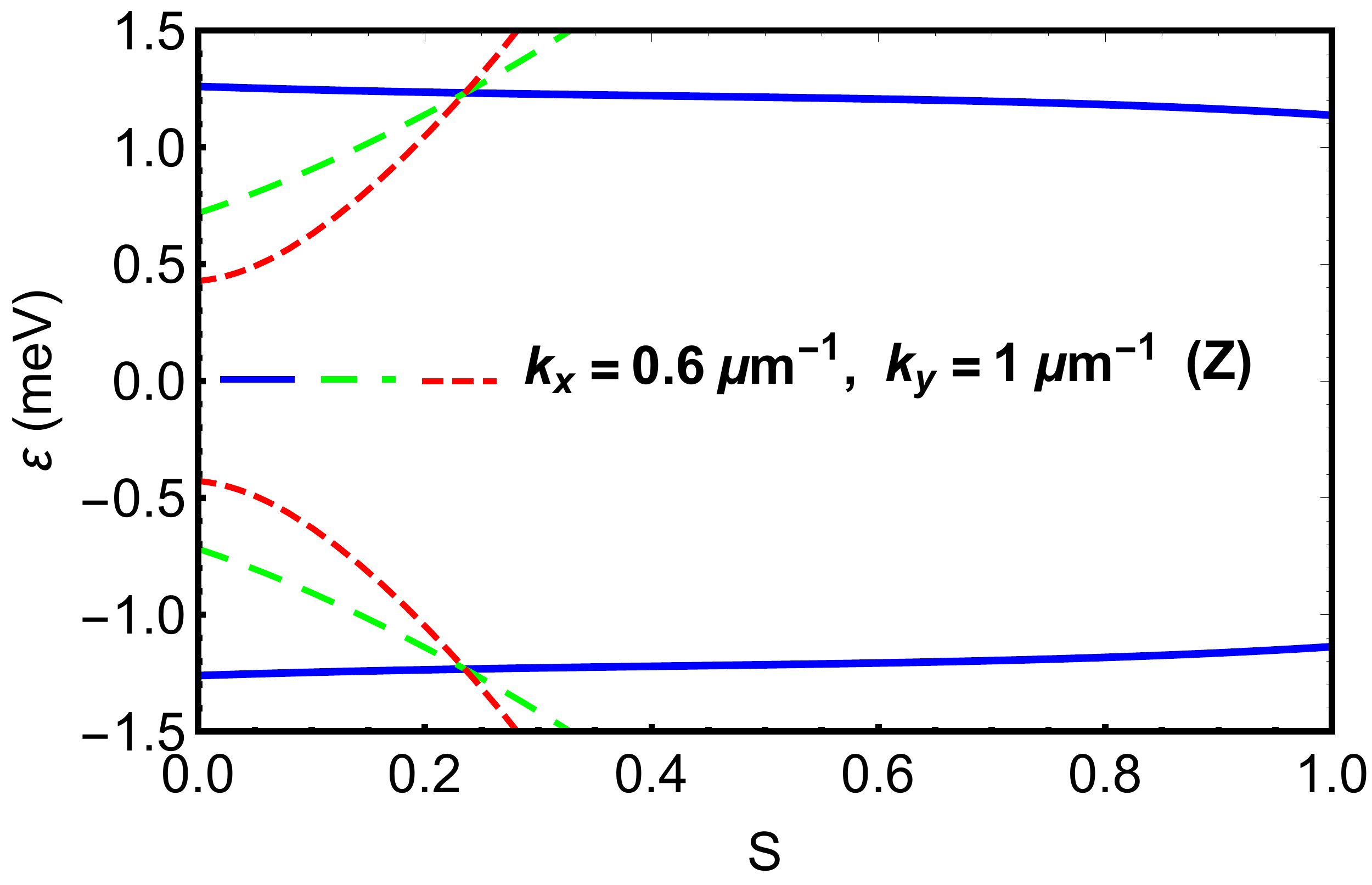}\hspace{2cm}
		\label{figur5}}
	\subfloat[]{
		\centering
		\includegraphics[scale=0.23]{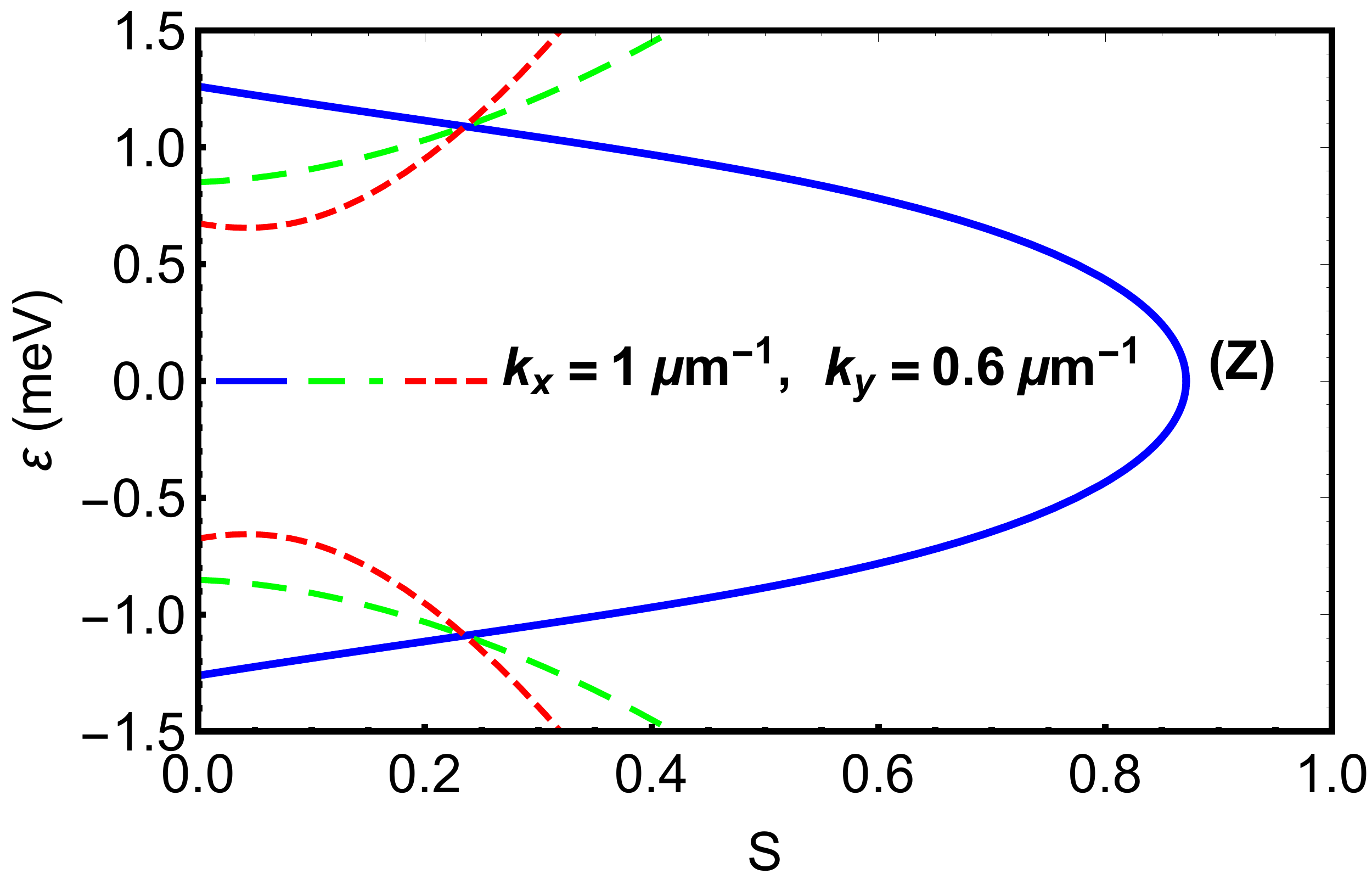}
		\label{figure}}
	\caption{\sf{(color online) The energy spectrum
			$\varepsilon$ of electron dressed by the linearly
			polarized field versus the strain amplitude $S$ along armchair and zigzag directions with {${\Delta_{g}}=2$  \text{meV}, $\hbar\omega=10$  \text{meV} and three values
				of the irradiation intensity $I=0.0$ (blue line), $I=13.3$ \text{kW/\text{cm$^{2}$}} (green line), $I=26.7$ \text{kW/\text{cm$^{2}$}} (red line)}\color{black}{.}  \textbf{\color{blue}{(a)-(e)}}\color{black}{:} ${k_{x}}=0.0$, ${k_{y}}=1$ \text{$\mu$m$^{-1}$}.
			\textbf{\color{blue}{(b)-(f)}}\color{black}{:} ${k_{x}}=1$ \text{$\mu$m$^{-1}$}, ${k_{y}}=0.0$. \textbf{\color{blue}{(c)-(g)}}\color{black}{:} ${k_{x}}=0.6$ \text{$\mu$m$^{-1}$}, ${k_{y}}=1$ \text{$\mu$m$^{-1}$}.
			\textbf{\color{blue}{(d)-(h)}}\color{black}{:} ${k_{x}}=1$ \text{$\mu$m$^{-1}$}, ${k_{y}}=0.6$ \text{$\mu$m$^{-1}$}.}}\label{fig6}
\end{figure}

\begin{figure}[H]
	\centering
	\subfloat[]{
		\centering
		\includegraphics[scale=0.185]{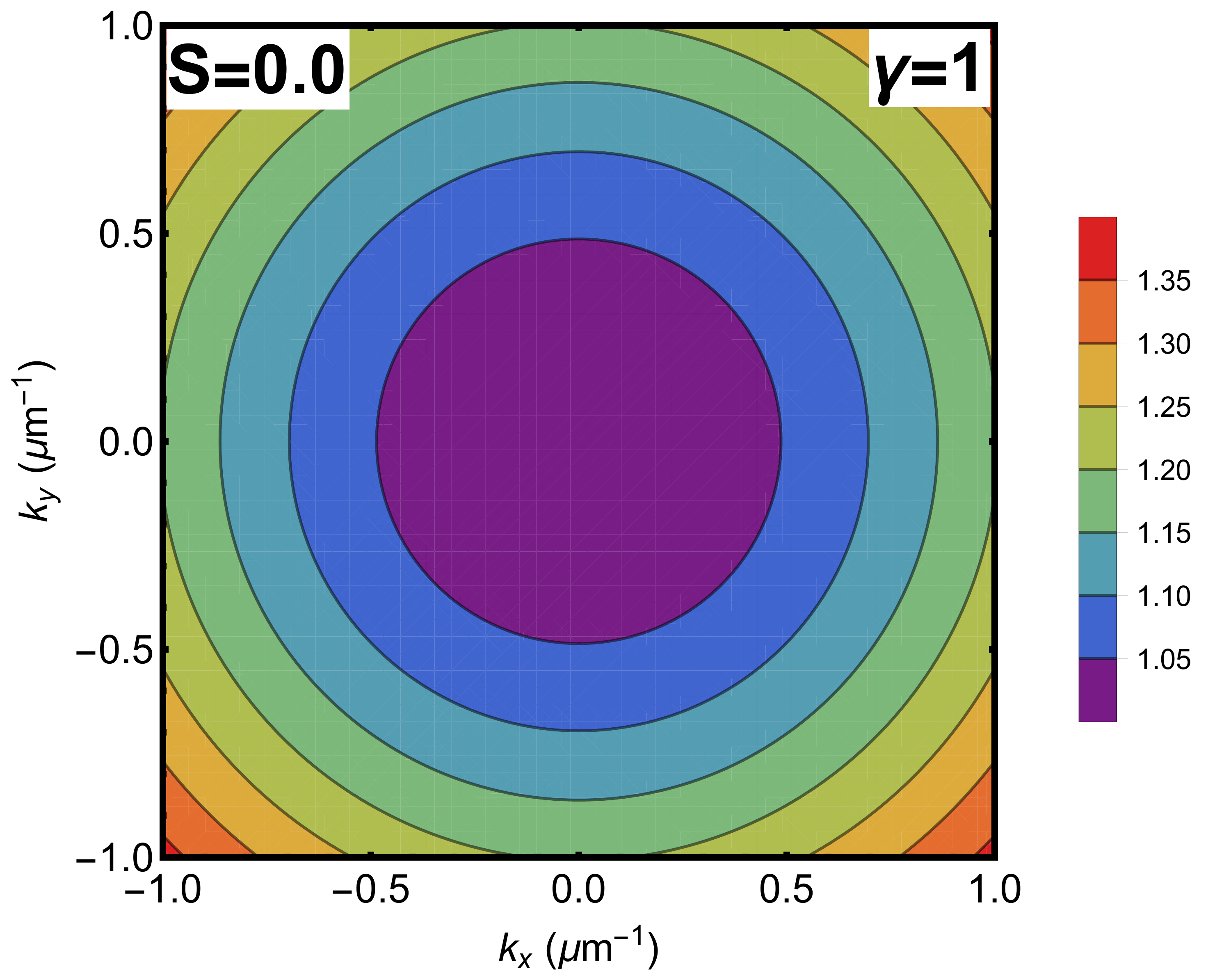}
		\label{figurb5}}
	\subfloat[]{
		\centering
		\includegraphics[scale=0.185]{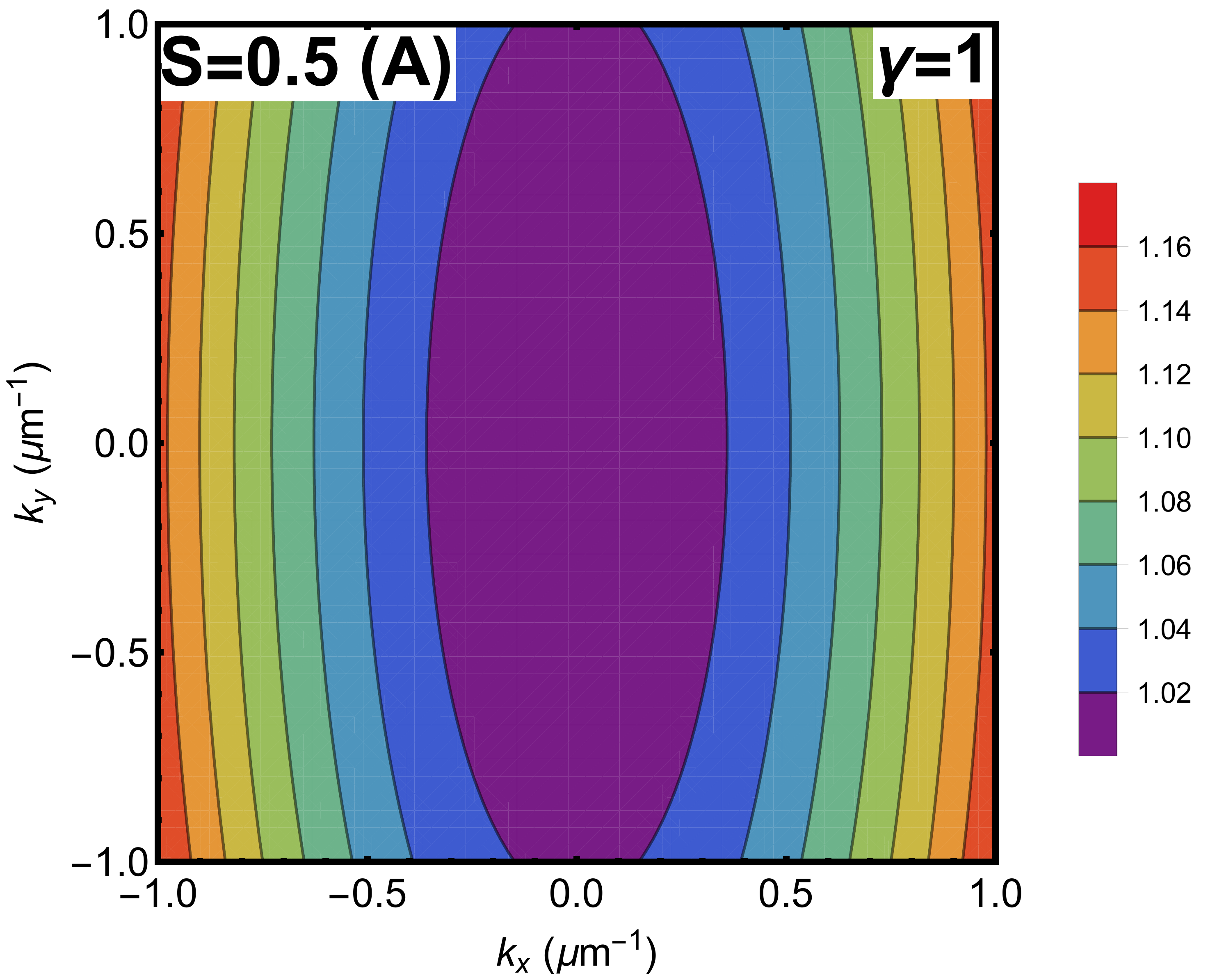}
		\label{figu5}}
	\subfloat[]{
		\centering
		\includegraphics[scale=0.185]{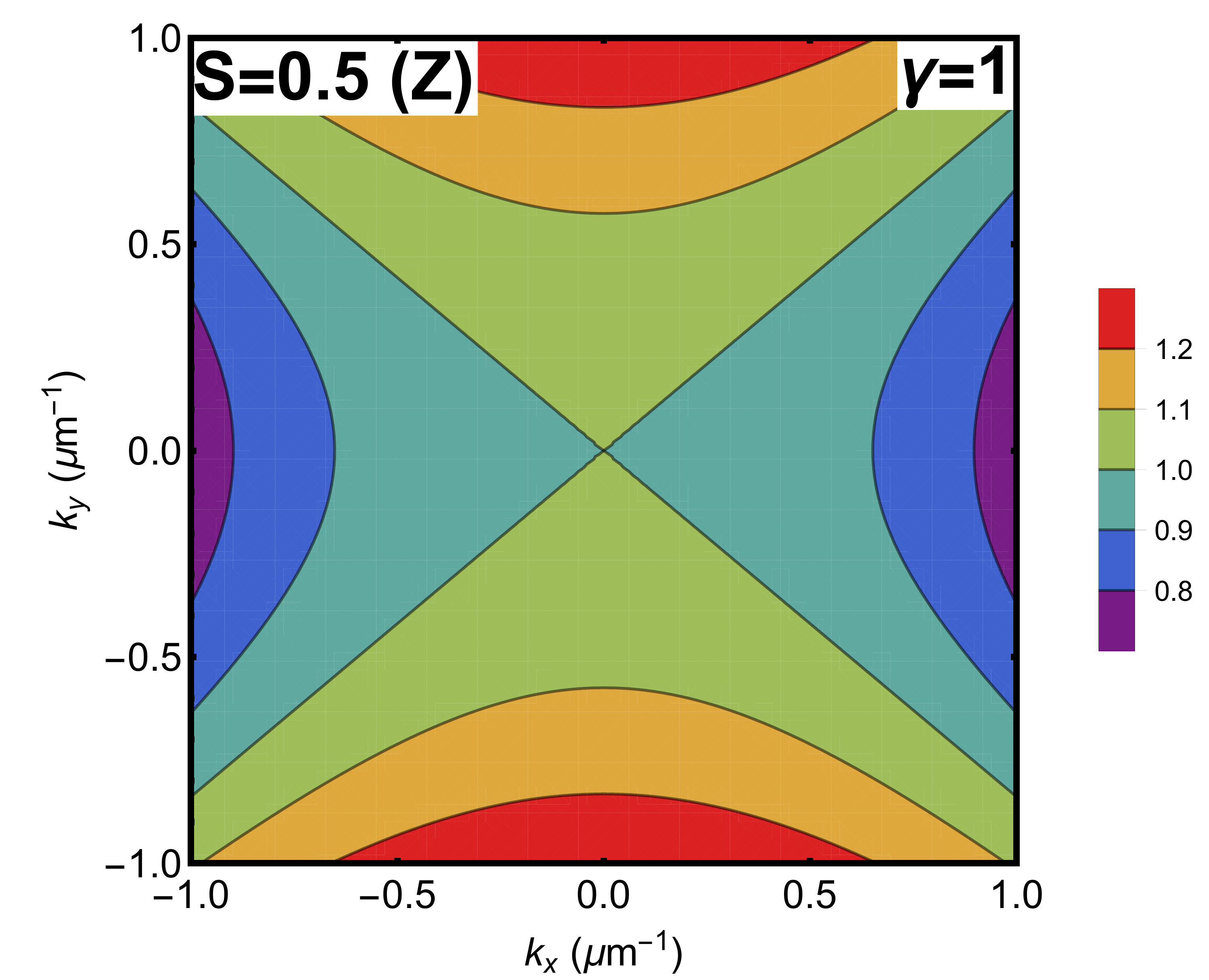}
		\label{figubre}}\\
	\subfloat[]{
		\centering
		\includegraphics[scale=0.185]{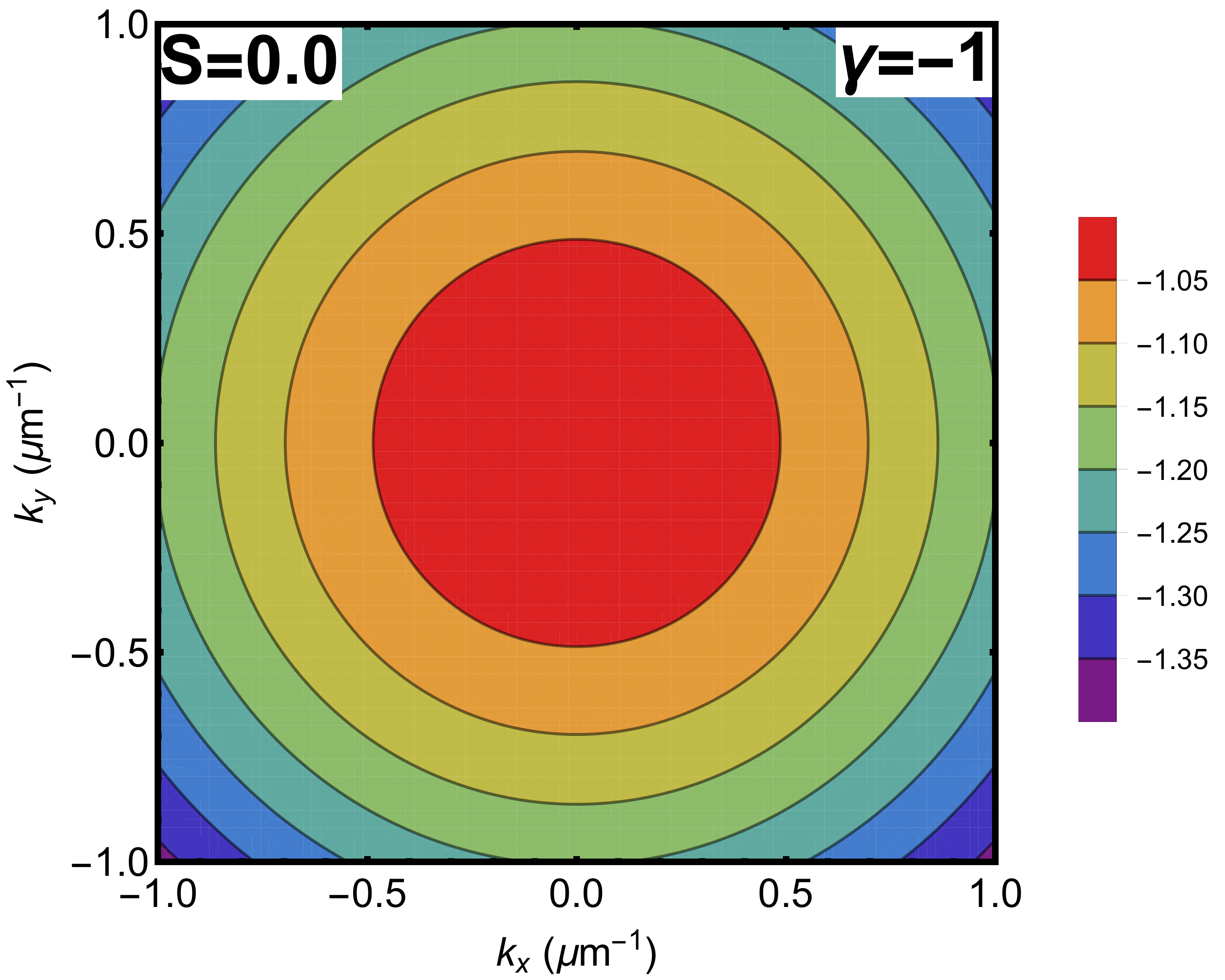}
		\label{figuj5}}
	\subfloat[]{
		\centering
		\includegraphics[scale=0.185]{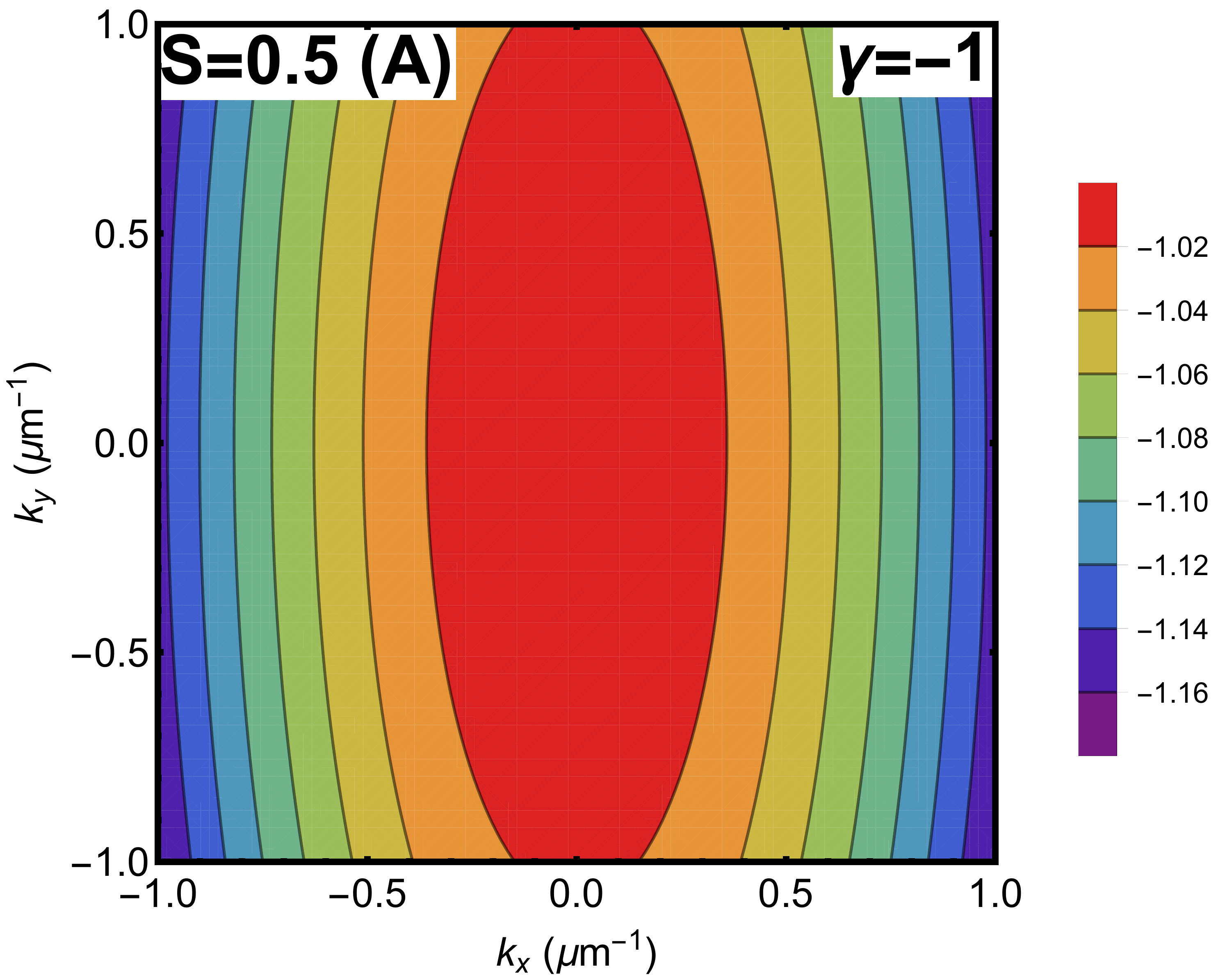}
		\label{figuj}}
	\subfloat[]{
		\centering
		\includegraphics[scale=0.185]{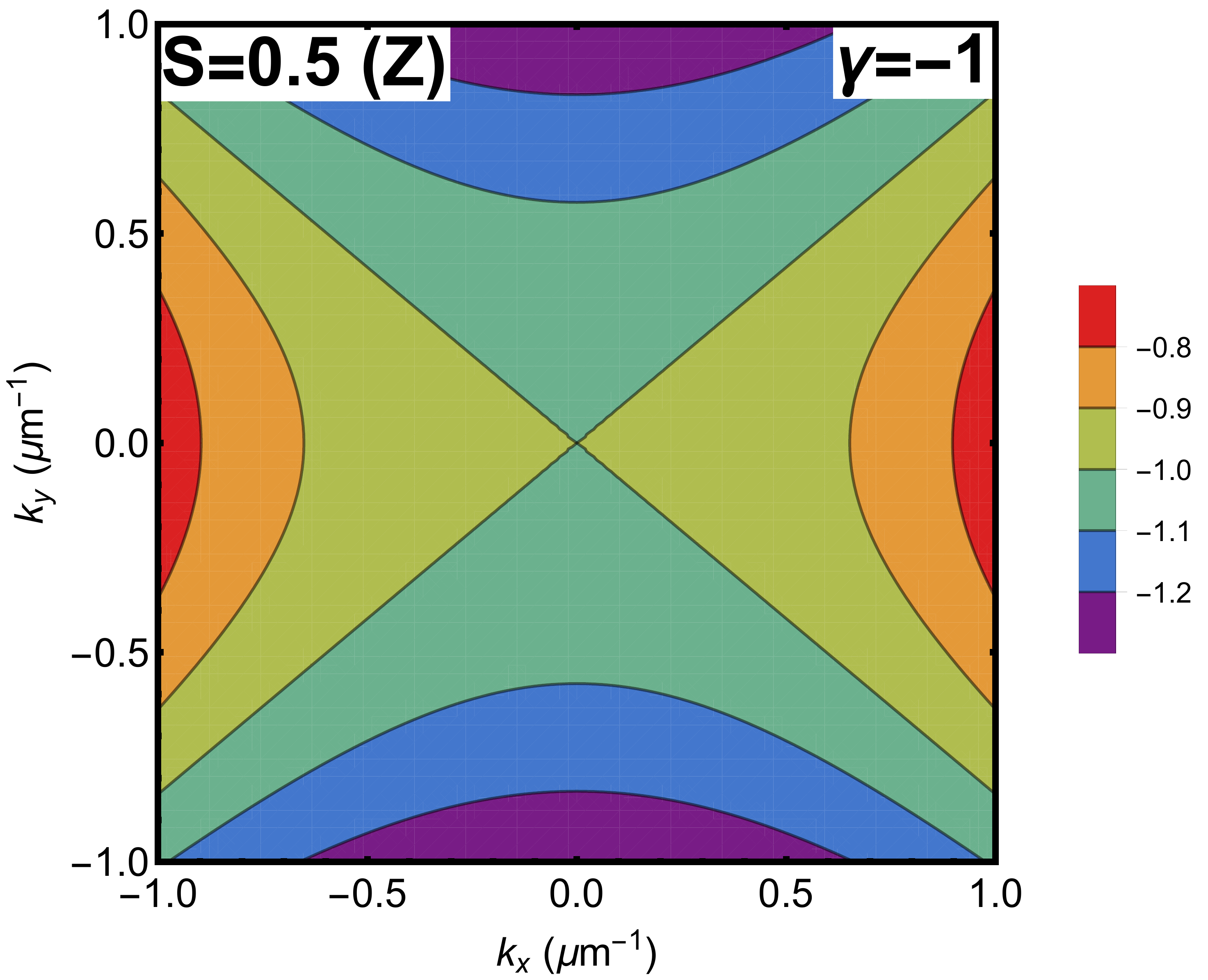}
		\label{fig}}
	\caption{\sf{(color online) Contour plot of the energy spectrum $\varepsilon$ of electron dressed by the linearly polarized field versus the  wave vector components (${k_{x}},{k_{y}}$) for {$I=0.0$, ${\Delta_{g}}=2$  \text{meV}, $\hbar\omega=10$  \text{meV}}\color{black}{.}
			\textbf{\color{blue}{(a-d)}}: Without strain ($S=0.0$). \textbf{\color{blue}{(b-e)}}:
			Effect of armchair strain direction $S=0.5$.
			\textbf{\color{blue}{(c-f)}}:
			Effect of zigzag strain direction with $S=0.5$.}}\label{fhig6}
\end{figure}

{To underline the effect of the strain along armchair and zigzag directions in the absence of the dressing field ($I=0.0$) on the energy spectrum  $\varepsilon$ of dressed electron described by  (\ref{rt18}), we present  Figure \ref{fhig6} showing a comparison of the contour plot for $\varepsilon$ at $\gamma=\pm1$ versus the wave vector components ($k_{x}, k_{y}$) with ${\Delta_{g}}=2$  \text{meV}, $\hbar\omega=10$  \text{meV}}.
Indeed,  we observe that in Figures
\text{\ref{fhig6}}\textbf{\color{blue}{(a)}},
\text{\ref{fhig6}}\textbf{\color{blue}{(b)}} for the strainless case ($S=0.0$), $\varepsilon$ has a circular form {which is similar to that observed for gapped graphene (${\Delta_{g}}\neq0.0$)}. When the strain ($S=0.5$) is applied along armchair direction, $\varepsilon$ grows up {perpendicularly} between the negative and positive values of $k_{x}$ taking an elliptic   {and anisotropic forms. Now for a strain ($S=0.5$) applied  along the zigzag direction, it is  clearly see that {the obvious change that we have $\varepsilon$} takes   hyperbolic form and  its band gap becomes very large along the wave vector components ($k_{x},k_{y}$), but it becomes linear and isotropic for   $\varepsilon=1$ \text{meV} and $1.2$ \text{meV} as in the case of pristine graphene}, also it presents a symmetry at {($k_{x}=k_{y}=0.0$)} \color{black}{as} shown in Figures
\text{\ref{fhig6}}\textbf{\color{blue}{(c)}},
\text{\ref{fhig6}}\textbf{\color{blue}{(f)}}. Moreover, Figures \text{\ref{fhig6}}\textbf{\color{blue}{(a)}},
\text{\ref{fhig6}}\textbf{\color{blue}{(b)}},
\text{\ref{fhig6}}\textbf{\color{blue}{(c)}} show the same behaviors as those in Figures
\text{\ref{fhig6}}\textbf{\color{blue}{(d)}},
\text{\ref{fhig6}}\textbf{\color{blue}{(e)}},
\text{\ref{fhig6}}\textbf{\color{blue}{(f)}} except that the first Figures correspond to $\gamma=1$ where $\varepsilon$ decreases inside for different values of $k_x$ and $k_y$, but it increases in the second Figures for $\gamma=-1$. The interesting results is that the effect of strain effect 
	for $I=0.0$ modifies the dispersion relation into an anisotropic and isotropic forms. In addition,  $\varepsilon$ can be exhibited an inter-valley spectrum symmetry and controlled from negative to positive values by changing the sign of $\gamma$.

\begin{figure}[H]
	\centering
	\subfloat[]{
		\centering
		\includegraphics[scale=0.1886]{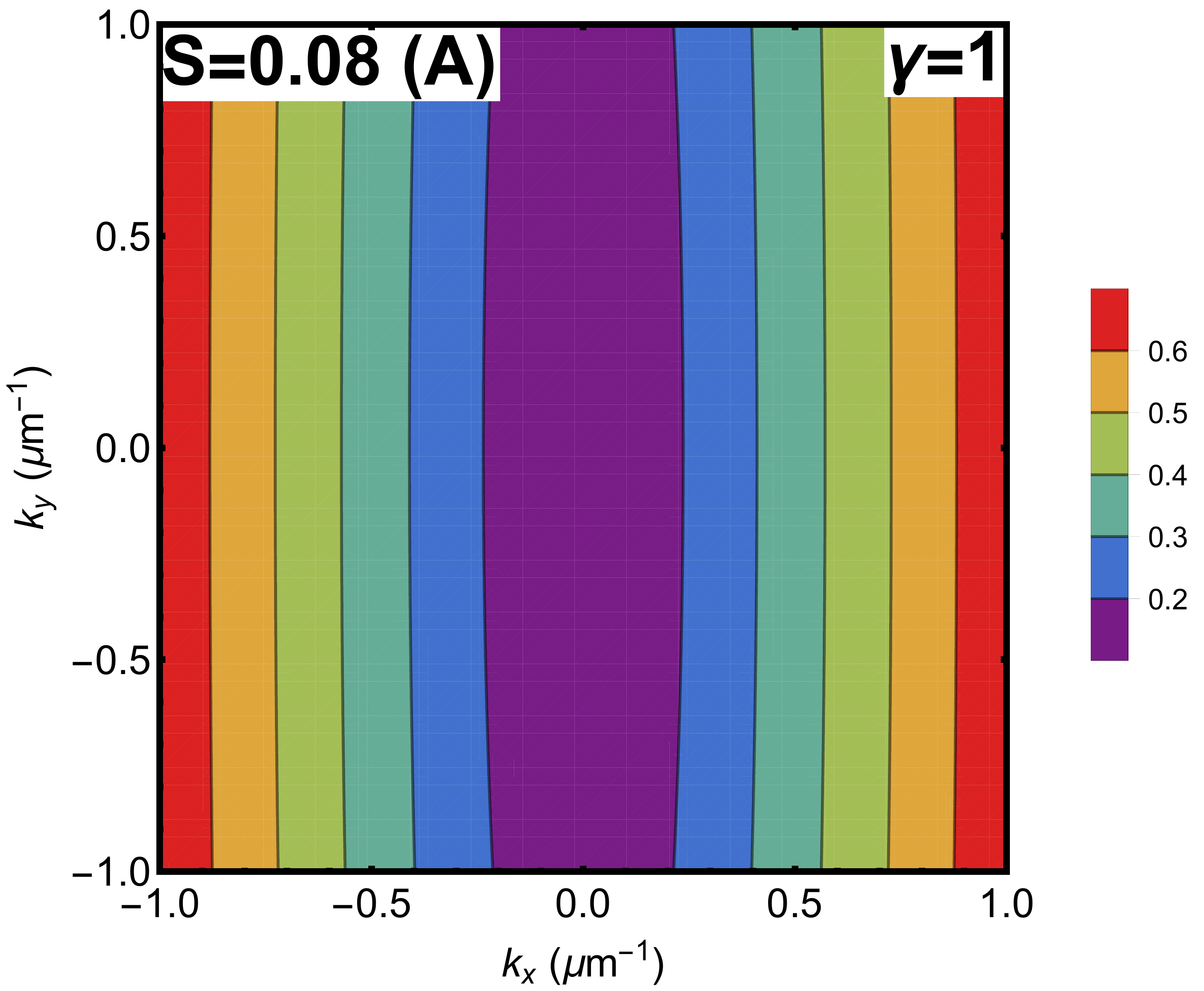}\label{faBg}}
	\subfloat[]{
		\centering
		\includegraphics[scale=0.1886]{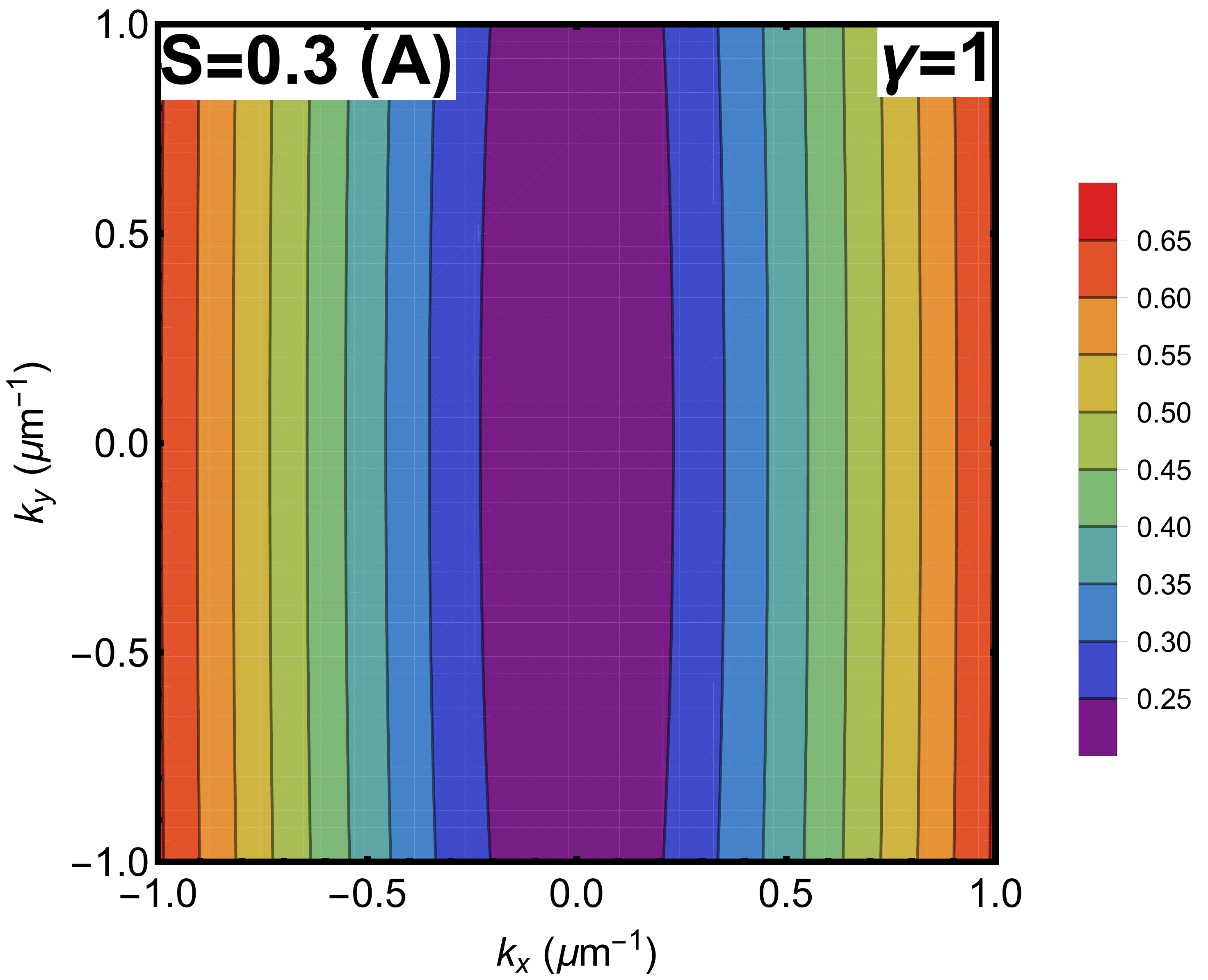}
		\label{fiaT}}
	\subfloat[]{
		\centering
		\includegraphics[scale=0.1886]{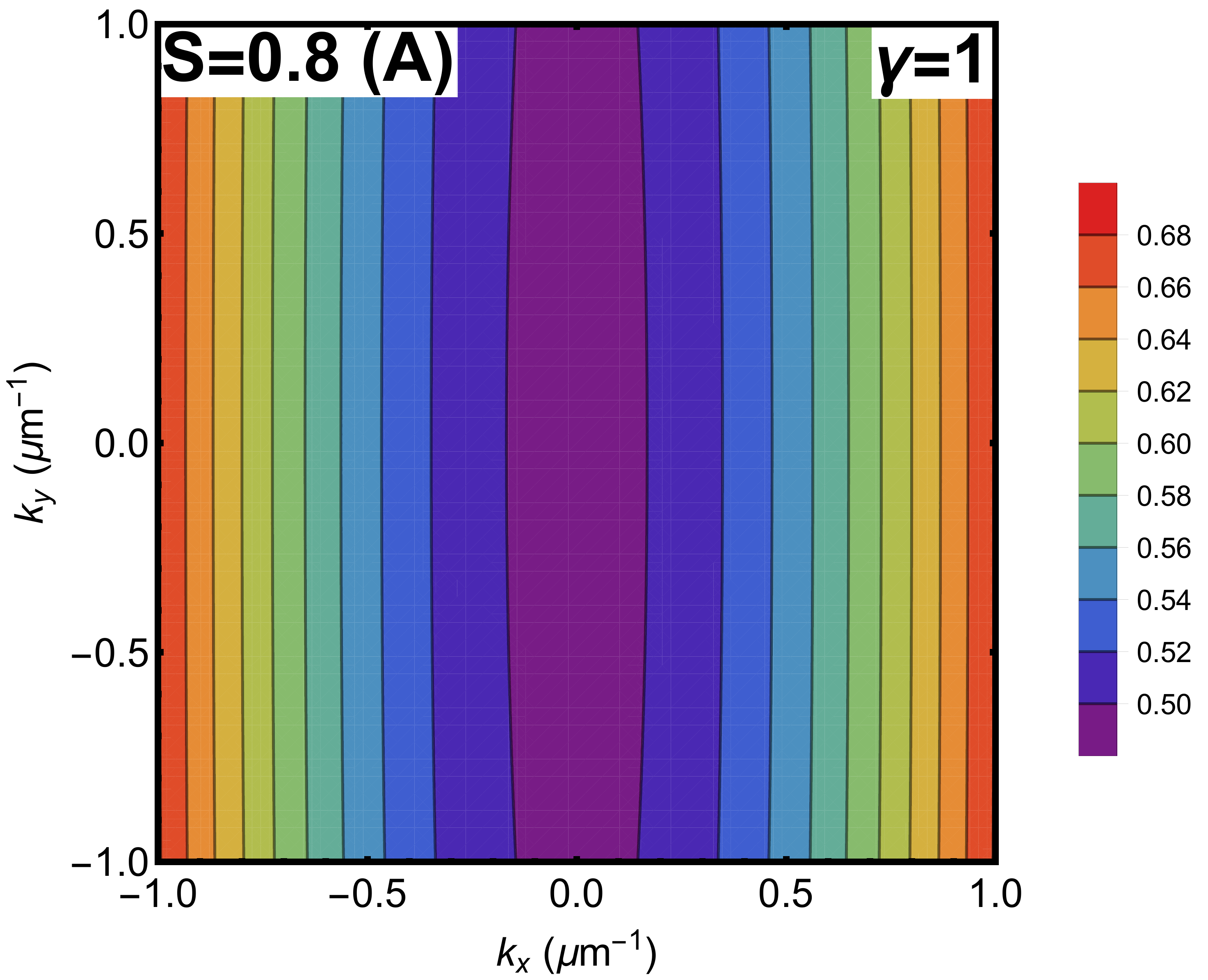}
		\label{fia}}\\
	\subfloat[]{
		\centering
		\includegraphics[scale=0.187]{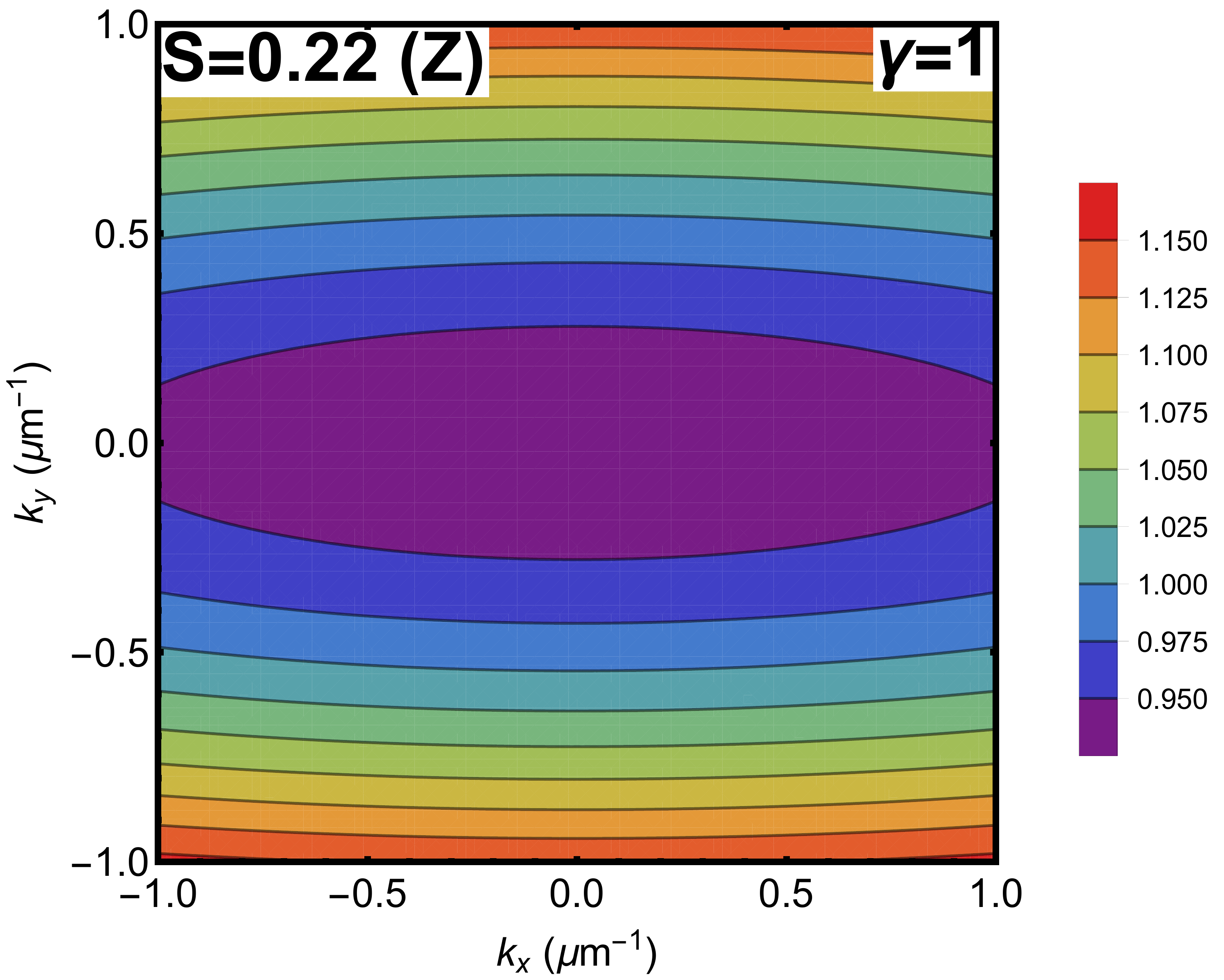}
		\label{fiBg}}
	\subfloat[]{
		\centering
		\includegraphics[scale=0.187]{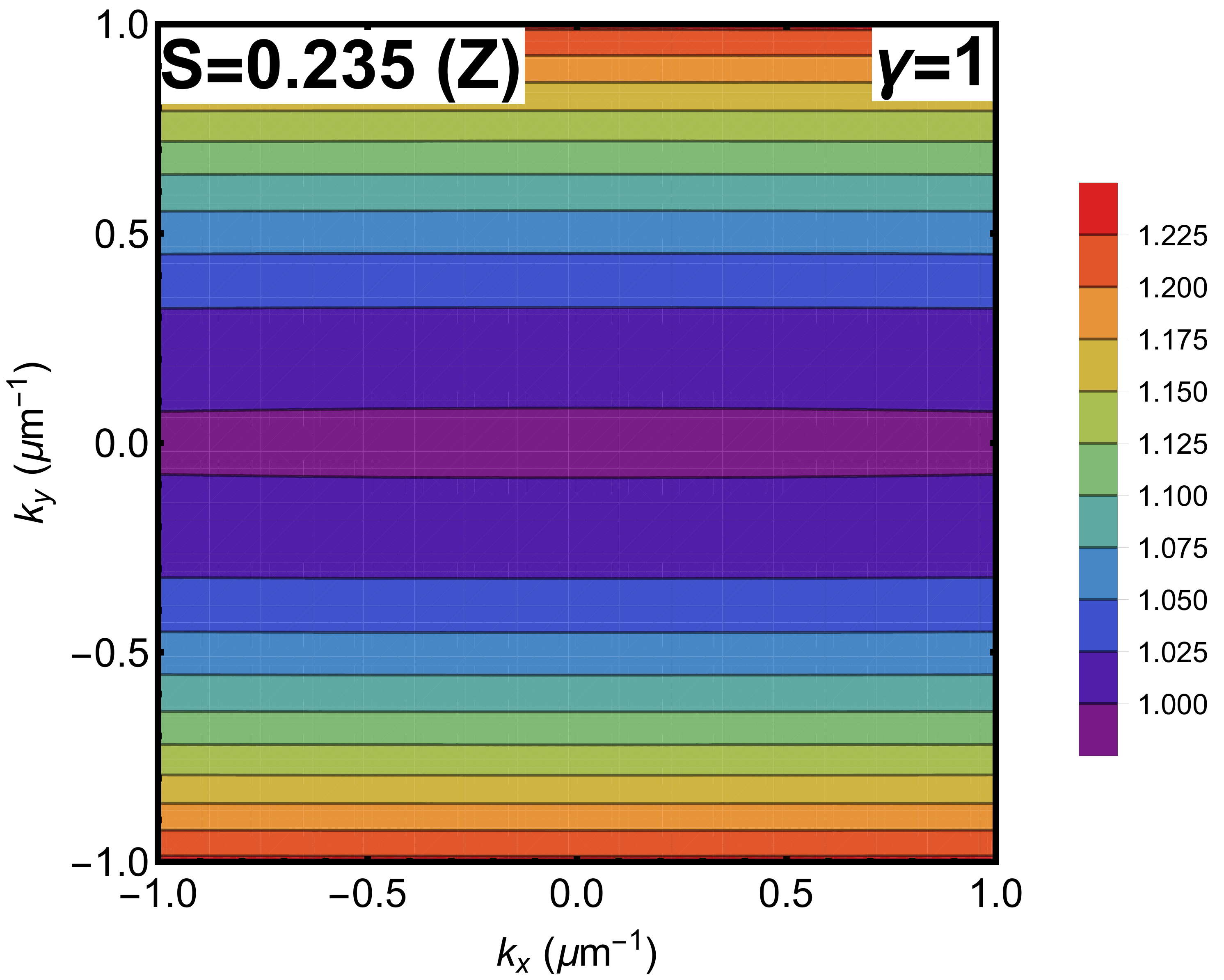}
		\label{fiT}}
	\subfloat[]{
		\centering
		\includegraphics[scale=0.187]{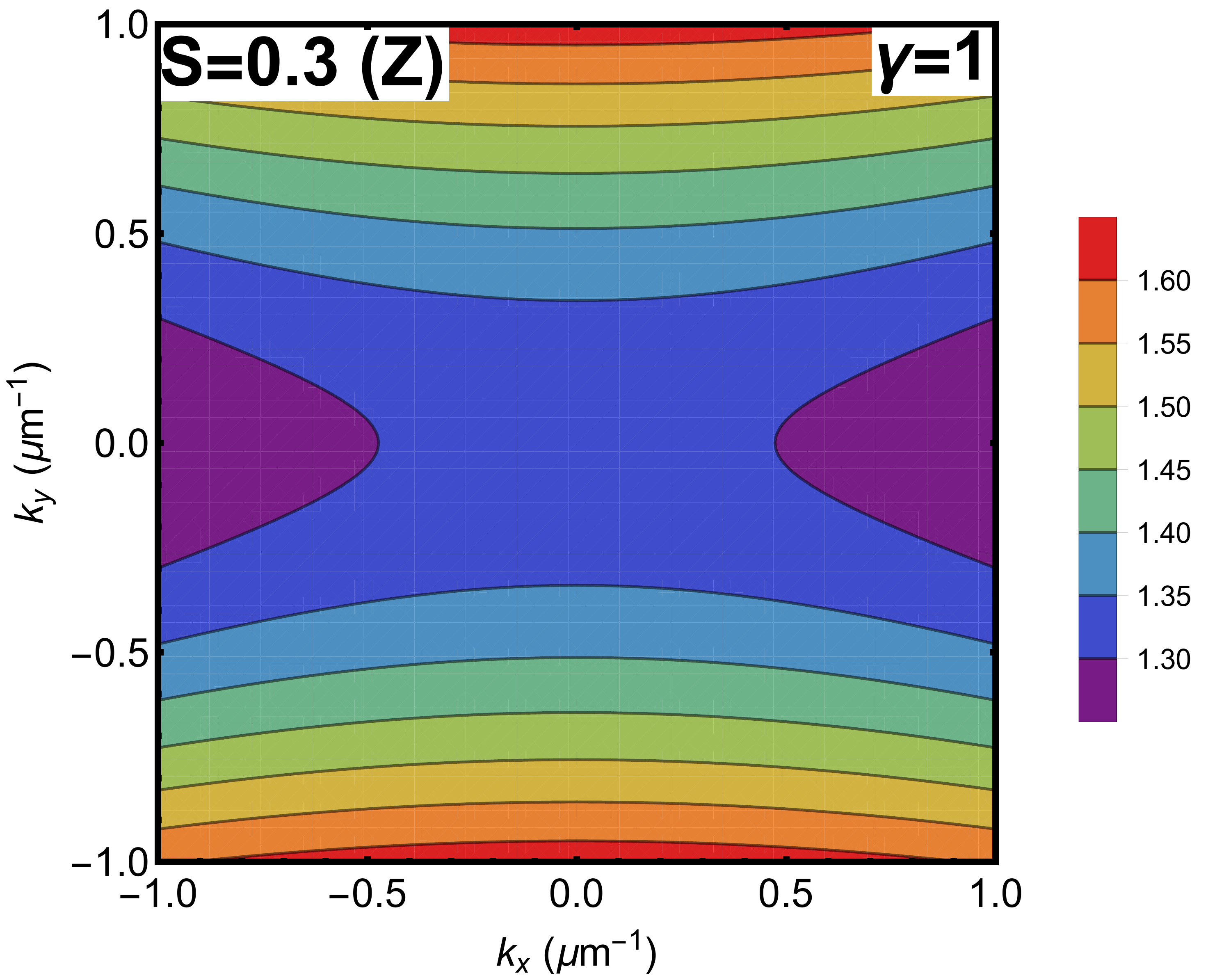}
		\label{fi}}
	\caption{\sf{(color online) Contour plot of the energy spectrum $\varepsilon$ of electron dressed by the linearly polarized field versus the wave vector components (${k_{x}},{k_{y}}$) for {$I=26.7$ \text{kW/\text{cm$^{2}$}}, ${\Delta_{g}}=2$  \text{meV}}, $\hbar\omega=10$  \text{meV},} \color{black}{showing the effects of armchair strain direction in  \textbf{\color{blue}{(a)}},\textbf{\color{blue}{(b)}},\textbf{\color{blue}{(c)}} and zigzag strain direction in
			\textbf{\color{blue}{(d)}},\textbf{\color{blue}{(e)}},\textbf{\color{blue}{(f)}}. }}\label{fig8}
\end{figure}

{To illustrate the effect of  some strain magnitudes in the presence of the dressing field ($I\neq0$) on the energy spectrum $\varepsilon$ of dressed electron for a linear polarization, we present in Figure \ref{fig8} the contour plot of $\varepsilon$ as function of the wave vector components ($k_{x}, k_{y}$) with  $I=26.7$ \text{kW/\text{cm$^{2}$}},  ${\Delta_{g}}=2$  \text{meV}, $\hbar\omega=10$  \text{meV}, $\gamma=1$. It is interesting to note that in Figure \text{\ref{fig8}}\textbf{\color{blue}{(a)}
	$\varepsilon$ is} {anisotropic}\color{black}{,} but it  develops in a slow manner in Figures \text{\ref{fig8}}\textbf{\color{blue}{(b)}} and
\text{\ref{fig8}}\textbf{\color{blue}{(c)}} {where it starts to increase} \color{black}{under} the change of values of the strain applied along the armchair direction and it has minima in the center. 
Now for the zigzag strain direction, we notice that in Figure
\text{\ref{fig8}}\textbf{\color{blue}{(d)}} for $S=0.22$, the energy spectrum is elliptic as shown in Figure
\text{\ref{fig8}}\textbf{\color{blue}{(a)}} but {it takes another form in Figure \text{\ref{fig8}}\textbf{\color{blue}{(b)}}}\color{black}{.} Moreover, it is clearly seen  in Figure
\text{\ref{fig8}}\textbf{\color{blue}{(f)}} for $S=0.3$, $\varepsilon$ is symmetric between the valence and conduction bands and isotropic} \color{black}{along the different values of} {${k_{x}}$ but parabolic along ${k_{y}}$ for 
$\varepsilon=0$}.  {Note that as long as $S$ increases {\color{blue} with $I\neq0$},  $\varepsilon$ 
increases rapidly in one direction either perpendicularly (Figures \text{\ref{fig8}}\textbf{\color{blue}{(a)}},\text{\ref{fig8}}\textbf{\color{blue}{(b)}},\text{\ref{fig8}}\textbf{\color{blue}{(c)}}) or horizontally (Figures \text{\ref{fig8}}\textbf{\color{blue}{(d)}},\text{\ref{fig8}}\textbf{\color{blue}{(e)}},\text{\ref{fig8}}\textbf{\color{blue}{(f)}}) along  $k_{x}$-axis. Additionally, $\varepsilon$  takes different forms when a strain is applied along zigzag direction, which is not the case for the armchair one.
In conclusion, we notice that the energy spectrum of fermions in gapped graphene can be adjusted by applying the strain  $S$ and irradiation intensity $I$}.

Figure \ref{elip00} shows the energy spectrum
$\varepsilon$ of electron dressed by the elliptically polarized field versus the wave vector $k_x$ for {${\Delta_{g}}=2$ \text{meV}, $\hbar\omega=10$ \text{meV}, $\theta=(\frac{\pi}{6},\frac{\pi}{2})$ and two values of the irradiation intensity $I=0.0$ (purple and blue lines), $I=2.53$ \text{kW/\text{cm$^{2}$}} (red and black lines) and (orange and green lines) correspond to $\tau=-1$ and $\tau=1$, respectively, with $S=(0.0, 0.5)$ for (A) and $S=0.1$ for (Z)}\color{black}{.} 
{We observe that for the strainless case ($S=0.0$) on contrary to the linearly polarized electromagnetic wave, $\varepsilon$ is still isotropic whatever the value of $I$ and does not induces the anisotropy, see Figure \ref{fig4}, as well as the difference becomes very important for the both valleys of the Brillouin zone (valley indices $\tau=\pm1$) which are in agreement with those obtained in \cite{Kibis}. We also notice that by increasing the values of the irradiation intensity for $\tau=-1$, the band gap of $\varepsilon$ becomes large for $\theta=\pi/6$ but small for $\theta=\pi/2$ and vice verse for $\tau=1$. Moreover, we clearly see that in Figures \ref{elip00}\textbf{\color{blue}{(a)}}, \ref{elip00}\textbf{\color{blue}{(b)}} for a strain $S=0.5$ applied along armchair direction, the energy spectrum presents the same behavior as for the case $S=0.0$ except that it is increased for $\tau=-1$ and decreased for $\tau=1$. Now for   zigzag direction with strain $S=0.1$ as shown in Figures \ref{elip00}\textbf{\color{blue}{(c)}}, \ref{elip00}\textbf{\color{blue}{(d)}}, $\varepsilon$ takes an anisotropic form for $I=2.53$ \text{kW/\text{cm$^{2}$}}, $(\tau=\pm1, \theta=\pi/6)$ and $(\tau=1, \theta=\pi/2)$ while it is linear and similar to that of pristine graphene for $(\tau=-1, \theta=\pi/2)$\color{black}{.} {These results show that} \color{black}{the energy spectrum can be adjusted from positive to negative  values  by changing the irradiation intensity $I$,  polarization $\theta$, sign of $\gamma$ and valley index $\tau$}.

\begin{figure}[H]
	\centering
	\subfloat[]{
		\centering
		\includegraphics[scale=0.23]{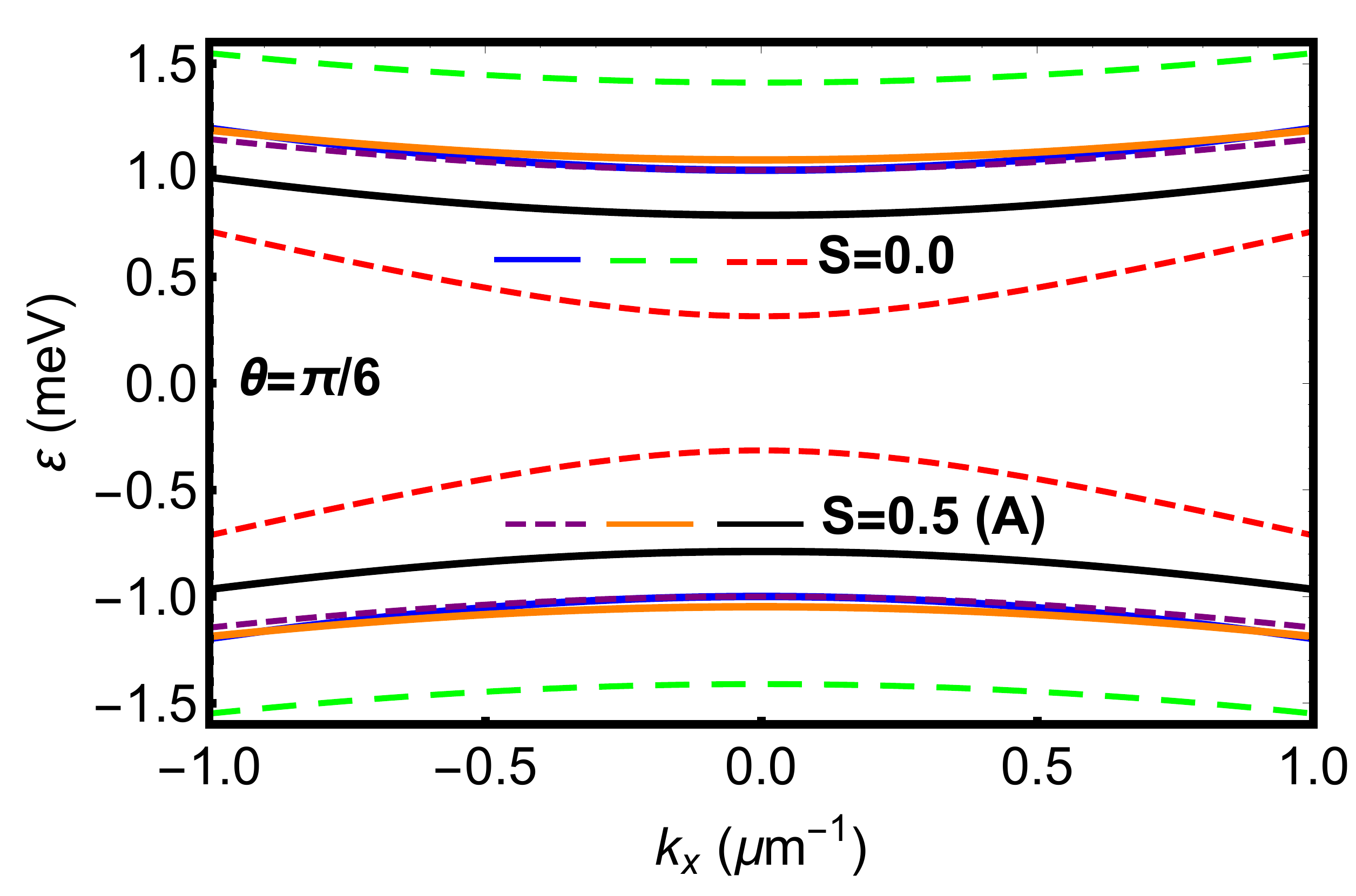}
		\label{figure5}}\hspace{2cm}
	\subfloat[]{
		\centering
		\includegraphics[scale=0.23]{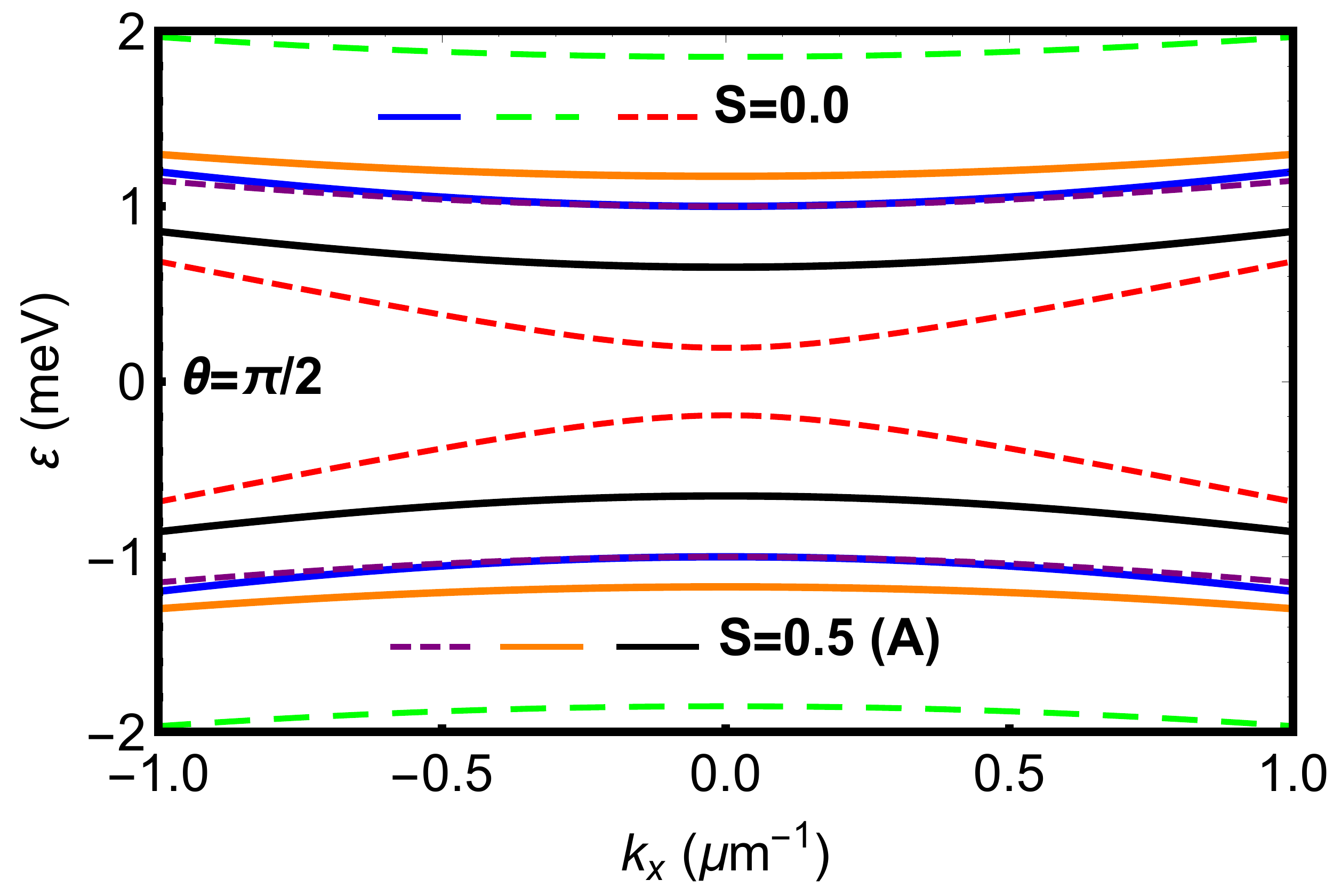}
		\label{figure6}} \\
	\subfloat[]{
		\centering
		\includegraphics[scale=0.23]{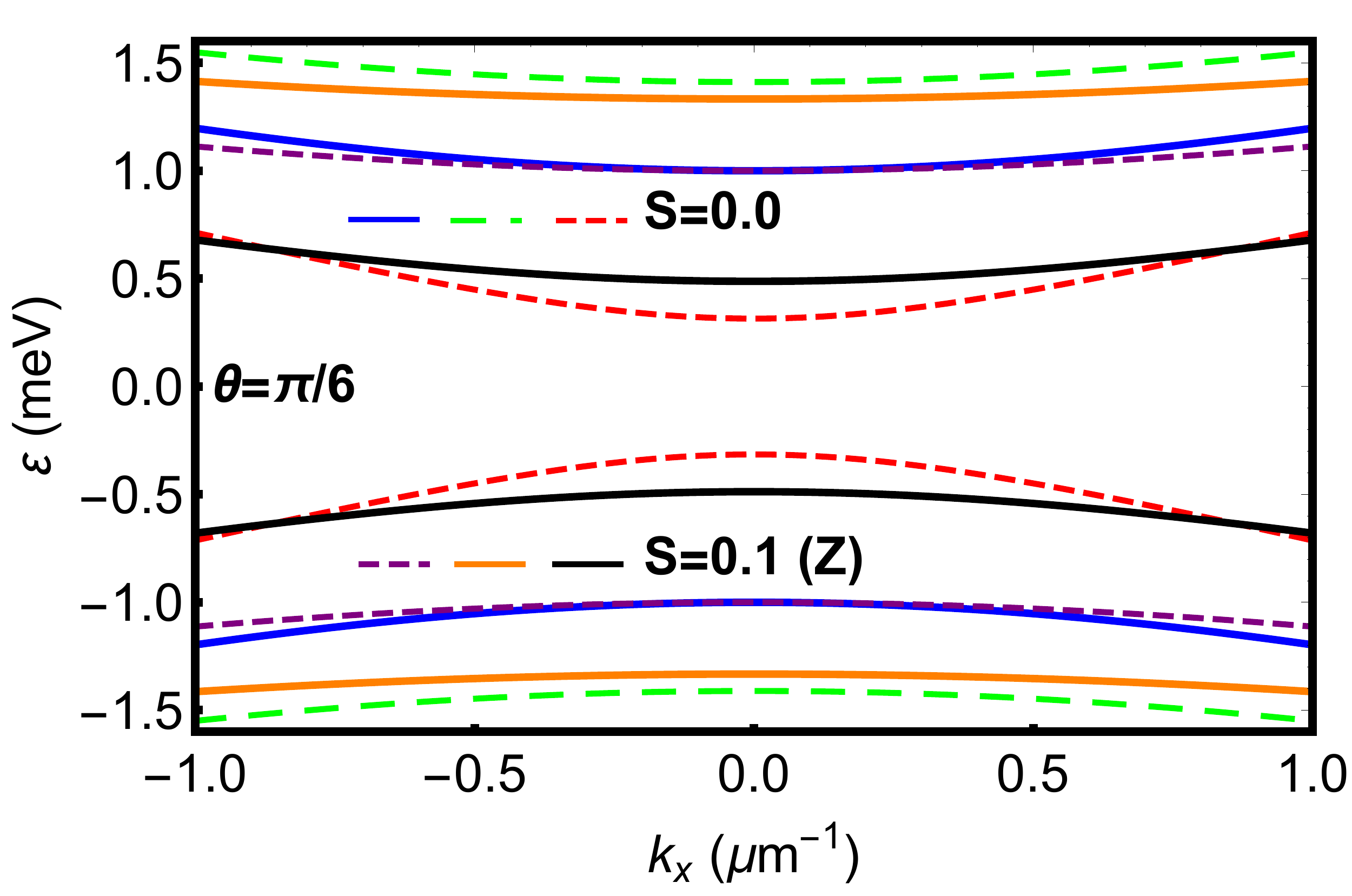}
		\label{figurje5}}\hspace{2cm}
	\subfloat[]{
		\centering
		\includegraphics[scale=0.23]{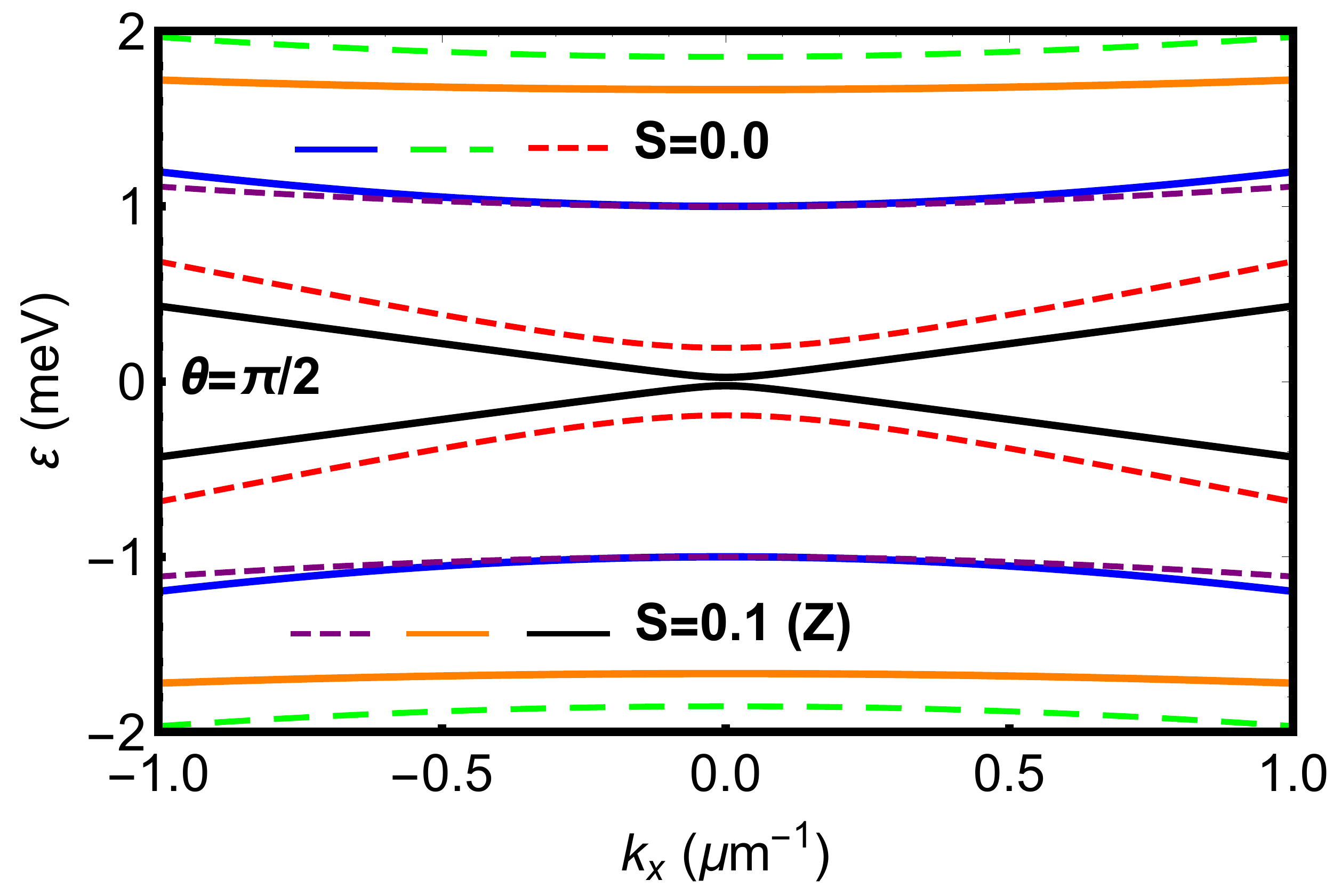}
		\label{figurej6}}
	\caption{\sf{(color online) The energy spectrum
			$\varepsilon$ of electron dressed by the elliptically polarized field versus the wave vector component $k_x$ for ${\Delta_{g}}={\color{blue}2}$ \text{meV}, $\hbar\omega=10$ \text{meV} and polarization angle $\theta=\frac{\pi}{6}/\frac{\pi}{2}$ with different values of the irradiation intensities {$I = 0.0$} \color{black}{(purple and blue lines)}, {$I=2.53$ \text{kW/\text{cm$^{2}$}} (red and black lines) and (orange and green lines) correspond to $\tau=-1$ and $\tau=1$, respectively}\color{black}.
			\textbf{\color{blue}{(a)}},\textbf{\color{blue}{(b)}}\color{black}{:}
			Effect of armchair strain direction with {$S=0.5$}\color{black}{.} \textbf{\color{blue}{(c)}},\textbf{\color{blue}{(d)}}\color{black}{:}
			Effect of zigzag strain direction with
			{$S=0.1$}\color{black}{.}}}\label{elip00}
\end{figure}

Figure \ref{fig11} illustrates the normalized band gap $|\Lambda_g/\Delta_g|$ for the linearly and circularly polarized fields, versus the irradiation intensity $I$ corresponds  to {$\tau\xi=1$ (green  and orange lines) and  $\tau\xi=-1$ (red and purple lines) for ${\Delta_{g}}=2$  \text{meV}, $\hbar\omega=10$  \text{meV}}\color{black}{,} $\theta={\pi}/{6}$. Note that for $S=0.0$,   $|\Lambda_g/\Delta_g|$ decreases and turns to zero by dressing field, then it increases but becomes also null at {$I=9$ \text{kW/\text{cm$^{2}$}}}\color{black}{,} which is due to zero of the Bessel function {$J_0\left[\frac{2E_{0}v_{x}(S)\mid e\mid}{\hbar\omega^{2}}\right]$}\color{black}{.} We also observe that two curves depend of the clockwise/counterclockwise circularly polarization field
(polarization indices $\xi=\pm 1$) and different valleys of the Brillouin zone (valley indices $\tau=\pm 1$). Indeed, we find for the
case of $\tau\xi=-1$,
$|\Lambda_g/\Delta_g|$
monotonously increases with irradiation intensity. However, for the case of $\tau\xi=1$,  $|\Lambda_g/\Delta_g|$ decreases to zero and then starts to grow up. We notice that in {Figures \text{\ref{fig11}}\textbf{\color{blue}{(a)}}, 
	\text{\ref{fig11}}\textbf{\color{blue}{(b)}}} \color{black}{when} strain is applied along the armchair direction, 
{for $\tau \xi=\pm1$ we observe that 
	$|\Lambda_g/\Delta_g|$ shifts to the right and intersects at different values of $I$ by increasing the magnitude of the tensional strain.} 
\color{black}{The band gap of the linearly polarized electromagnetic wave presents the same behavior compared} {to the case  $S=0.0$} \color{black}{except that it is equal zero at} {$I=9.7$ \text{kW/\text{cm$^{2}$}} and $I=11$ \text{kW/\text{cm$^{2}$}} for $S=0.2$ and $S=0.4$, respectively.} \color{black}{In} Figures \text{\ref{fig11}}\textbf{\color{blue}{(c)}}, \text{\ref{fig11}}\textbf{\color{blue}{(d)}}, 
it is showed that the strain along the zigzag direction produces obvious change on $|\Lambda_g/\Delta_g|$ {because as long as $I$ increases}\color{black}{,} the renormalized gap when the polarization is linear decreases slowly {for $S=0.2$ but increases rapidly for $S=0.4$} \color{black}{and non null in the interval} }{$0<I<20$ \text{kW/\text{cm$^{2}$}}}\color{black}{,} while it {displaces to the left and is the same} {for circular polarization ($\tau \xi=\pm1$) for $S=0.4$ as presented in Figure \text{\ref{fig11}}\textbf{\color{blue}{(d)}}}\color{black}{.}

\begin{figure}[H]
	\centering
	\subfloat[]{
		\centering
		\includegraphics[scale=0.23]{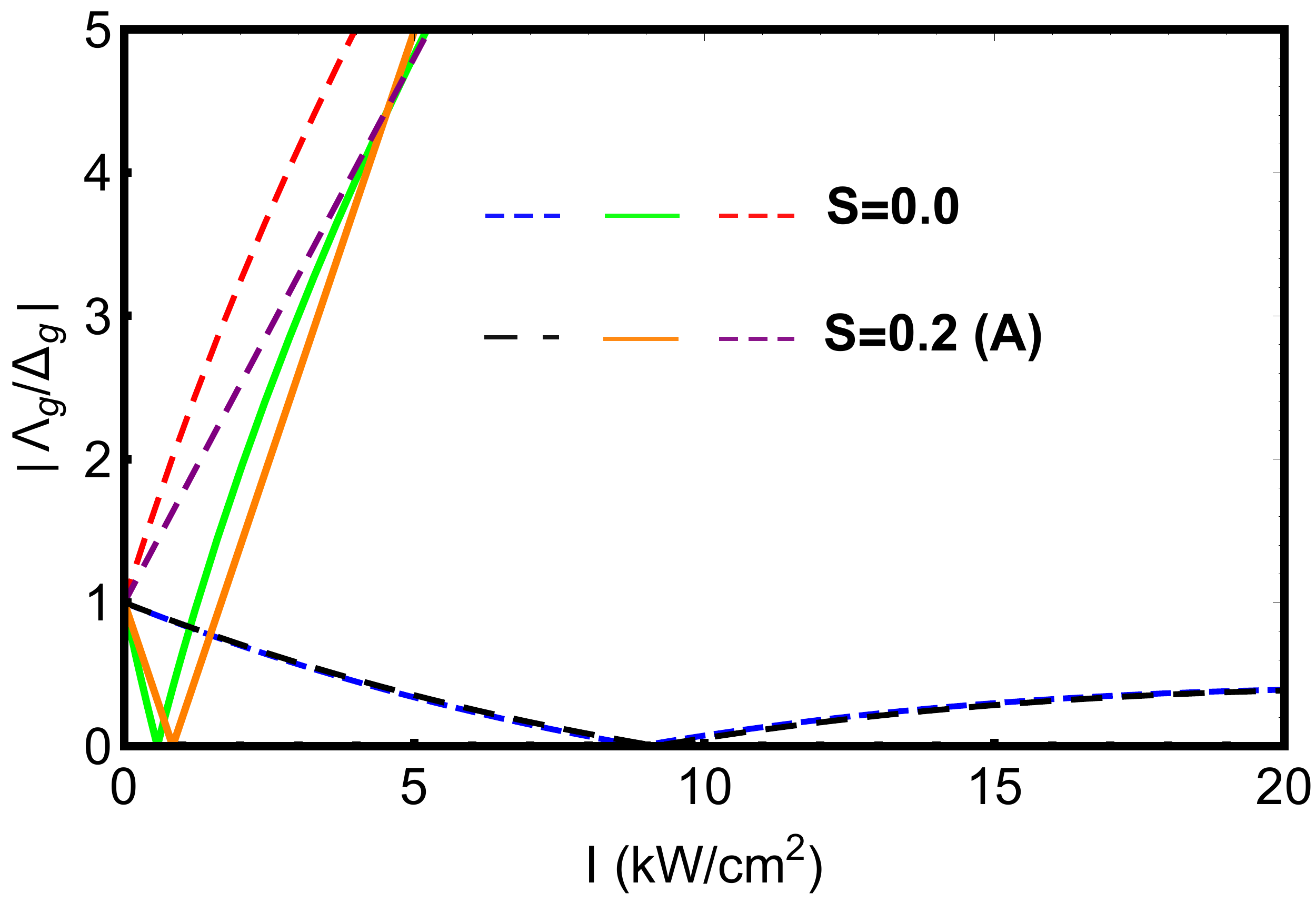}
		\label{fatr}}\hspace{2cm}
	\subfloat[]{
		\centering
		\includegraphics[scale=0.23]{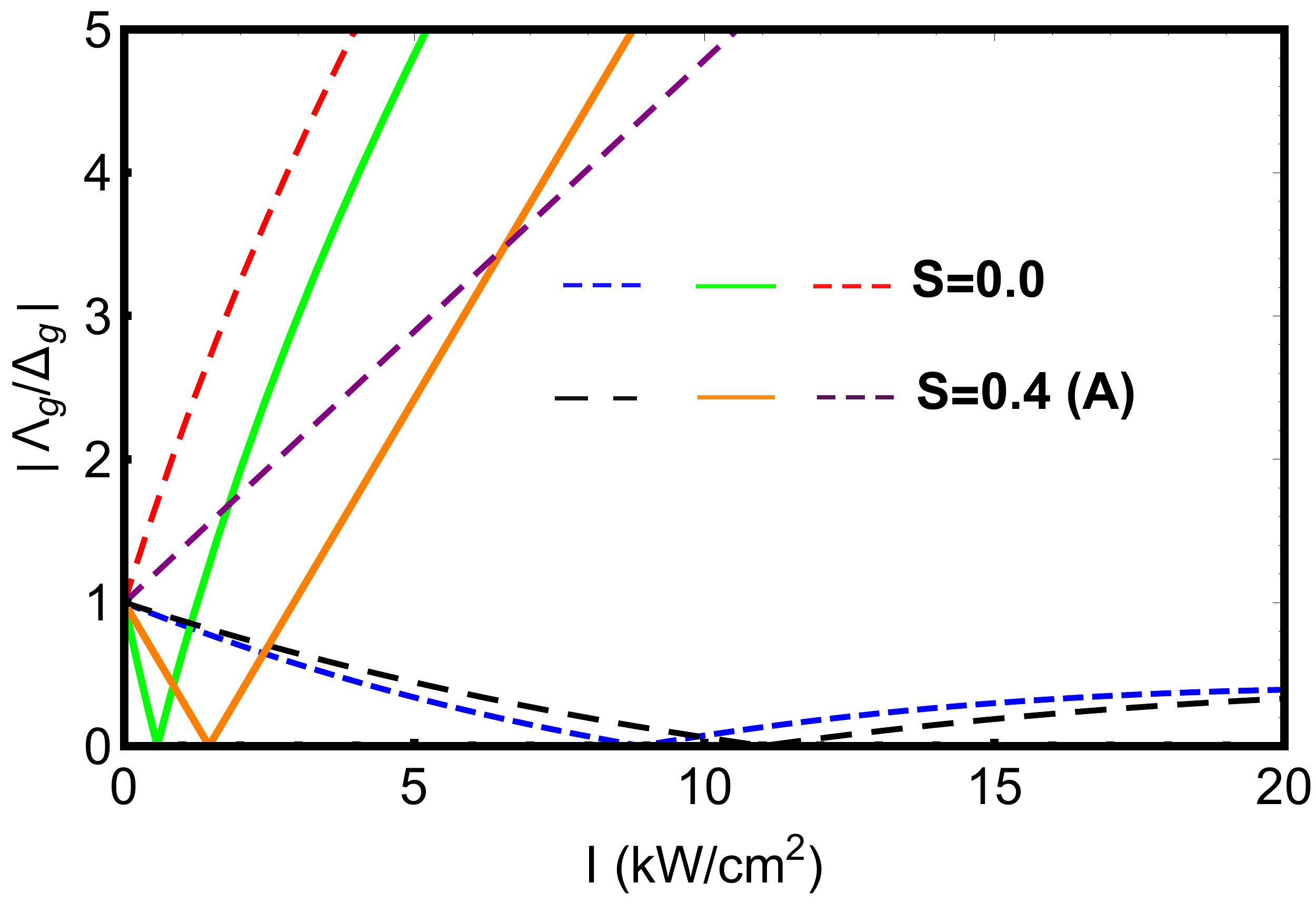}
		\label{firl}}\\
	\subfloat[]{
		\centering
		\includegraphics[scale=0.23]{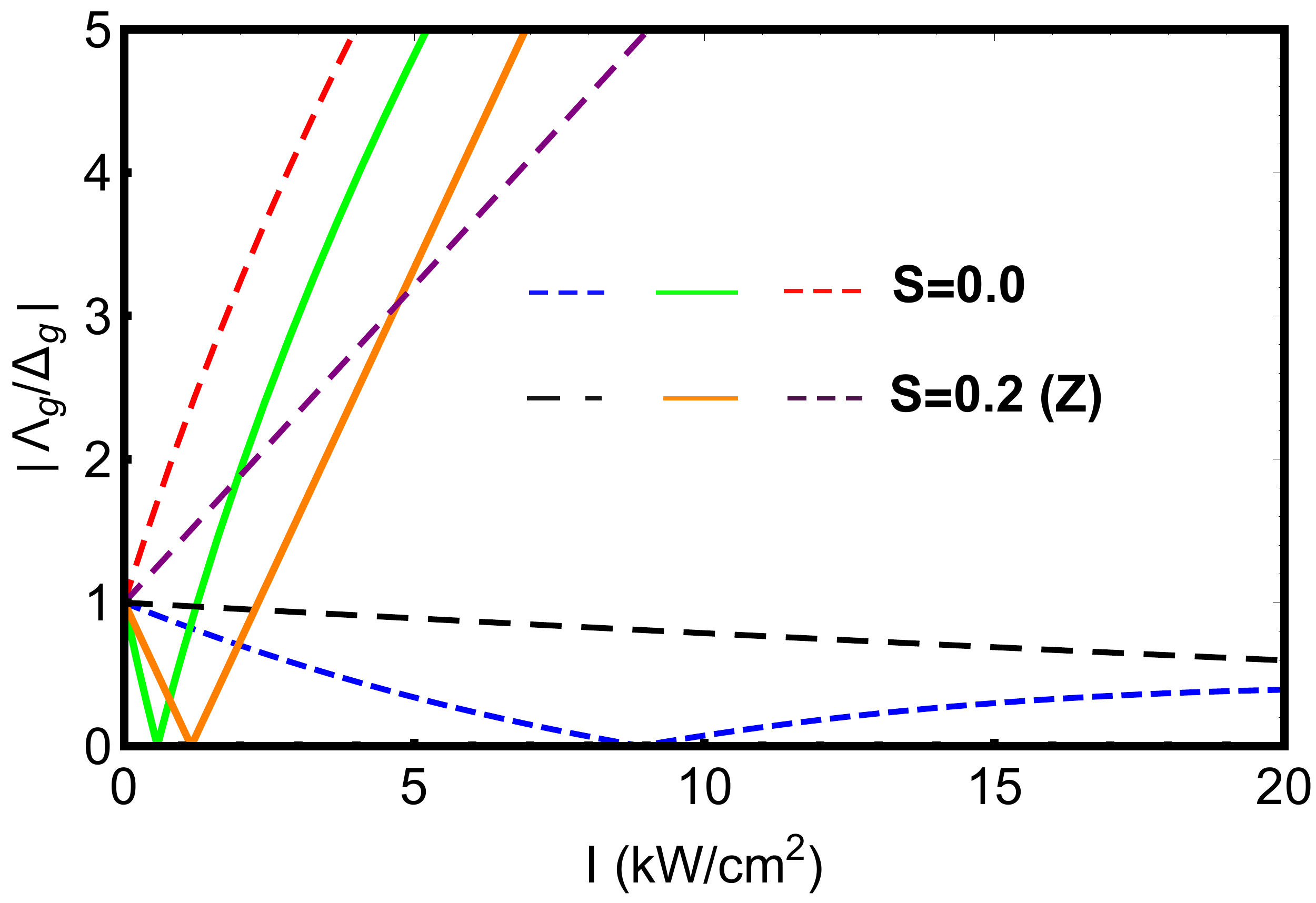}
		\label{fia1}}\hspace{2cm}
	\subfloat[]{
		\centering
		\includegraphics[scale=0.23]{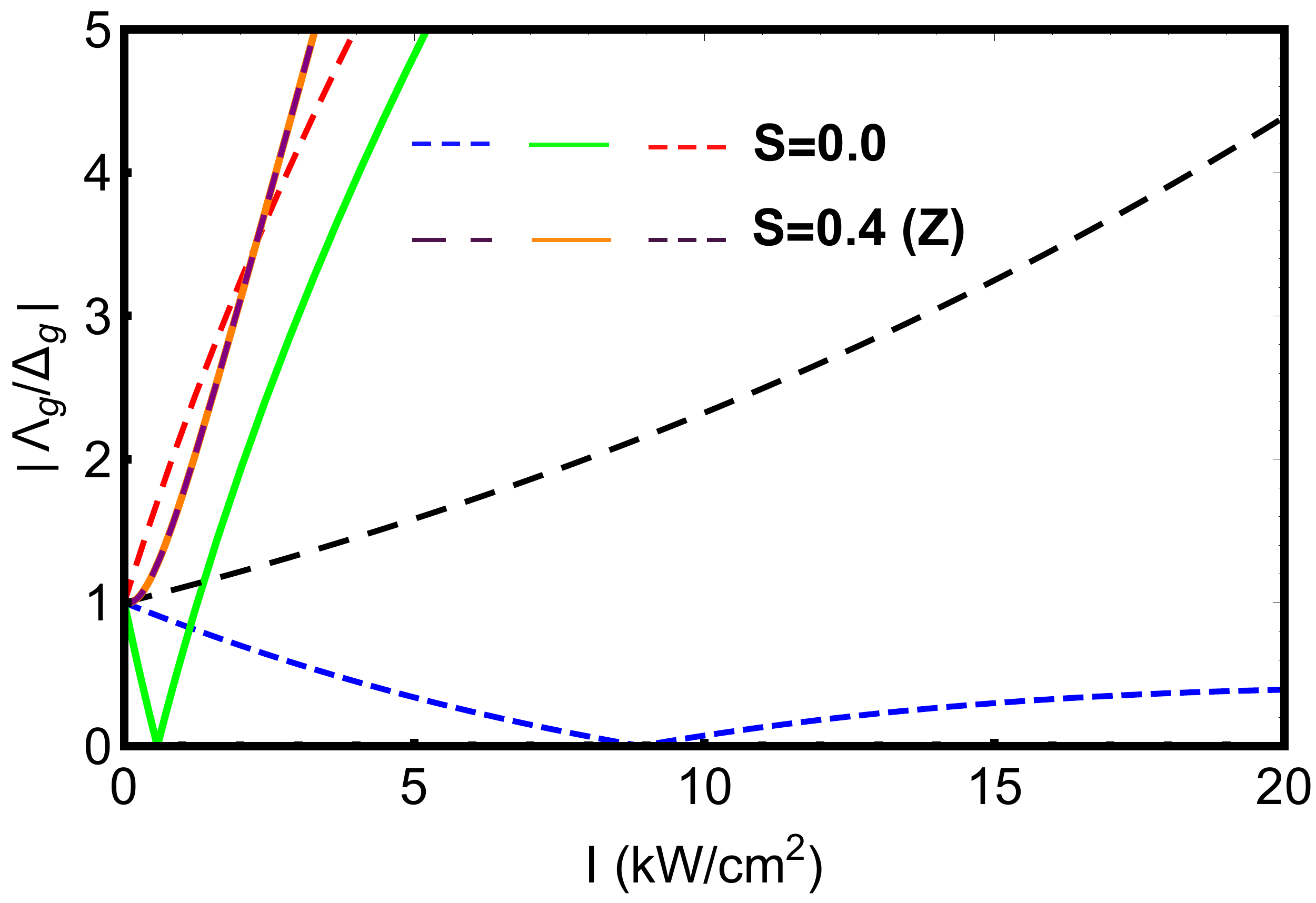}
		\label{fatyr}}
	\caption{\sf{(color online) The band gap $|\Lambda_g/\Delta_g|$
			versus the irradiation intensity $I$ with {${\Delta_{g}}=2$  \text{meV}, $\hbar\omega=10$  \text{meV}} {for linear (blue and black lines) and circular polarization corresponds to  $\tau\xi=1$ (green  and orange lines) and  $\tau\xi=-1$ (red and purple lines)}.
			\textbf{\color{blue}{(a),(b)}}\color{black}{:} Effect of armchair
			strain direction with {$S=0.2,0.4$}\color{black}{.}
			\textbf{\color{blue}{(c),(d)}}\color{black}{:} Effect of zigzag 
			strain direction with  { $S=0.2,0.4$}\color{black}{.}
	}}\label{fig11}
\end{figure}

\begin{figure}[H]
	\centering
	\subfloat[]{
		\includegraphics[scale=0.186]{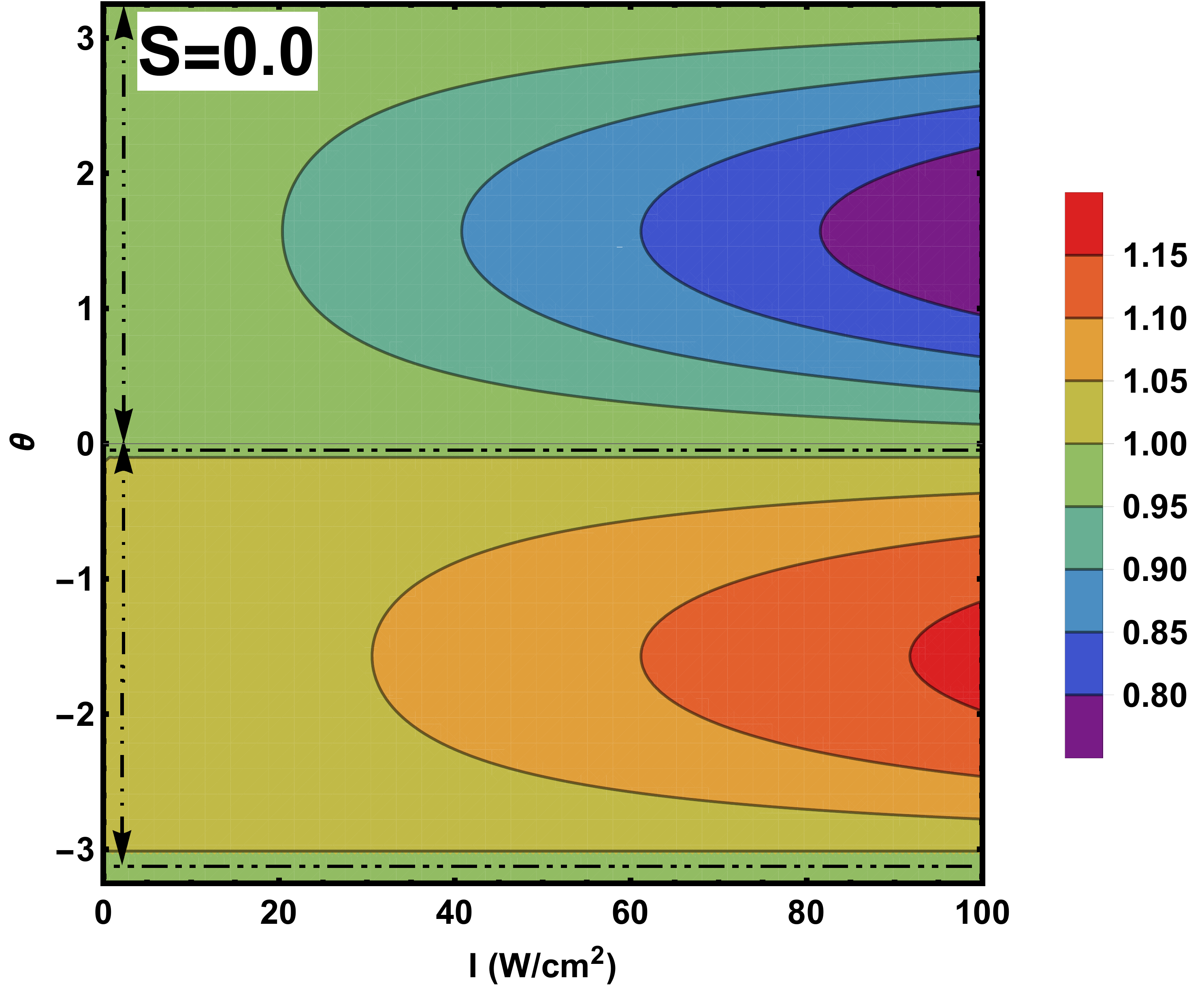}
		\label{figu}}\\
	\subfloat[]{
		\centering
		\includegraphics[scale=0.186]{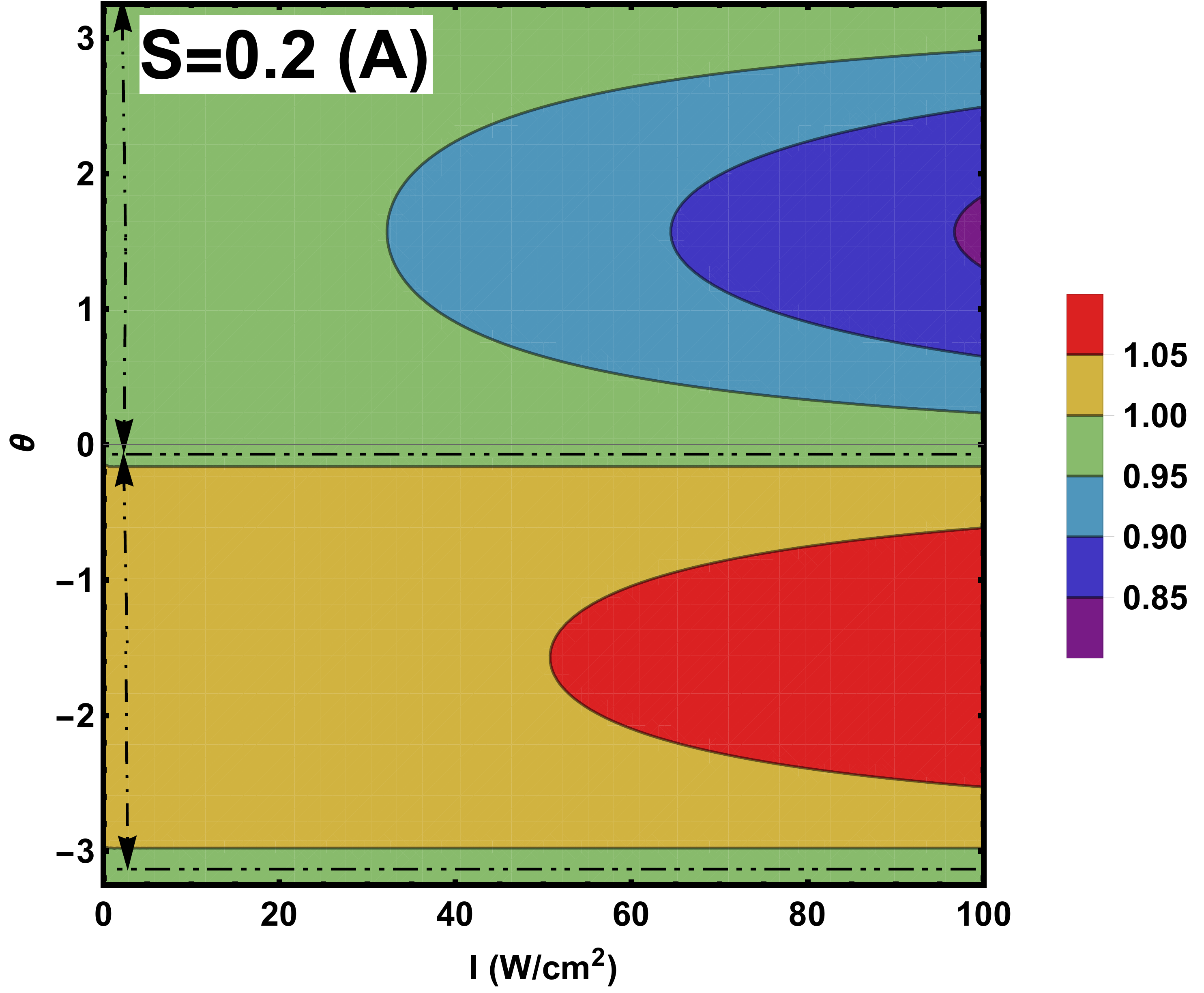}
		\label{fir}}
	\subfloat[]{
		\centering
		\includegraphics[scale=0.186]{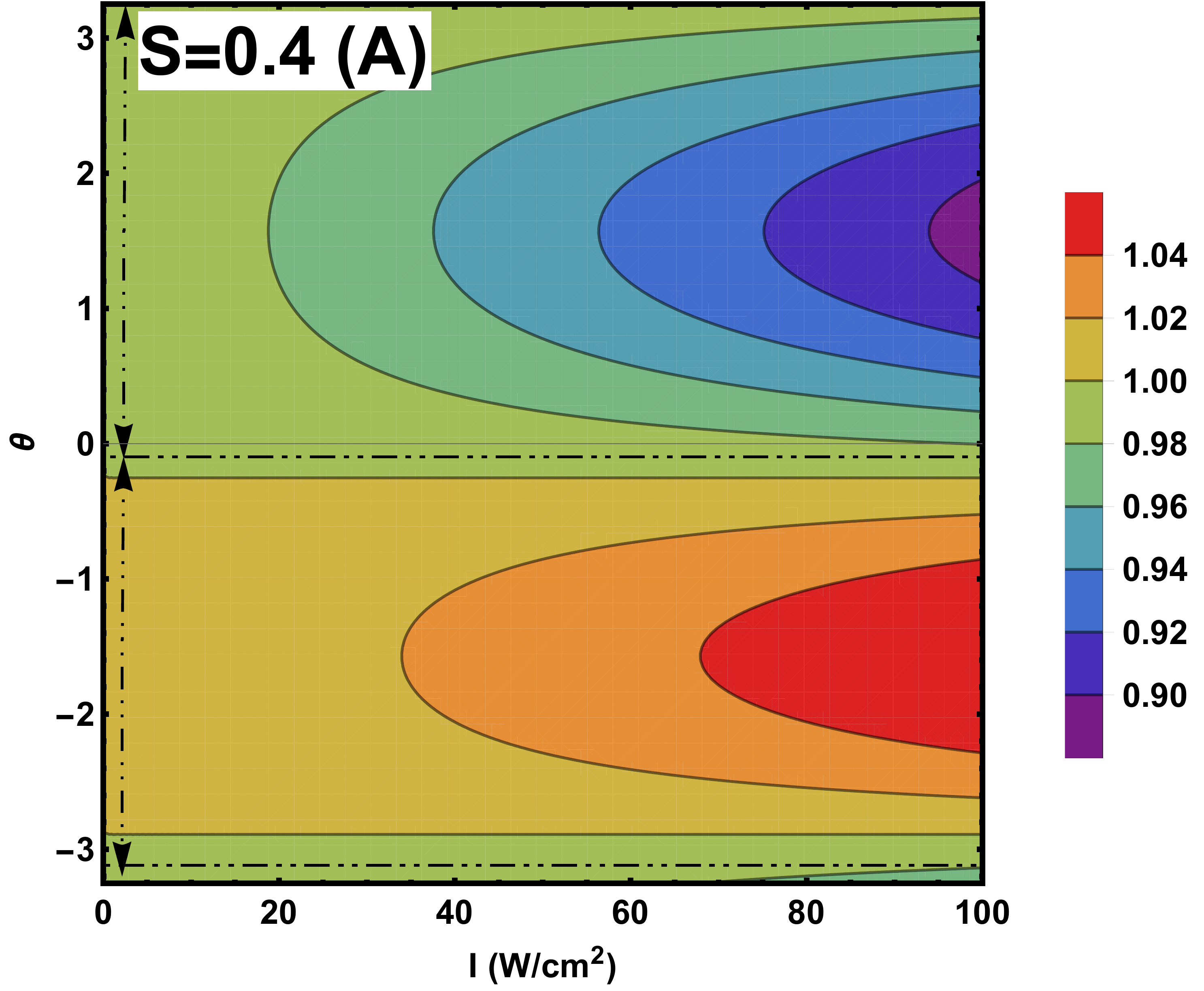}
		\label{fio}}
	\subfloat[]{
		\centering
		\includegraphics[scale=0.186]{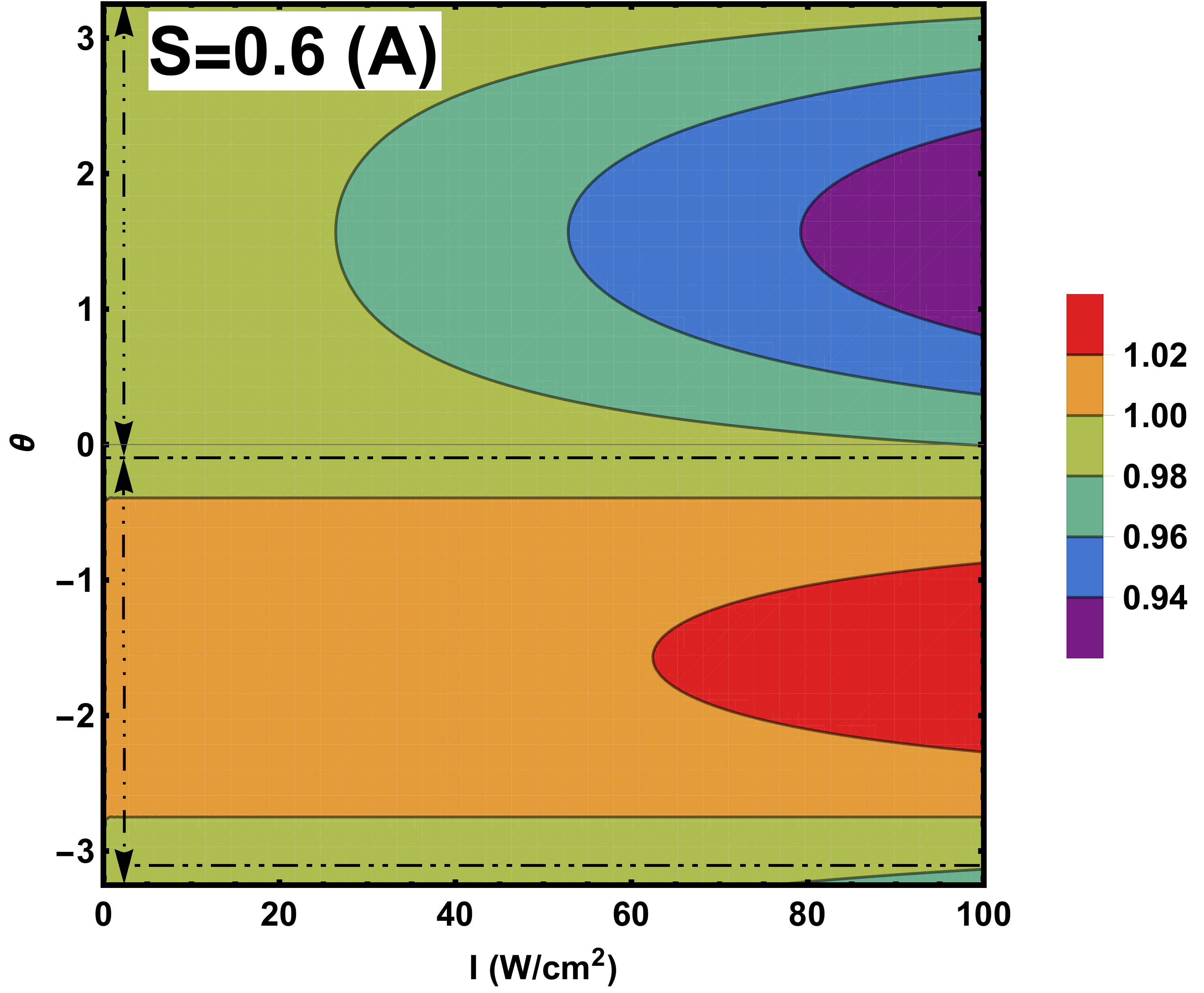}
		\label{firu}}\\
	\subfloat[]{
		\centering
		\includegraphics[scale=0.186]{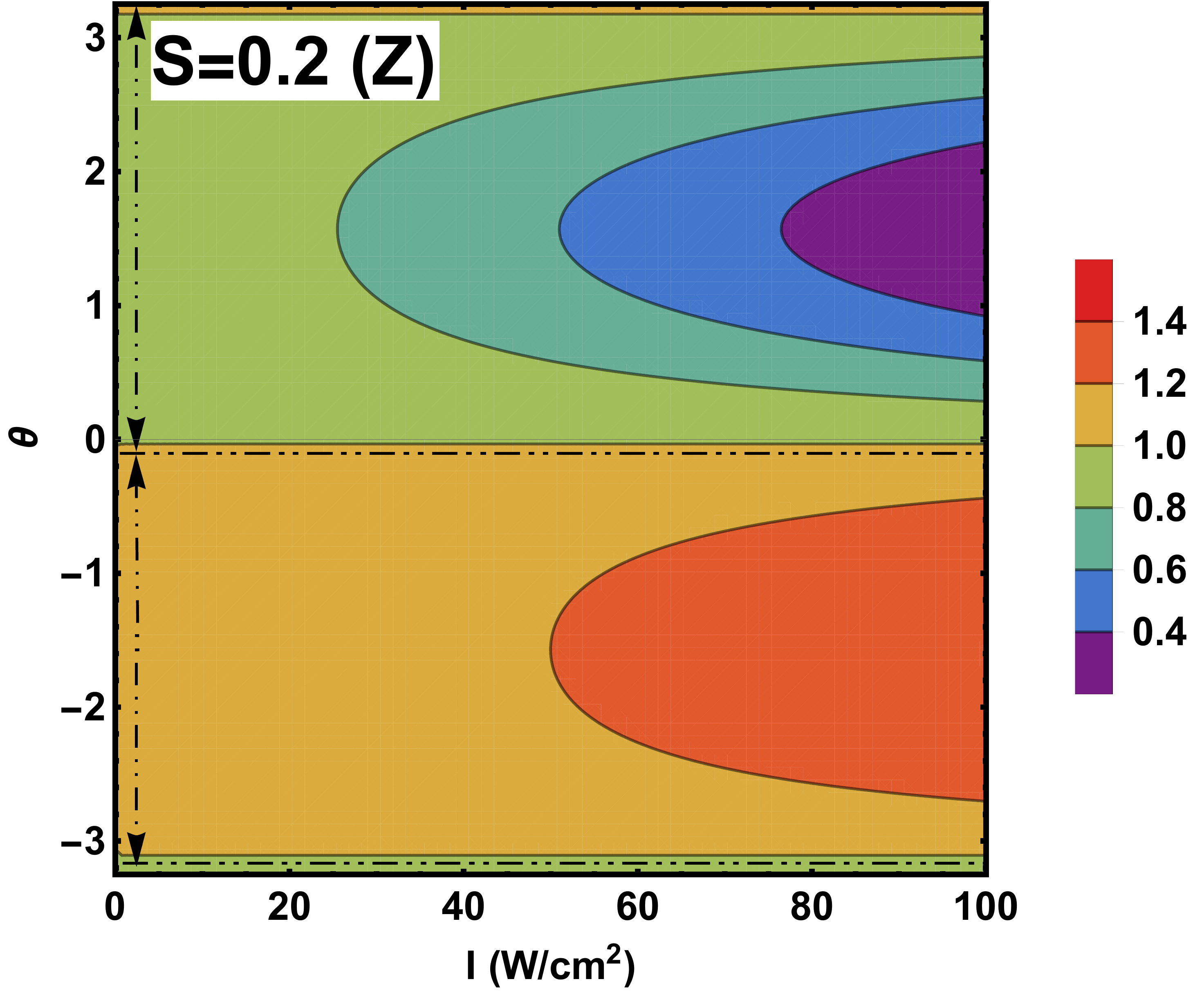}
		\label{fip}}
	\subfloat[]{
		\centering
		\includegraphics[scale=0.186]{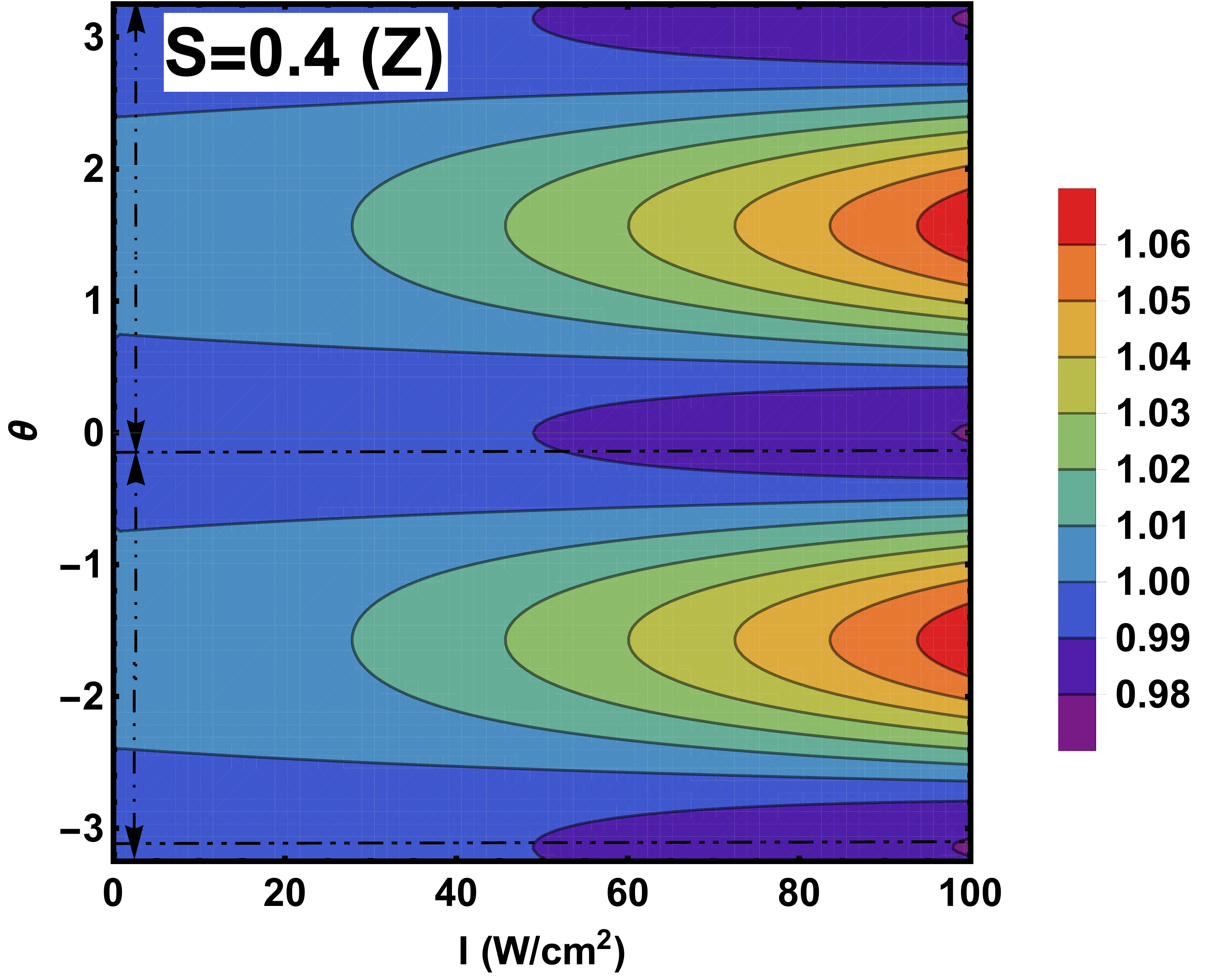}
		\label{fim}}
	\subfloat[]{
	\centering
	\includegraphics[scale=0.186]{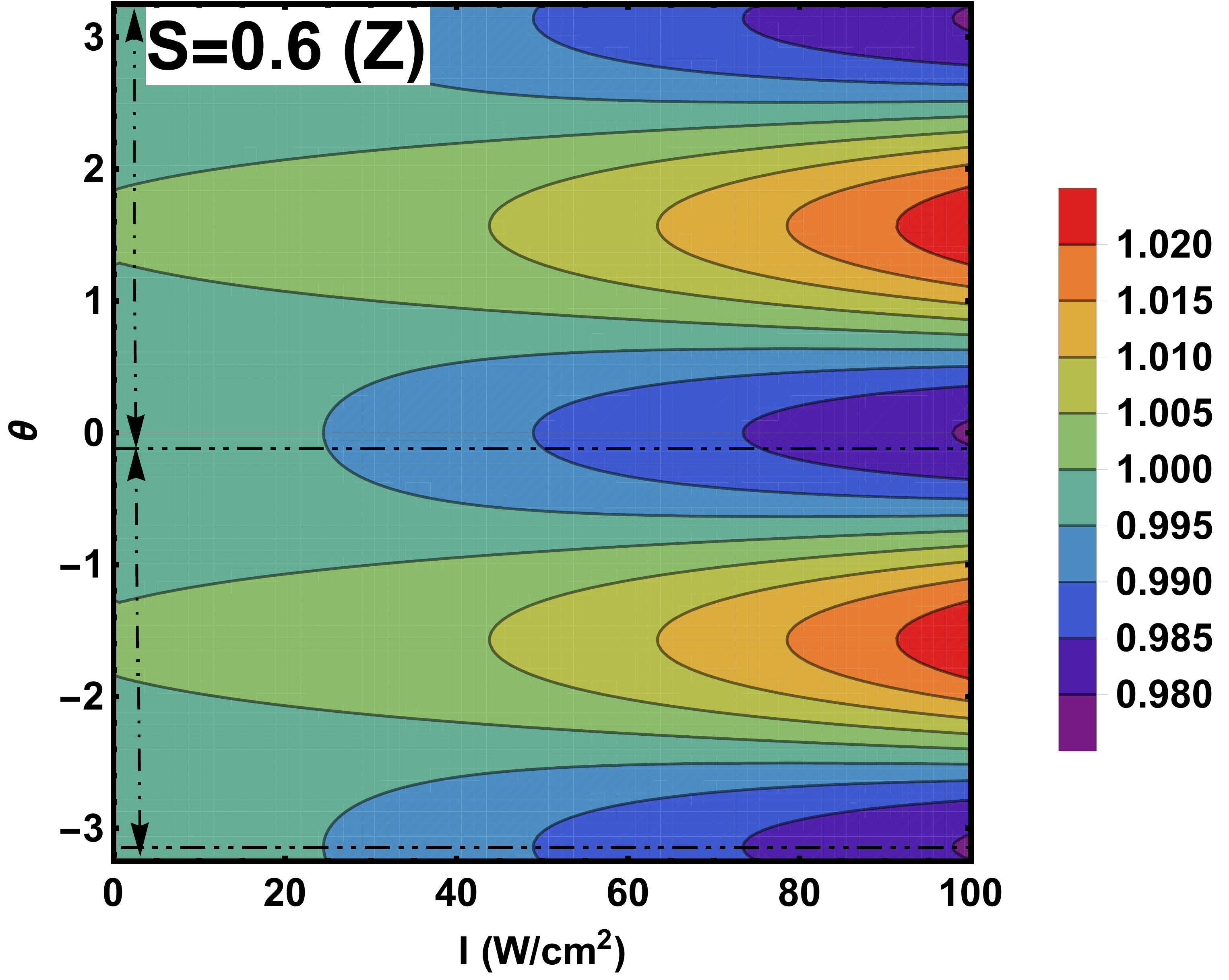}
	\label{fioy}}
\caption{\sf{(color online) Contour plot of the band gap
	$|\Lambda_g/\Delta_g|$ versus the irradiation intensity $I $ and the polarization $\theta$ for {${\Delta_{g}}=2$  \text{meV}, $\hbar\omega=10$  \text{meV}, $\tau=1$}\color{black}{.}}}\label{fig10}
\end{figure}

Figure \ref{fig10} presents the band gap $|\Lambda_g/\Delta_g|$ versus the irradiation intensity $I$ and the polarization $\theta$
for {${\Delta_{g}}=2$  \text{meV}, $\hbar\omega=10$ \text{meV}, $\tau=1$} \color{black}{correspond} to elliptically polarized dressing field. The different colors from purple to red correspond to different values of $|\Lambda_g/\Delta_g|$ from $0$ to a maximum value which varies along the propagation direction induced by the zigzag and armchair strains. Indeed, {in Figure \text{\ref{fig10}}\textbf{\color{blue}{(a)}} for $S=0.0$  and by increasing the  irradiation intensity, we observe two types of the band gap $|\Lambda_g/\Delta_g|$ showing different behaviors such as the first one decreases slowly for positive values of $\theta$, i.e. $\theta>0$ and the second one increases for $\theta<0$. Note  that the form of the band gap does not change whatever the sign of the polarization phases $\theta$ and these are marked by the dotdashed lines, see Figure \text{\ref{fig10}}\textbf{\color{blue}{(a)}}. It is important to mention that this results is similar to that obtained in \cite{Kibis}. When we introduce a armchair strain with $S=(0.2,0.4,0.6)$, we clearly see the disappearance (Figures \text{\ref{fig10}}\textbf{\color{blue}{(b)}}, \text{\ref{fig10}}\textbf{\color{blue}{(d)}})} and appearance (Figures \text{\ref{fig10}}\textbf{\color{blue}{(c)}}) of some band gaps. In addition,  as long as $S$ and $I$ increase, $|\Lambda_g/\Delta_g|$ increases quickly in the interval $0<\theta<3.25$ while it decreases slowly for $\theta<0$ compared to Figure \text{\ref{fig10}}\textbf{\color{blue}{(a)}}. Applying the strain along zigzag direction, one can see that in Figure \text{\ref{fig10}}\textbf{\color{blue}{(e)}} $|\Lambda_g/\Delta_g|$ presents the same behavior as those in Figures \text{\ref{fig10}}\textbf{\color{blue}{(b)}}, \text{\ref{fig10}}\textbf{\color{blue}{(d)}} except that it decreases rapidly for $\theta>0$ but it increases for a large irradiation intensity and takes  $|\Lambda_g/\Delta_g|=1.4$ as a
maximum value. Moreover,  from Figures \text{\ref{fig10}}\textbf{\color{blue}{(g)}}, \text{\ref{fig10}}\textbf{\color{blue}{(h)}} for $S=(0.4,0.6)$ we observe a symmetry at normal incidence ($\theta=0$) separating the absolute values of $|\Lambda_g/\Delta_g|$.  It is clearly see that the increase in strain is accompanied by appearance of other band gaps. \color{black}{We conclude that the band gap can be changed along the armchair and zigzag directions and not changed for the polarization phases $\theta$ compared to linearly and circularly polarized fields which change the gap values oppositely like Figure \ref{fig11}}. 

\begin{figure}[H]
\centering
\centering
\subfloat[]{
\centering
\includegraphics[scale=0.23]{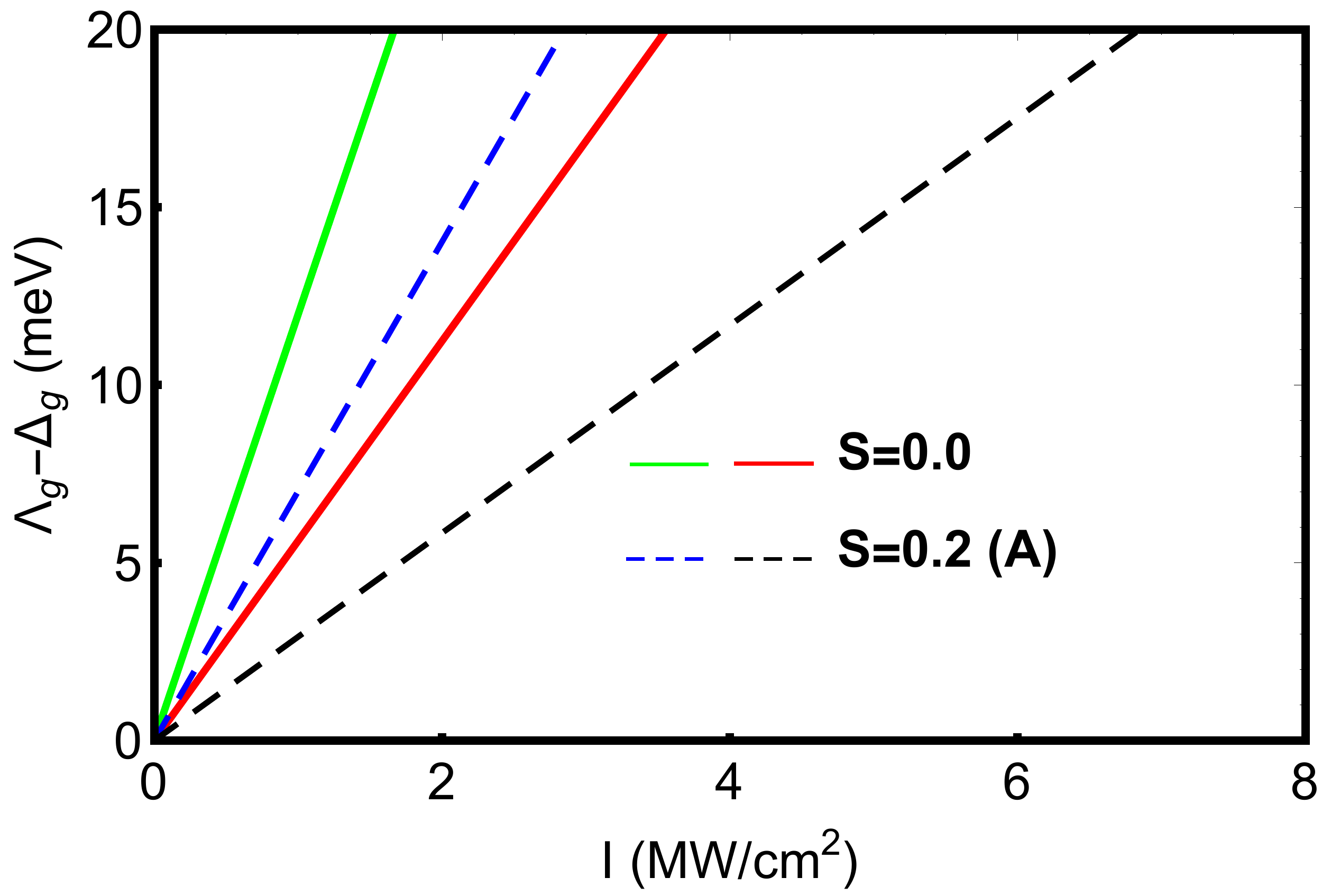}
\label{fartr}}\hspace{2cm}
\subfloat[]{
\centering
\includegraphics[scale=0.23]{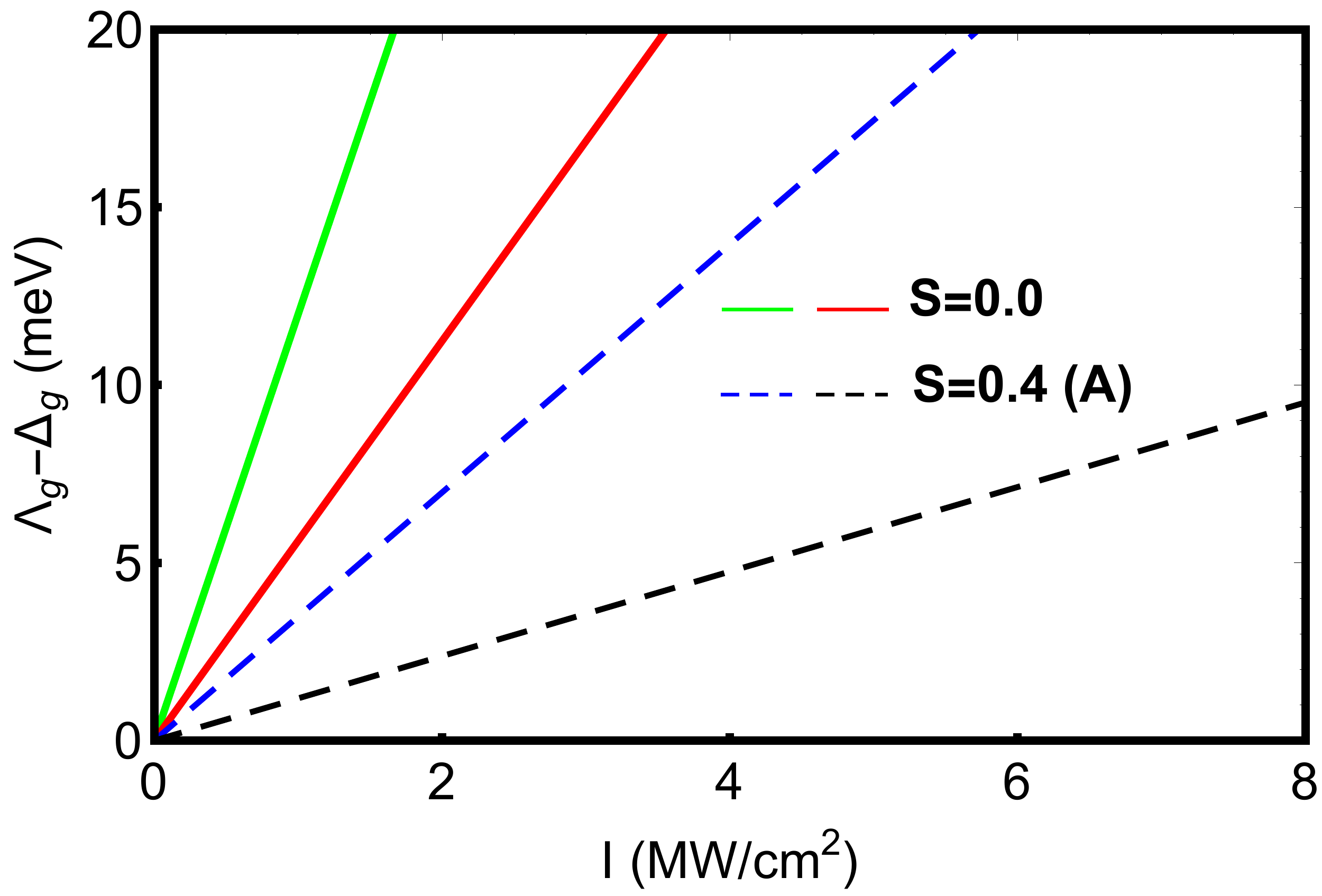}
\label{firrl}}\\
\subfloat[]{
\centering
\includegraphics[scale=0.23]{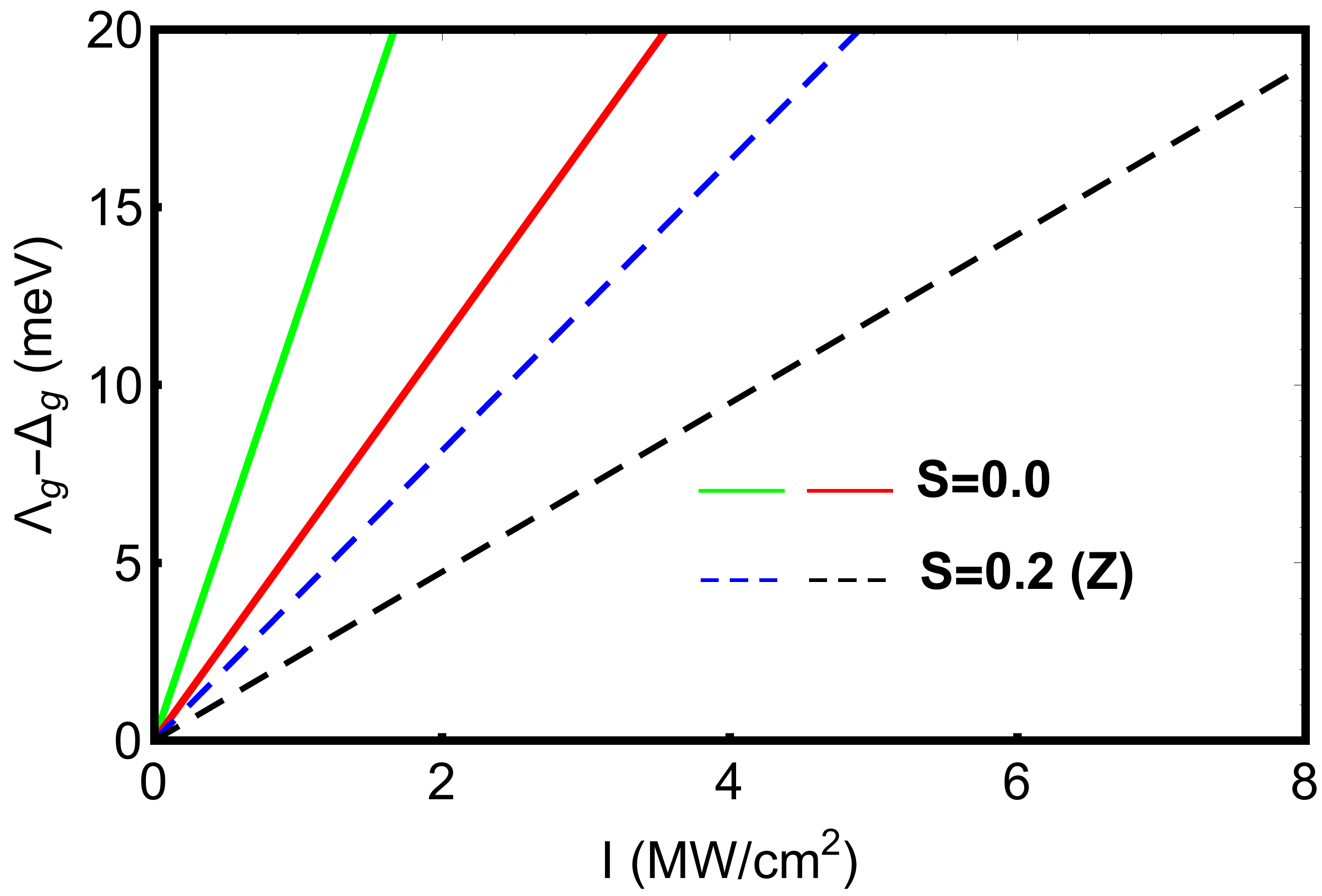}
\label{fira1}}\hspace{2cm}
\subfloat[]{
\centering
\includegraphics[scale=0.23]{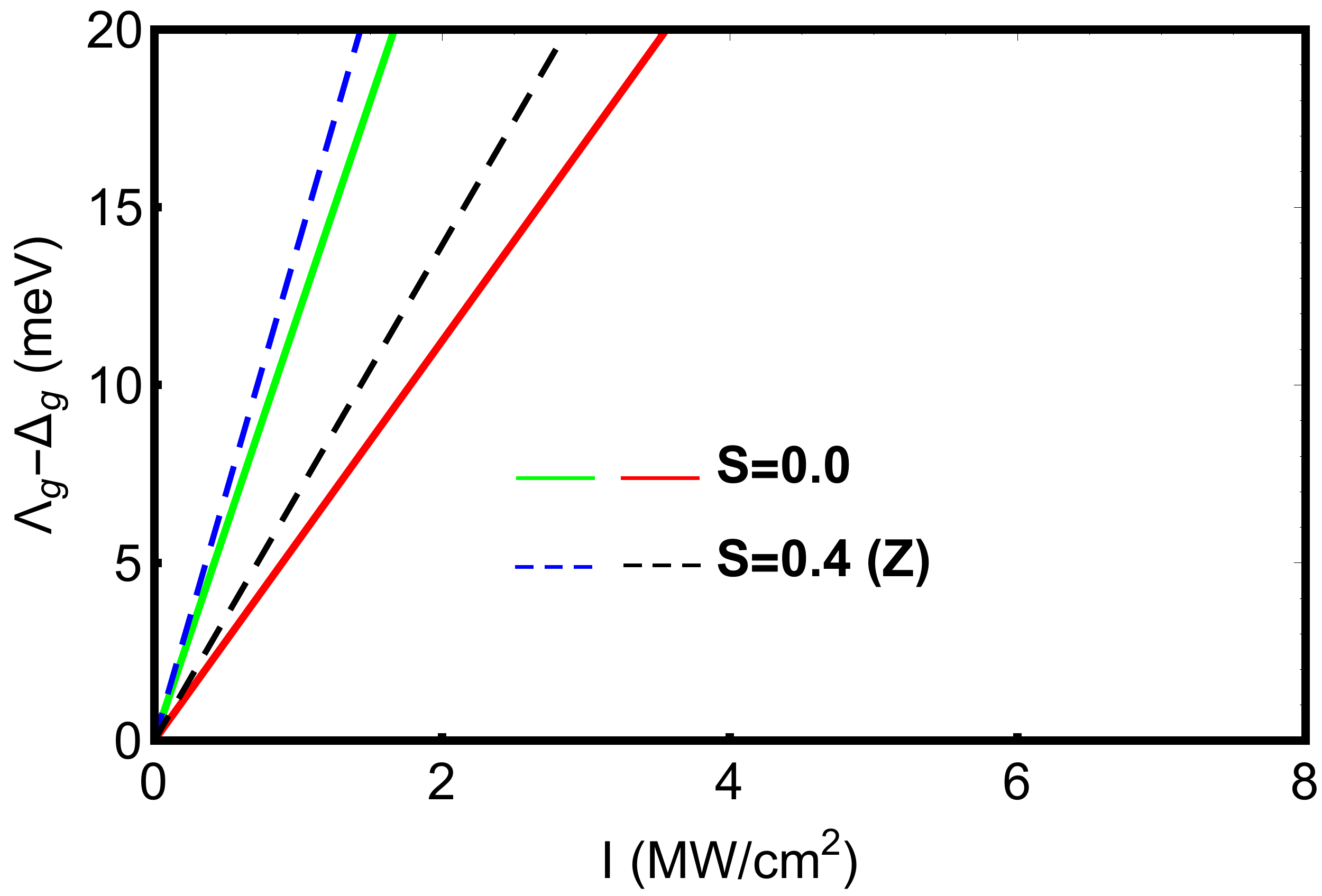}
\label{faty}}
\caption{\sf{(color online) The changing gap $\Lambda_{g}-\Delta_{g}$ versus the irradiation intensity $I$ for {${\Delta_{g}}=2$  \text{meV}, $\hbar\omega=10$  \text{meV}, $\theta={\pi}/{6}$ with $\tau\xi=-1$ (green and blue lines), $\tau=-1$ (red and black lines)}\color{black}{.} 
	\textbf{\color{blue}{(a),(c)}}\color{black}{:} Effect of armchair strain direction with {$S=0.2,0.4$}\color{black}{.}
	\textbf{\color{blue}{(b),(d)}}\color{black}{:} Effect of zigzag strain direction with {$S=0.2,0.4$}\color{black}{.}}}\label{fig1}
\end{figure}

In Figure \ref{fig1}, we investigate the changing gap versus the irradiation intensity $I$ correspond to circularly and elliptically polarized field for {${\Delta_{g}}=2$  \text{meV}, $\hbar\omega=10$  \text{meV}}\color{black}{,} 
$\theta={\pi}/{6}$ with $\tau\xi=-1$ ({green and blue} lines), $\tau=-1$ ({red and black} lines). It is clearly shown that for ($S=0.0$), the values of $\Lambda_{g}-\Delta_{g}$ increase as long as $I$ increases. In Figures \text{\ref{fig1}}\textbf{\color{blue}{(a)}}, {\ref{fig1}\textbf{\color{blue}{(b)}}} \color{black}{when} the strain is applied along the armchair direction we observe that  the two curves are
shifted {to the right} \color{black}{and} for the elliptically polarized
electromagnetic wave {for $S=0.4$,} \color{black}{$\Lambda_{g}-\Delta_{g}$} decreases rapidly by increasing the values of
the $I$  where it becomes equal {$\Lambda_{g}-\Delta_{g}=9.5$ \text{meV} for the value $I=8$  \text{MW/\text{cm$^{2}$}}}\color{black}{.} In Figures
\text{\ref{fig1}}\textbf{\color{blue}{(c)}}, \text{\ref{fig1}}\textbf{\color{blue}{(d)}} there is significant change {if the strain is along zigzag direction}  \color{black}{because}  $\Lambda_{g}-\Delta_{g}$ starts increasing for both polarization but {for $S=0.4$, it exhibit a translation to left as we increase the irradiation intensity}\color{black}{.}

\section{Conclusion}

We have studied the electron-field interaction in gapped graphene  subjected to the tensional strain applied along armchair and zigzag directions. 
By applying the  Floquet
theory, we have  analytically determined the effective Hamiltonian in terms of strain, band gap and valley index for linearly,
circularly and elliptically polarized dressing field. Solving Dirac equation,
we have obtained the solutions of  energy spectrum as function of the physical parameters characterizing our system.

Subsequently, we have discussed our results numerically for various choices of the physical parameters. Indeed, we have investigated the energy spectrum for two directions of strain
including zigzag and armchair as function of the wave vectors and the irradiation intensity. It is observed that for the linear polarization there is a symmetry separating positive and negative behavior of $\varepsilon$ as in the case of the pristine graphene. Also the energy spectrum {for the strainless case takes an isotropic and anisotropic forms} {and} {increases}{slowly} when strain is along the armchair direction but rapidly for the zigzag case with increasing the values of irradiation intensity.
The contour plot of the energy spectrum was illustrated and we also have noticed through it that $\varepsilon$ decreases for $\gamma=1$ but increases for $\gamma=-1$. As a results, we have found that the armchair strain direction causes some changes  on the energy spectrum of dressed electron while the  zigzag strain direction  produces remarkable influence.

Interesting numerical results concerning the renormalized band gap have been reported. It was shown that the band gap decreases and turns to zero by dressing field, then it increases slowly but becomes null at certain values of the irradiation intensity. Whereas, for the circular polarization we have observed that in the case $\tau \xi=-1$, the band gap monotonously
increases but decreases to zero and starts to grow up for $\tau\xi=1$. By applying the zigzag strain direction, we have showed that as long as $I$ increases the renormalized band gap for the both polarization decreases and equal  zero. Furthermore, for the elliptically polarized dressing field the band gap drops
dramatically by altering the strain magnitude and did not change for the polarization phase $\theta$.

\section*{Acknowledgment}
The generous  support provided by the Saudi Center for Theoretical Physics (SCTP) is highly appreciated by all authors. {We are  indebted to the referee’s for their instructive comments.}

\section*{Author contribution statement}

	All authors contributed equally to the paper.


\end{document}